\documentclass[preprint, times]{elsarticle}

\usepackage{epsfig} 
\usepackage{graphics}
\usepackage{xspace}
\usepackage{amsmath}
\usepackage{subfigure}
\usepackage{booktabs}
\usepackage{color}
\usepackage{algorithm}
\usepackage{algpseudocode} 
\usepackage{hyperref} 
\usepackage{soul}
\usepackage{amssymb}

\definecolor{Red}{rgb}{0.7,0.0,0.0}
\definecolor{Green}{rgb}{0.0,0.7,0.0}
\definecolor{Blue}{rgb}{0.0,0.0,0.7}

\newcommand{\ie}{\textit{i.e.}\xspace}

\newcommand{\et}{\emph{et al.}\xspace}

\renewcommand{\figurename}{Fig.\xspace}
\newcommand{\sectionname}{Sec.\xspace}
\newcommand{\equationname}{Eq.\xspace}

\journal{Physics Reports}

\begin{document}

\begin{frontmatter}

\title{Human Mobility: Models and Applications}

\author[ros]{Hugo Barbosa-Filho}
\ead{hbarbosafilh2011@my.fit.edu}

\author[cea,ehess]{Marc Barthelemy}
\ead{marc.barthelemy@cea.fr}

\author[ros]{Gourab Ghoshal\corref{cor}}
\ead{gghoshal@pas.rochester.edu}

\author[bri]{Charlotte R. James}
\ead{charlotte.james@bristol.ac.uk}

\author[isrtea]{Maxime Lenormand}
\ead{maxime.lenormand@irstea.fr}

\author[geocites]{Thomas Louail}
\ead{louail@parisgeo.cnrs.fr}

\author[fit]{Ronaldo Menezes\corref{cor}}
\ead{rmenezes@cs.fit.edu}

\author[ifisc]{Jos\'e J. Ramasco}
\ead{jramasco@ifisc.uib-csic.es}

\author[bri]{Filippo Simini}
\ead{f.simini@bristol.ac.uk}

\author[fit]{Marcello Tomasini}
\ead{mtomasini@my.fit.edu}

\cortext[cor]{Corresponding authors}

\address[ros]{Department of Physics \& Astronomy and Goergen Institute of Data Science, University of Rochester, Rochester, NY, USA}

\address[cea]{Institut de Physique Th\'eorique, CEA, CNRS-URA 2306, F-91191, Gif-sur-Yvette, France}

\address[ehess]{Centre d'Etudes et de Math\'ematique Sociales, EHESS, Paris 75006, France}

\address[bri]{Department of Engineering Mathematics, University of Bristol, UK}

\address[isrtea]{Irstea, UMR TETIS, 500 rue JF Breton, 34093 Montpellier, France}

\address[ifisc]{Instituto de F\'{\i}sica Interdisciplinar y Sistemas Complejos IFISC (CSIC-UIB), Campus UIB, 07122 Palma de Mallorca, Spain}

\address[geocites]{CNRS, UMR 8504 G\'eographie-cit\'es, 13 rue du four, F-75006 Paris, France}

\address[fit]{BioComplex Laboratory, School of Computing, Florida Institute
  of Technology, Melbourne, USA}

 \begin{abstract}
 Recent years have witnessed an explosion of extensive geolocated datasets related to human movement, enabling scientists to quantitatively study individual and collective mobility patterns, and to generate models that can capture and reproduce the spatiotemporal structures and regularities in human trajectories. The study of human mobility is especially important for applications such as estimating migratory flows, traffic forecasting, urban planning, and epidemic modeling. In this survey, we review the approaches developed to reproduce various mobility patterns, with the main focus on recent developments. This review can be used both as an introduction to the fundamental modeling principles of human mobility, and as a collection of technical methods applicable to specific mobility-related problems. The review organizes the subject by differentiating between individual and population mobility and also between short-range and long-range mobility. Throughout the text the description of the theory is intertwined with real-world applications. 
  
  \end{abstract}


\end{frontmatter}





\tableofcontents

\section{Introduction}
\label{sec:intro}

\subsection{Motivation and History}
While the term \emph{mobility} has multiple connotations, in the context of this review it refers to the movement of human beings (individuals as well as groups) in space and time and thus implicitly refers to \emph{human mobility}. Indeed, from the migration of {\it Homo sapiens} out of Africa around 70,000 years ago, through the European discovery of the ``New World", to the existence of expatriate populations in contemporary times, the existence of human beings has always been inextricably linked with their movement. While earlier movement patterns were primarily driven by factors such as climate change, inhospitable landscapes, conflict and food scarcity, in modern times, socio-economic factors such as wage imbalance, differences in welfare and living conditions and globalization play an increasing role. 

From hunters and gatherers of prehistoric times to present-day commuters of large metropolitan areas, humans are bound to move on a daily basis in order to earn their livelihood. However, daily trips are also undertaken to perform social and leisure activities. 
The temporal and spatial scales of these trips are much shorter than those of migratory flows, and they are often characterized by the regularities and periodicities that mark human lives. 
The daily movement of a large and growing population has important impact on the lives of the individuals and the environmental conditions. Studies conducted in Europe and the United States found that the average household spending on transportation is between 15 and 25 percent of the total expenditures, making transportation the second largest expenditure category after housing. Transportation is also the second source of greenhouse gas emissions to the atmosphere. 
From these few examples it should be clear that mobility has an enormous impact on human societies and an accurate quantitative description of human mobility is of fundamental importance to understand the processes related to human movement and their impact on the community and the environment. 
A quantitative theory of human mobility ought to be able to provide answers to relevant questions, such as, what determines the decision to start a trip? which factors determine the choice of the destination? to what extent is human movement predictable, and what is the intrinsic degree of randomness? is it possible to find general rules or laws to explain empirical patterns and regularities exhibited by travels in many diverse countries, such as the distribution of commuting times and distances, and the degree of predictability of future whereabouts?


While the study of human mobility currently spans several disciplines, arguably, geography was the first discipline to analyze mobility data and put forward corresponding theories to describe travel patterns. Indeed, 
the pioneers of quantitative and theoretical geography in the 50's defined ``geography as (the scientific discipline of) \emph{spatial interaction}" \cite{ullman_1980_geography}. Early quantitative studies of the movements of people and vehicles were held in large US metropolitan areas \cite{helvig_1964_chicago} (see the first chapters of  \cite{boyce_2015_forecasting} for details), and initial studies of human travel were of scales ranging from international migrations \cite{reilly_1929_methods,stouffer_1940_intervening,zipf_1946_p1,anderson_1955_intermetropolitan} to journey-to-work commuting \cite{hanson_1980_importance,huff_1986_repetition,kitamura_2000_micro,bhat_2004_comprehensive,pendyala_2005_florida}. Indeed, the elucidation and understanding of these patterns was motivated by its relation to several real-world applications such as traffic forecasting \cite{boyce_2015_forecasting, nagel_1995_emergent, wang_2012_understanding}, urban planning \cite{hillier_2009_metric,kitamura_2000_micro}, internal security \cite{krebs_2002_mapping,clauset_2010_strategic} and epidemic modeling \cite{colizza_2007_modeling,vespignani_2012_modelling,tizzoni_2014_use} to name but a few. 

The first systematic analysis of the concept of distance as a constraint to movement was proposed in the 19$^\text{th}$ century:
in his 1965 review~\cite{olsson_1965_distance} Gunnar Olsson cites Henry C. Carey's \textit{Principles of Social Science} (1858) as the first work to explicitly make the observation about the  amount of interaction between two cities being proportional to their population size and inversely proportional to the intervening distance. Few decades later, the geographer Ernst Ravenstein further developed and popularized the idea in a seminal work where he formulated his \emph{laws of migration}~\cite{ravenstein_1885_laws}. 
Further refinements on this theme were made in the 1940's by the American sociologist Samuel Stouffer~\cite{stouffer_1940_intervening, bright_1941_interstate} in his \emph{law of intervening opportunities}, and by the American linguist George Kingsley Zipf~\cite{zipf_1946_p1}.
Zipf's formulation led to what is now conventionally referred to as the \emph{gravity law}. The increasing availability of datasets on population movements at various levels of granularity, coupled with the \emph{quantitative revolution} in geography \cite{schaefer_1953_exceptionalism,berry_1993_geographys,adams_2001_quantitative}, led to the introduction of more sophisticated mathematical methods such as hidden Markov models and diffusion processes. Gender and socio-economic factors behind population movement were further analyzed thanks to richer datasets resulting from surveys and interviews \cite{ericksen_1977_analysis,hanson_1981_travel,hanson_1985_gender}, and through theories of labor economics focusing on wage differential between locations~\cite{jennissen_2007_causality}. Thus through the 20$^\text{th}$ century, contributions to the theories of human movement were made chiefly in geography, 
sociology, and economics, while the scale at which this was primarily studied was at the population level.


To provide context for what is to follow, we briefly describe a selection of influential historical contributions, keeping in mind ``Stigler's law of eponymy'' which states: ``No scientific discovery is named after its original discoverer''
~\cite{stigler_1997_statistics}). However, if we restrict ourselves to the recognition of \emph{distance} as a primary factor in determining movement and interactions between places, then it is reasonable to start with the work of Ernst Georg Ravenstein~\cite{ravenstein_1885_laws}.\\

\noindent {\bf Laws of Migration.}~Ravenstein was a German-English geographer who made important contributions to cartography as well as providing one of the first rough estimates of a ``maximal'' global population based on resource consumption. He was also one of the first to attempt an explanation and prediction of migration patterns within and between countries. Considering the effect of distance as well as the type of migrant (male or female, old or young) as primary factors, he posited the following seven laws:
\begin{enumerate}
\item Most migrants only travel short distances, and ``currents of migrations'' are in the direction of the great centers of commerce and industry given that these can absorb the migrants.
\item The process of absorption occurs in the following manner: inhabitants of the areas immediately surrounding a rapidly growing town flock to it, thus leaving gaps in the rural areas that are filled by migrants of more remote districts, creating migration flows that reach to ``the most remote corner of the kingdom''.
\item The process of dispersion is inverse to that of absorption, and exhibits similar features.
\item Each main current of migration produces a compensating counter-current.
\item Migrants traveling long distances generally go by preference to one of the great centers of commerce or industry.
\item The natives of towns are less migratory than those of the rural parts of the country.
\item Females are more migratory than males.
\end{enumerate}
Ravenstein added another two laws in 1889 \cite{ravenstein_1889_laws}: 
\begin{enumerate}
  \setcounter{enumi}{7}
\item Towns grow more by immigration than by natural increase.
\item The volume of migration increases as transport improves and industry grows.
\end{enumerate}
%


While the laws are non-quantitative and observational in character, Ravenstein correctly identified socio-economic factors as well as distance-constraints to be the essential ingredients behind modern population movement. Consequently, his laws stimulated an enormous volume of work, and although they have been refined and adjusted over the years, the essential ingredients of his formulation remain relevant even today.\\

\noindent{\bf Law of Intervening Opportunities.} One of the most important refinements was made in the 1940's by the American sociologist Samuel Stouffer~\cite{stouffer_1940_intervening}. Roughly speaking, Stouffer was looking to expand upon Ravenstein's observations regarding migrants moving shorter distances and flocking to commercial centers. To account for this, he proposed that \emph{the number of people going a given distance is directly proportional to the number of opportunities at that distance and inversely proportional to the number of intervening opportunities}. In other words, trips between two locations are driven primarily by relative accessibility of socio-economic opportunities that lie between those two locations. In this context, \emph{opportunity} is defined as a potential destination for the termination of a traveler's journey, whereas an \emph{intervening opportunity} is one that the traveler rejects in favor of continuing on. In Stouffer's original formulation, this can be mathematically expressed as
\begin{equation}
\frac{d y(r)}{dr} \propto \frac{1}{x}\frac{d x(r)}{dr},
\label{io_eq1}
\end{equation}
where $y(r)$ is the cumulative number of migrants that move a distance $r$ from their original location, and $x(r)$ is the cumulative number of intervening opportunities. 
Assuming that $x$ itself is a continuous function $x(r)$ of distance, then the expression above can be integrated to yield
\begin{equation}
y(r) = \log x(r) + C,
\label{io_eq2}
 \end{equation}
where $C$ is some constant denoting the number of opportunities at the origin location. Thus, the relationship between mobility and distance is indirect and is established only through an auxiliary dependence via the intervening opportunities: the higher the number of intervening opportunities between two locations at distance $r$, the smaller the number of migrants that would travel that distance.
This may explain why rural migrants may flock to urban centers over large distances, whereas those already in commercial centers are comparatively more stationary.\\

\noindent {\bf Distance-Decay and the Gravity Law.} Around the same time as Stouffer, the Harvard Philologist, George Kingsley Zipf, was expanding upon his famous observation of the rank-frequency dependence in linguistics; the eponymous Zipf's law, where the frequency of a word ranked $z$---in terms of usage---has the statistical dependence $f_z \sim 1/z$~\cite{zipf_1936_psychobiology}. Zipf found that this relation was expandable to other realms of society, specifically the size of cities~\cite{zipf_1940_generalized}, where the occurrence of a city with population $P$ and consequent rank $z$ also follows the relation 
\begin{equation}
P_z \sim 1/z^{\alpha}.
\label{eq:zipflaw}
\end{equation}
Broadly speaking, Zipf's argument for this relation was due to the tension between two competing factors. The first, which he called \emph{Force of Diversification} relates to the likelihood of populations living close to the source of raw materials (commodities) in order to minimize the cost of transportation to production centers. The second effect, referred to as the \emph{Force of Unification}, is the tendency of populations to aggregate in urban centers due to the minimization of work required to transport finished products to consumers. While the former leads to the formation of multiple urban centers (given that the commodity sources are not localized in one part of a country) with smaller populations, the latter has the competing effect of urban agglomeration in a \emph{few} centers of large population. Assuming some kind of equilibrium between these quasi-forces, the rank-frequency relation of \equationname~\eqref{eq:zipflaw} with $\alpha = 1$ naturally follows~\cite{zipf_1941_national}. Deviations from equilibrium, where one force dominates over the other, then leads to a change in the exponent.  

Carrying the argument further, under somewhat unrealistic assumptions of equitable share of national income as well as urban centers being self-sufficient (i.e. production and consumption at equal levels), the share of any center $i$ in the total flow of goods is proportional to its population $P_i$. Therefore, the flow of goods between two centers is proportional to the product of their populations. Finally, if one would like to minimize the cost and work associated with transportation of goods, this flow must be inversely proportional to the distance between centers. Putting all this together we arrive at the relation
\begin{equation}
w_{ij} \propto \frac{P_i P_j}{r_{ij}},
\label{eq:dist-decay}
\end{equation}
where $w_{ij}$ represents the flow of goods between two population centers $i,j$ and $r_{ij}$ is the distance between these two centers. Zipf tested his theory on both freight and population movement data and got good qualitative agreement. This formulation is of course quite different form Stouffer's explanation as the effect of distance is quite explicit in \equationname~\eqref{eq:dist-decay}. Indeed, the form of \equationname~\eqref{eq:dist-decay} is such that it leads to a distance-decay effect suppressing long-range movement. Furthermore, the flow of populations and goods is seen to be as a result of some dynamic equilibrium between the cost of transportation, manufacture, and distribution of goods and services. The references to ``forces'' as well as its functional dependence on distance eventually led to Zipf's formulation being dubbed the \emph{Gravity law} in analogy to Newtonian mechanics.

The common theme connecting Ravenstein, Stouffer and Zipf of course is the geographical distance, though its functional effect on movement is quite different in the Intervening Opportunities and Gravity Law models. Nevertheless, both these models were quite influential on subsequent work, setting off two major strains of parallel research as well as attempts at unification.\\ 

\begin{figure}[tpb]
  \centering
\includegraphics[width=.8\textwidth]{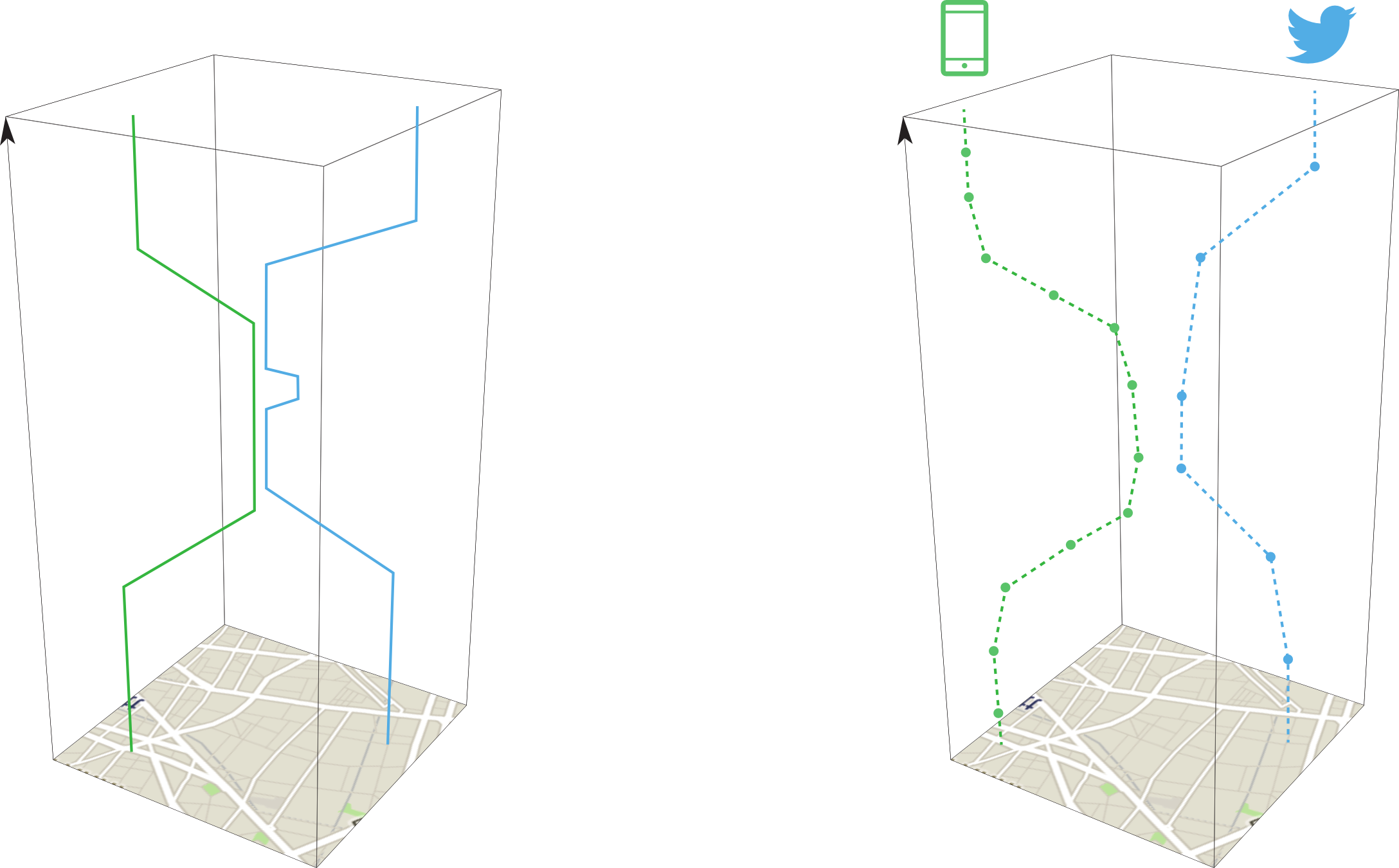}
  \caption{The cubes of time geography, as first proposed by Torsten H\"agerstrand in \cite{hagerstraand_1970_what}. The geographical space is represented by the 2D plan, while time is figured by the vertical axis. (Left) The two curves represent the daily space-time trajectories of two individuals living in the same neighborhood and working in the same place. (Right) The geographical footprints continuously and passively produced by individuals through the use of their ICT devices allow to approximate their trajectories. While these re-constructed trajectories are partial and contain errors that might mislead the understanding of underlying trajectories, they are nonetheless more precise nowadays than they were 10 years ago, and produced by a constantly growing number of individuals worldwide.}
\label{fig:timegeo-cube}
\end{figure}

\noindent {\bf Time geography and ICT Data.} Measuring, understanding and forecasting the displacements of individuals in space and time has long been part of the program of quantitative and theoretical geography, a branch of geography ``born'' academically in the 1950's \cite{berry_1993_geographys,adams_2001_quantitative}. While the first efforts in capturing human displacements focused on the aggregated levels of flows between spatial units, some also focused on individual trajectories. Torsten H\"agerstrand, a Swedish geographer, laid down in the early 1950's the basis of \emph{time geography}, and brought a number of conceptual and graphical tools to formalize the trajectories of individuals through space and time. His seminal work \cite{hagerstraand_1970_what} remains famous for its proposal to represent individual trajectories in a cube (also known as the ``space-time aquarium''), in which the horizontal plane represents the geographical space, while the vertical axis represents time (as depicted in Fig.~\ref{fig:timegeo-cube}). 
H\"agerstrand proposed a number of graphical conventions (a ``notation system" in his own words) to picture the constraints imposed by social life on individual daily trajectories. He also provided means to represent the co-presence and synchronization of several individuals in space, and more generally a set of (essentially graphical) conventions useful to represent the structure and behavior of individual human mobility. 

Time geography was naturally invoked (and somewhat rediscovered) in the 1990's, when the modeling of human mobility shifted towards individual-based simulation (micro/multi-agent/agent-based simulation)~\cite{chardonnel_2007_time}. This can be understood in the twin contexts of increased computing power and the development of more expressive programming frameworks, allowing for semantically richer and more ambitious models of human dynamics. However, while the models increased in complexity, they were somewhat artificial;
relevant data of comparable complexity and resolution was not available for their calibration. Indeed, while the models progressed, the data lagged behind and the best one had to work with was longitudinal survey data collected since the 1970's. 

The beginning of the 21$^\text{st}$ century saw the introduction and subsequent widespread adoption of mobile phones, as well as the pervasive usage of Global Positioning System (GPS). This led to an exponential increase in data-generation on human movement. Coupled with further progress in computing power and sophisticated data-mining methods, it enabled to capture the movement not just of populations at finer levels of spatial granularity, but potentially of \emph{individuals}. In particular, the large volume and frequency of Call Detail Records (CDRs) from mobile phones (see section \ref{sec:cdr}) enabled the analysis of human movement at very fine \emph{temporal} scales. Thus statistical information on mobility became available, on the scale of hours to decades; from the individual, through communities to the level of country-wide populations.

\medskip

\begin{figure}[t!]
\includegraphics[width=0.9\textwidth]{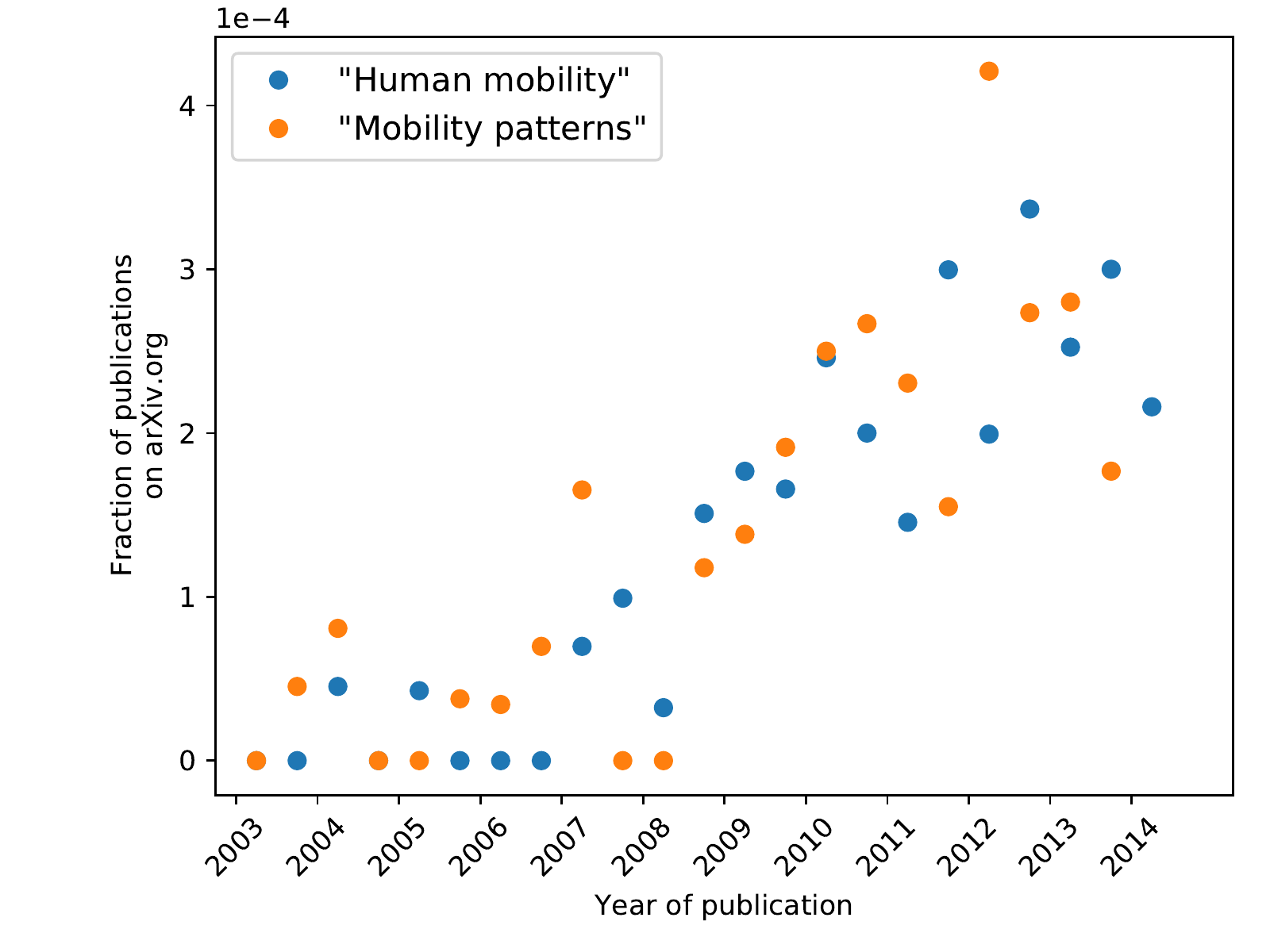}
\caption{Fraction of papers on arXiv.org with mention to the terms "Human Mobility or "Mobility Patterns" from 2004 to 2015. The growth in the number of published papers displays a dramatic increase around 2008. Data source: arXiv.org (last date accessed: Feb 16, 2017).}
\label{fig:pubs}
\end{figure}

This review predominantly focuses on these later developments. As we will see throughout the text, these new sources of data have rejuvenated the scientific interest for human mobility, opening the door to new questions and measures, as well as enhancing and validating the insight gained from studying traditional data sources. For reasons which are probably technical, historical, and societal, research teams in physics and computer science have broadly ``invested" in these new georeferenced (meta)data resulting from human activity~\cite{osullivan_2015_do}. This is evidenced by a quick search on the online website arXiv\footnote{http://arxiv.org} for physics papers that include the terms ``Human Mobility'' or ``Mobility Patterns''. As seen in \figurename~\ref{fig:pubs}, since about 2004-2005, the number of such papers displays an almost exponential increase. One must keep in mind that this is probably an incomplete sample, and is underestimating the volume of research in this field. It is likely that one would see an even more dramatic trend if a more comprehensive list of publications 
were to be included.

While this growing interest of the Physics community in human mobility has multiple reasons, it is somewhat to be expected given the traditional interest of statistical physicists in studying emergent collective properties of systems of many interacting particles and in describing the dynamics of randomly moving particles undergoing (anomalous) diffusion. The availability of data at multiple spatio-temporal scales is helping uncover robust statistical patterns as well as aiding the development of phenomenological theories of human mobility, that are well suited to the tools, methods and paradigms of \emph{statistical physics}. 

\subsection{Scope and Limitations}

The recent pace of developments and the volume of work published are such that a comprehensive review of the findings appeared necessary. Moreover, the study of human mobility is a highly interdisciplinary endeavor and progress has occurred in parallel across different academic communities, sometimes with one unaware of the others' works. This is reflected in different terminology of common metrics and similar results being cast in different and seemingly unrelated contexts. Consequently with this review, we also aim at \emph{(i)} bringing disparate communities together by unifying the findings in a common context, and \emph {(ii)} providing new researchers in the field with
an accessible starting point and a minimal set of tools, metrics, concepts and models.



One must note that humans, of course, do not move in a vacuum. Just as the diffusion of molecules in materials is mediated by the structural and thermodynamic properties of the material, similarly, spatio-temporal patterns of human movement are necessarily shaped by spatial constraints and limitations of geography. Examples of this are the topography of urban centers or the pattern of roads and transportation infrastructure, the properties of which are studied under the aegis of \emph{spatial networks}~\cite{barthelemy_2011_spatial}. Yet, this survey is \emph{not} about Spatial Networks, nor do we give details on the so-called ``science of cities" (e.g., city forms, properties of urban transport networks, urban scaling laws, etc. \cite{barthelemy_2016_structure}); these two aspects will be considered only when made necessary for the reader to understand the concepts related to human mobility.


\subsection{Organization}

We begin in \sectionname~\ref{sec:data} with a comprehensive (but by no means exhaustive) list of data sources used for empirical studies. In \sectionname~\ref{sec:scaling} we introduce key metrics, measures, spatio-temporal scales, as well as an overview of the fundamental physics behind the study of human mobility.  In \sectionname~\ref{sec:models} we describe the state-of-the-art families of models (both generative and phenomenological)  that best describe the empirical observations of human mobility. The models are categorized according to scale, starting from the level of individuals, through to the collective level of populations, and finally a mixture of the two incorporating the concept of (inter) modality.  Continuing with the theme of scales, in \sectionname~\ref{sec:applications} we describe some selected applications of the framework and models ranging from intra-urban movement to flows between countries and continents including the case of epidemic spreading, transportation systems, and a brief digression on new results on virtual mobility (web browsing). Finally, we conclude in \sectionname~\ref{sec:conclusion} with challenges and future directions for the field. We also added Appendix~\ref{sec:comp} where we provide descriptions of some basic tools and algorithms (agent-based modeling, random walkers, etc.) that may be of use to researchers making initial forays into the field.

\section{Data Sources}
\label{sec:data}


A natural starting point is to describe the nature of empirical data which has been used in mobility research. Indeed, empirical data has been vital to both aggregate and individual mobility investigation, by providing means of parameter calibration as well as model validation. In this section, we outline the main sources available for mobility research and the relevant information that can be extracted from them. The data sources are presented in (rough) chronological order of their historical availability and consequent use in research.

\subsection{Census Data \& Surveys}
\label{sec:census}

Census data is collected in periodical national surveys in which householders are asked questions about the socio-demographic and economic status of the household members. Of particular interest in terms of mobility are questions related to the location of the workplace, or the place of current and previous residence. Collectively, this information can then be used to estimate commuting flows or internal migration flows within a country.

Different countries introduced national censuses at different times, and the type of information collected also varies across countries and time periods. In the United Kingdom, for example, the first national census was done in 1841 and contained the names, ages, sexes, occupations, and places of birth of each individual living in a household. In 1921, the location of workplaces was added to the questions asked, and since 1961 the location of previous residence was also included. In the United States, the first population census was taken in 1790, while information on workplace and transportation activities was incorporated in 1940~\cite{potter_2010_new}. Nowadays, national censuses are held in most countries typically every ten years. Below we outline some main sources of census and survey data along with the methodology used to extract commuting and migration flows.

\subsubsection{Census Data}

The data on commuting trips between United States' counties is available from the United States Census Bureau\footnote{~\url{http://www.census.gov/hhes/commuting/data/commutingflows.html}}. These county-to-county work flow files are available for each state as well as at the aggregated level across the country. Furthermore, files are available in either county of residence format---containing work destinations of people who reside in each county---or county of work format---which contain the county of residence of people working in each county. The inclusion of a commuter in the data is preconditioned upon them working during the week leading up to the census. Due to the questionnaire design, any individual not satisfying this condition is precluded from entering a workplace location. A worker is defined as someone who has spent at least 1 hour in paid employment or 15 hours in a voluntary role. Any individual required to provide their workplace location was asked further questions about their mode of transport and travel time.

From this information, the census Bureau compiles and publishes aggregated flows corresponding to the residence and workplace location of everyone classed as a worker. In order to validate models and determine statistics such as trip distance and relationship to the origin, destination and intervening population density, further information on the spatial distribution of counties and socio-demographic data is required. This supporting data is available online\footnote{~\url{ftp://ftp2.census.gov/geo/tiger/tiger2k/}}. 
Among other things, such census data is invaluable for validating generative models of commuter flow and migration. For instance, Census data from the 2000s (consisting of county level work flow data, corresponding to 3,141 counties and 34,116,820 commuters within the United States) was used to evaluate the so-called radiation model (described in sec.~\ref{sec:Intervening}) by Simini \et\cite{simini_2012_universal}

\begin{figure}[t!]
\centering
\includegraphics[width=\textwidth]{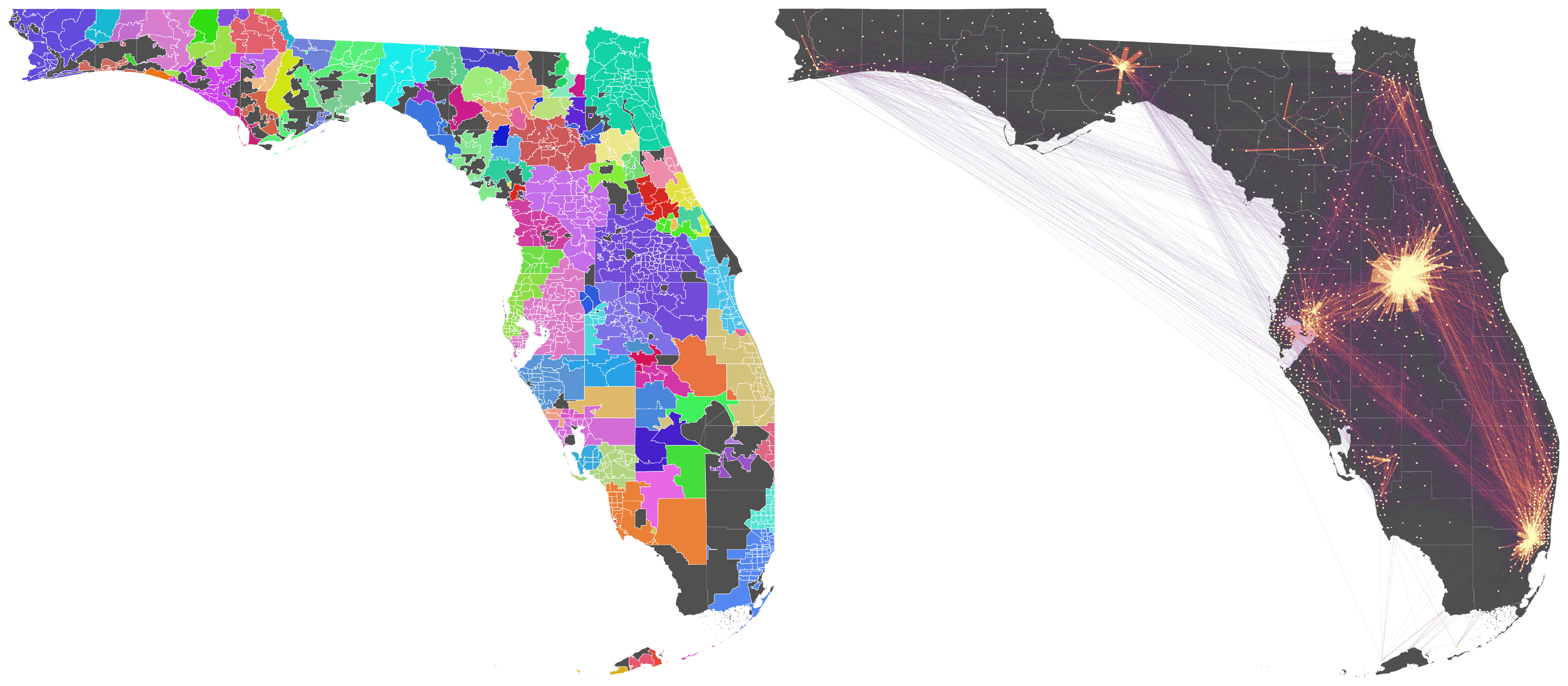}
\caption{Commuting flows compiled from census data. Left panel: The state of Florida partitioned according to its counties. Right panel: Commuting flows between counties, where thickness of lines correspond to volume of flow. Data compiled from the United States Census Bureau.}
\label{fig:Census_example}
\end{figure}

Commuting and migration flows data for the United Kingdom is also available online\footnote{~\url{http://webarchive.nationalarchives.gov.uk/20160105160709/}\\ \url{http://www.ons.gov.uk/ons/guide-method/census/2011/census-data/2011-census-data-catalogue/origin-destination/index.html}}. The information is available at different spatial resolutions: ranging from the geographical area covered by a single postcode, to areas covered by the 7,201 electoral wards. Files contain origin and destination flows divided into subsets of age and gender allowing for a deeper socio-demographic analysis if required. There have been many different countries whose commuting data has been analyzed at the level of municipalities (or smaller administrative units such as zipcodes or electoral tracks) including France\footnote{\url{https://www.insee.fr/fr/statistiques/2022117}} and most European countries. In most cases, when not directly available on an open data portal, commuting data is available to researchers on request from national statistics institutes.


\subsubsection{Tax Revenue Data}

An alternative source for estimating aggregated migration flows is the Statistics of Income Division (SOI) of the Internal Revenue Service (IRS) in the United States. The IRS maintains records of all individual income tax forms filed in each year. Using individual tax return files, it is possible to determine who has, or has not, moved residence or workplace locations in the intervening fiscal year. To do this, first, coded returns for the current filing year are matched to coded returns filed during the prior year. The mailing addresses on the two returns are then compared to one another. If the two are identical, the return is labeled a non-migrant. Any relevant information change during the fiscal year results in the return labeled as that of a migrant. US migration data from 1992-1993 to 2006-2007 is available online\footnote{~\url{http://www.irs.gov/uac/SOI-Tax-Stats-Migration-Data}}. Alongside tax returns, estimated migrations flows are also included in each Census\footnote{~\url{https://www.census.gov/hhes/migration/data/acs/county-to-county.html}}, based on yearly surveys carried out by the American Community Survey (ACS).

\subsubsection{Local Travel Surveys}

An important source of data used to construct trips, and therefore flows between two locations, is local surveys. It provides an advantage over censuses as it is possible to collect a more detailed data set that includes information such as the purpose of the trip and mode of transport used. However, the increase in accuracy, resolution, and variety of meta-data leads to a corresponding sacrifice in scale. As such, surveys typically have a much lower number of respondents compared to a census. Additionally, these typically cover smaller areas such as a single city (or indeed neighborhoods within cities), as opposed to the level of a state, or an entire country.


A representative example is provided by the household Travel Tracker Survey for the Chicago metropolitan area, conducted between 2007 and 2008, and carried out by the Chicago Metropolitan Agency for Planning\footnote{~\url{http://www.cmap.illinois.gov/data/transportation/travel-tracker-survey}}. This survey contains information on travel activities of household members, among other details. Information was collected from a total of 10,552 households over a one or two-day period, and households were required to provide a detailed travel inventory for each member, including details such as trip purpose, transport mode, departure and arrival times, and public transport information such as boarding location, distance to final destination, and fare paid. 
A similar survey is available for the Los Angeles region\footnote{~\url{http://www.scag.ca.gov/travelsurvey}}
and was analyzed by Liang \et\cite{liang_2013_unraveling}, where 46,000 trips between 2,017 zones within the Los Angeles county were extracted.
The use of small scale surveys is often augmented by other forms of data (for example GPS tracks, see \sectionname~\ref{sec:gps}) in order to have accurate records of an individual's position combined with an annotated description of the purposes of each trip~\cite{liang_2013_unraveling, schneider_2013_unravelling}.

While historically most prevalent, census data provides us only with coarse-grained patterns of human migration and movement. Surveys, on the other hand, provide far more detail but on a much smaller scale, thus limiting their use in terms of statistical validation of models. While the census reveals little information about where people spend their time, or where and how they travel, at the survey level, data reliability is restricted by scale and self-reporting errors. Thus both sources lack the ability to provide a dynamic picture of human mobility~\cite{palmer_2013_new}.

\subsection{Dollar Bills}
\label{sec:dollar}

\begin{figure}[t!]
\centering
\includegraphics[width=0.9\textwidth]{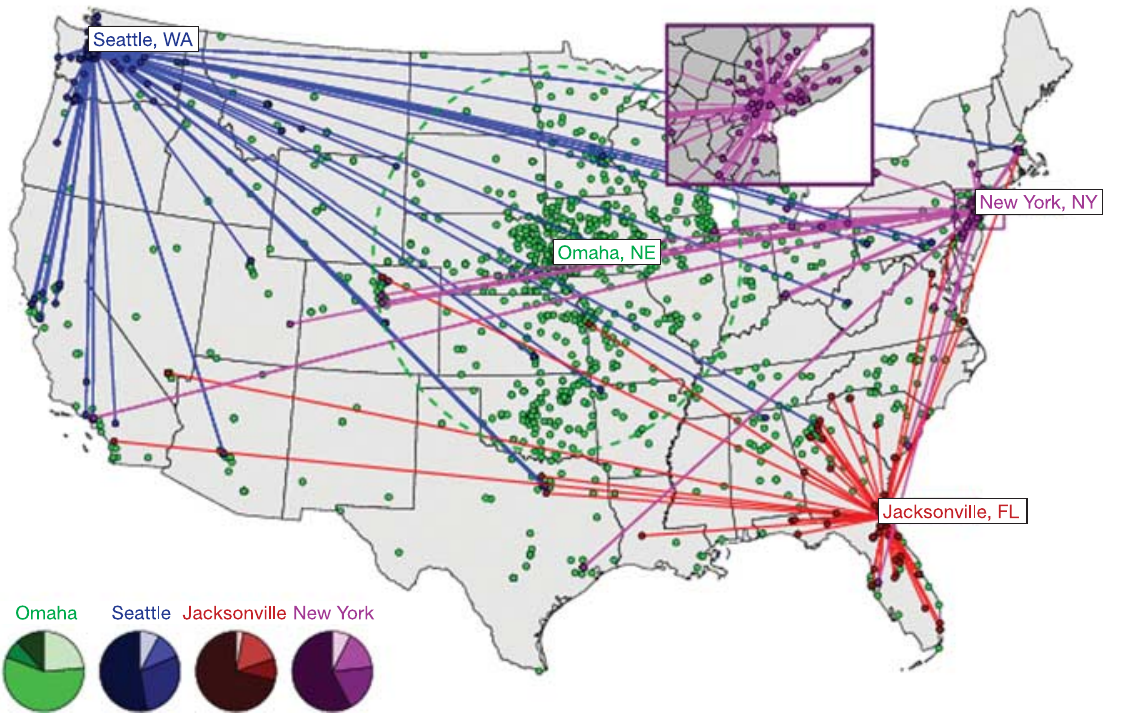}
\caption{Trajectories of bank notes originating from four different locations. Tags indicate initial, symbols secondary report locations. Lines represent short time trajectories with traveling time $T < 14$ days. Lines are omitted for the long time trajectories (initial
entry: Omaha) with $T > 100$ days. The inset depicts a close-up of the New York area. Pie charts
indicate the relative number of secondary reports coarsely sorted by distance. The fractions of secondary
reports that occurred at the initial entry location (dark), at short ($0 < r < 50$km), intermediate ($50 < r < 800$km) and long ($r > 800$km) distances are ordered by increasing brightness of hue. The
total number of initial entries are $N = 2,055$ (Omaha), $N = 524$ (Seattle), $N = 231$ (New York), $N = 381$ (Jacksonville). Figure from~\cite{brockmann_2006_scaling}.}
\label{fig:2006_Brockmann_Nature_1ab}
\end{figure}

An unusual (and perhaps not immediately apparent) source of mobility data is related to the tracking of currency. An archetypal example of this are online bill-tracking websites, which are designed to monitor the dispersal of individual bank notes worldwide. The movement of a bank note between geographical locations occurs when it is carried by an individual, therefore data collected from such websites may be used to infer  trajectories and patterns of human mobility. Brockmann \emph{et al.}~\cite{brockmann_2006_scaling} obtained trajectories of 464,670 dollar bills from the website \url{www.wheresgeorge.com}, allowing for an analysis of bank note dispersal in the United States (excluding Hawaii and Alaska). The data set contains entries corresponding to time-stamped reports of the geographic location of individual bills, numbering around 250 million as of 2017.  All bills registered on the site are initially entered by a user which involves the input of their ZIP code and a serial number printed on the bill. While the majority of the bills are not entered more than once, around 11\% of the bills are reported multiple times, with 3-5 entries per bill being fairly common (Any more than 5 entries is considered rare). Once a bill has been registered, any future hits allow for time and distance between reports to be determined and recorded in the database (see \figurename~\ref{fig:2006_Brockmann_Nature_1ab}). 

While a convenient and large-scale source of mobility data, the use of currency to infer patterns in human mobility is problematic. For instance, bank note dispersals do not contain information on the number of individuals that have carried a given note during two instances of measurement (i.e., when it appears in the database). Consequently, the trajectories of bank notes are likely a convolution of the mobility of several individuals; coupled with the relatively few samples per dollar bill,  this makes it a rather inaccurate and  problematic source for inferring individual mobility patterns. Nevertheless, similar to census data, currency tracking provides a convenient, and relatively easily accessible picture of the coarse-grained patterns of human mobility.

\subsection{Mobile Phone Records}
\label{sec:cdr}

\begin{figure}[t!]
\centering
\includegraphics[width=0.9\textwidth]{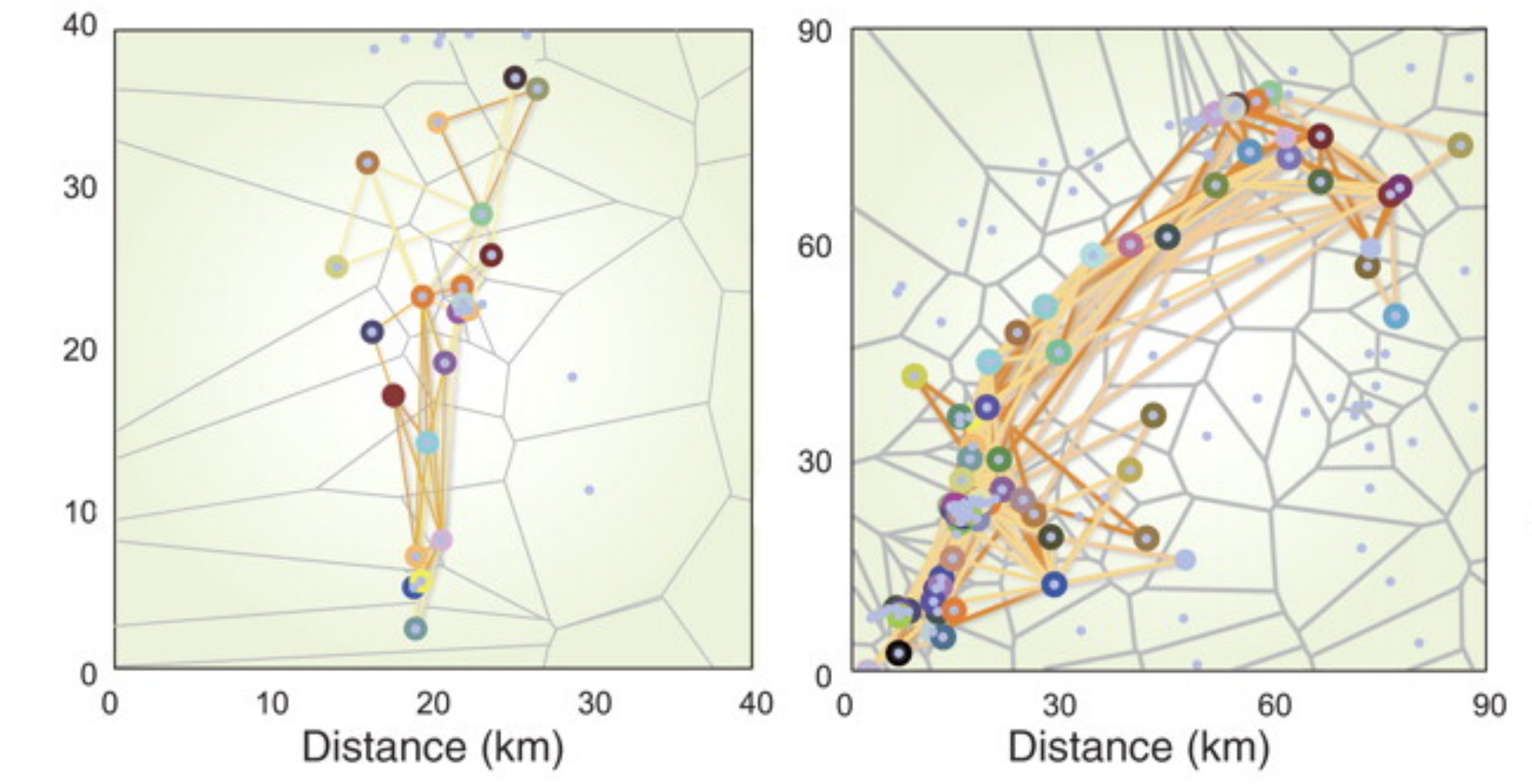}
\caption{Trajectories of two anonymized mobile phone users traveling in the vicinity of $N = 22$ and $76$ different cell phone towers during a 3-month-long observational period. Each dot corresponds to a mobile phone tower, and each time a user makes a call, the closest tower that routes the call is recorded, pinpointing the users approximate location. The gray lines represent the Voronoi lattice, approximating each towers area of reception. The colored lines represent the recorded movement of the user between the towers. Figure from~\cite{Song_2010_Limits}.}
\label{fig:2010_Song_Science_fig1a}
\end{figure}

Perhaps the most important, game-changing data of the last decade for inferring human mobility are the mobile phone Call Detail Records (CDRs). Most cell phone companies maintain detailed records of customer information related to their calls and text messages. The information contained in these CDRs include the time of the call/message as well as the company's cell tower routing the communication. Knowing the exact geographic coordinates of the cell tower makes it possible to estimate the approximate location of the user. As mobile phones are (typically) personal devices and are mostly carried by a single person, the corresponding location data can be used to infer a single individual's movements over the recorded period (an example is shown in \figurename~\ref{fig:2010_Song_Science_fig1a}). Unlike the census, surveys, or currency tracking, CDRs allow for the characterization of  \emph{individual mobility patterns}, and -- depending on the dataset -- at an unprecedented spatio-temporal resolution. Furthermore, due to the tremendous trend of global cell-phone adoption, one can conduct multi-scale studies ranging from the level of neighborhoods to the country-wide level (including international travel and movement across borders).

CDR data concern both phone calls and SMS exchange and always include the following information: time stamp, caller ID, recipient ID, call duration and antenna (cellular tower) code. Due to privacy concerns, customer identifiers are anonymized before the data is made available to scientists for analysis. Quite often, additional clean up of the data is required to make the data more analyzable; for instance, one typically needs to restrict the dataset to users who visit a minimum threshold number of towers, as well as make calls at a high enough frequency to maintain statistical reliability for each user.

One of the first work using CDR's was performed by Gonzal\'ez \et\cite{gonzalez_2008_understanding} on two anonymized datasets provided by a major mobile operator in a large European country. The first set contained the locations of 100,000 randomly selected mobile users whose positions were recorded over a period of 6 months, each time a call or text was received or sent. The second set (collected for validation purposes) corresponded to the locations of 206 mobile users, recorded every two hours over a period of a week. From this data, traces of mobility patterns such as the distribution of displacements between the locations of consecutive calls (jump-lengths) were measured. The temporal period between consecutive calls was also measured to determine a distribution of wait times (i.e., characteristic time spent by users in a given location). The knowledge of users' consecutive locations also enabled the calculation of so-called return time distributions, measuring the probability of a user to return to a given location within a given time. Other more complex metrics, including the radius of gyration were also extracted. (For details on metrics see Sec.~\ref{sec:metrics}). 

The quality of mobility information extracted from mobile phone data depends on the frequency at which the location of the user is recorded. For example, the larger data set used in~\cite{gonzalez_2008_understanding} displayed an irregular call pattern; many calls were placed over a short time period, followed by long periods of inactivity. Consequently, the data displayed a highly inconsistent temporal frequency, which may confound the analysis of mobility patterns. The smaller-scale validation dataset was therefore collected to account for these irregularities. Other problematic issues also arise, such as the accuracy of recorded locations. In general, it is the position of the cell tower routing the call that is recorded and not the exact location of the individual carrying the phone. Since there is significant variability in the areas covered by cell towers---with coverage ranging from tens of meters in the densest neighborhoods of urban areas, to a few kilometers in rural areas---a recorded user in a rural area may transmit all communication via a single tower during their daily routine, while moving the exact same area as that of an urban user who may ping several towers. Thus, in the former case, despite possibly significant movement on part of the user, for purposes of analysis they are considered stationary, which is obviously misleading.

We must note that obtaining CDRs for research purposes poses significant challenges. Since they are not typically publicly available, one has to directly approach a mobile phone operator. As the data contains sensitive information~\cite{de_montjoye_2013_unique}, any data collected by a specific group of researchers may not be shared with other groups, making reproducibility problematic. Yet, recent initiatives have seen some phone companies release large scale CDR data within the context of "challenges" among researchers. The purpose of making this data available is to provide researchers some material allowing them to extract useful information to address major societal challenges. Such recent datasets are those that have been provided by Orange in the framework of their Data 4 Development (D4D) challenge. These contain four anonymized tables for mobile phone users in the Ivory Coast (C{\^o}te d'Ivoire) and Senegal. For example, in the case of Ivory Coast, data contained (i) hourly antenna-to-antenna traffic; (ii) individual trajectories of 50,000 users over two weeks with antenna location information; (iii) individual trajectories of 500,000 users over an entire year with sub-prefecture location information; (iv) communication graphs for 5,000 users~\cite{blondel_2012_data}. In particular (iii) contains information regarding individual trajectories for the entire observation period at the cost of reduced spatial resolution; position is recorded as a sub-prefecture rather than antenna location. The raw data consists of a user ID, date and time stamp of the communication and the sub-prefecture number (1-255) which can be linked to a file containing the latitude and longitude of the center of each sub-prefecture~\cite{lu_2013_approaching}.
%
%
%

Telecom Italia's ``Big Data Challenge'' is another example of a large scale CDR data set readily available to researchers. Alongside anonymized call records, additional metadata related to weather, news, Twitter activity, and electricity data from two areas in Italy (Milan and Trentino) was provided~\cite{barlacchi_2015_multi}. Data from this challenge has been used to investigate the relationship between individual daily trips and personal constraints~\cite{de_2015_personalized}, estimate urban population from call volume~\cite{douglass_2015_high} or determine the relationship between urban communication and happiness~\cite{alshamsi_2015_misery}. The increasing availability of such large-scale, anonymized datasets for research purposes is an encouraging trend for future research on human mobility. For more detailed information on CDR sources and some associated results (including other research topics than human mobility), see the review by Blondel \et~\cite{Blondel_2015_CDRreview}.

\subsection{GPS Data}
\label{sec:gps}

The greatest level of accuracy on movement trajectories is provided by Global Positioning System (GPS) data, which is capable of providing precise information on any location covered by at least four GPS Satellites. GPS trackers are units which receive signals from GPS satellites and compute the device's position at regular intervals. This technology allows researchers to trace the movement of individuals with a high degree of accuracy and temporal frequency, thus providing a rich source of data that can be analyzed and directly mapped to human mobility patterns.

\begin{figure}[t!]
\centering
\includegraphics[width=0.8\textwidth]{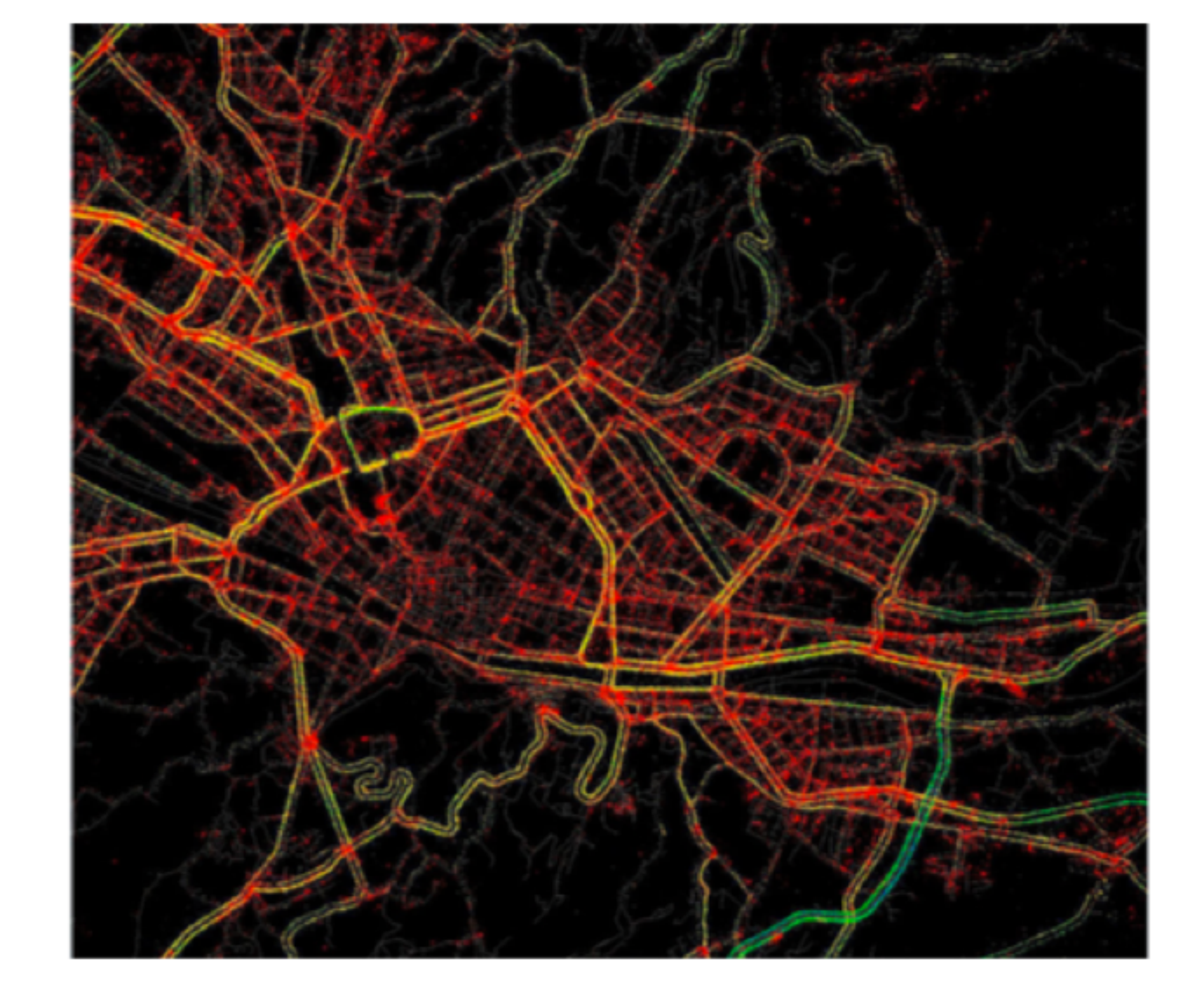}
\caption{Aggregated GPS position data (collected from vehicles) in a part of Florence, Italy,  measured during March 2008; the red dots correspond to a recorded instantaneous velocity $\leq30$ km h$^{-1}$, whereas the yellow dots correspond to velocities in the interval $30--60$ km h$^{-1}$ and the green dots to a velocity $\geq60$ km h$^{-1}$. Figure from~\cite{bazzani_2010_statistical_arxiv}.}
\label{fig:2010_Bazzani_JStatMech_fig1}
\end{figure}

Rhee \et\cite{shin_2008_levy} used GPS trackers with a position accuracy of less than three meters to record the movements of over $150$ individuals across two University campuses, a state fair (US), Disney World, and New York City. The position of the individuals carrying the trackers was recorded every 10 seconds while they were in these locations resulting in daily traces that had on average a temporal scale of 9-10 hours. Additionally, Microsoft has made GPS trajectories from 182 individuals freely available online via their GeoLife project\footnote{~\url{http://research.microsoft.com/en-us/downloads/b16d359d-d164-469e-9fd4-daa38f2b2e13}}. 
The dataset consists of 17,621 trajectories recorded by GPS trackers and GPS enabled mobile devices, and has a high spatial and temporal scale for over $90\%$ of individuals with positions recorded every 1-5 seconds or 5-10 meters for time periods that span several years. Since its release, GeoLife has been used in studies of both human mobility~\cite{zheng_2008_understanding, zheng_2009_mining} and social interactions~\cite{li_2008_mining}. For a review of potential applications see~\cite{zheng_2010_geolife}. 

Transmitters attached to vehicles~\cite{bazzani_2010_statistical, pappalardo_2013_understanding} are another source of GPS data. These are useful for studying questions related to urban traffic: monitoring, prediction, and prevention of congestion~\cite{pappalardo_2013_understanding}. In Italy, $\simeq 2\%$ of privately owned cars have a GPS system; each vehicle has assigned a unique identifying number, thus enabling the tracing of anonymous individual cars' trajectories with high precision. A GPS signal is transmitted approximately every 2km, as well as when the engine is switched on and off~\cite{bazzani_2010_statistical}. Each signal is converted to a data point containing information on position, velocity and distance covered. While the data suffers from inaccuracy due to variables such as satellite coverage and position precision, the temporal precision is of high quality, as is both instantaneous velocity and distance covered. From such GPS vehicle traces, path length distributions can be measured, which corresponds to the distribution of distances traveled in single trips, where a trip is defined as travel that occurs between instances of the engine being started and stopped. Activity distributions, which are equivalent to waiting time distributions in other studies, can also be determined from the elapsed time between consecutive trips. 
A further advantage of GPS traces over CDRs is that the locations are recorded to a much higher degree of accuracy and at a constant frequency. 
Yet, there are drawbacks: GPS transmitters do not work indoors and generally rely on batteries as their source of power. As a result, traces can contain periods during which no signal is received or may be terminated prematurely. Furthermore, GPS datasets typically feature a smaller number of individual users, up to several thousand, in comparison to mobile phone data which can provide information on the mobility of millions of users. 

\subsection{Online Data}
\label{sec:online}

\begin{figure}[t!]
\centering
\includegraphics[width=\textwidth]{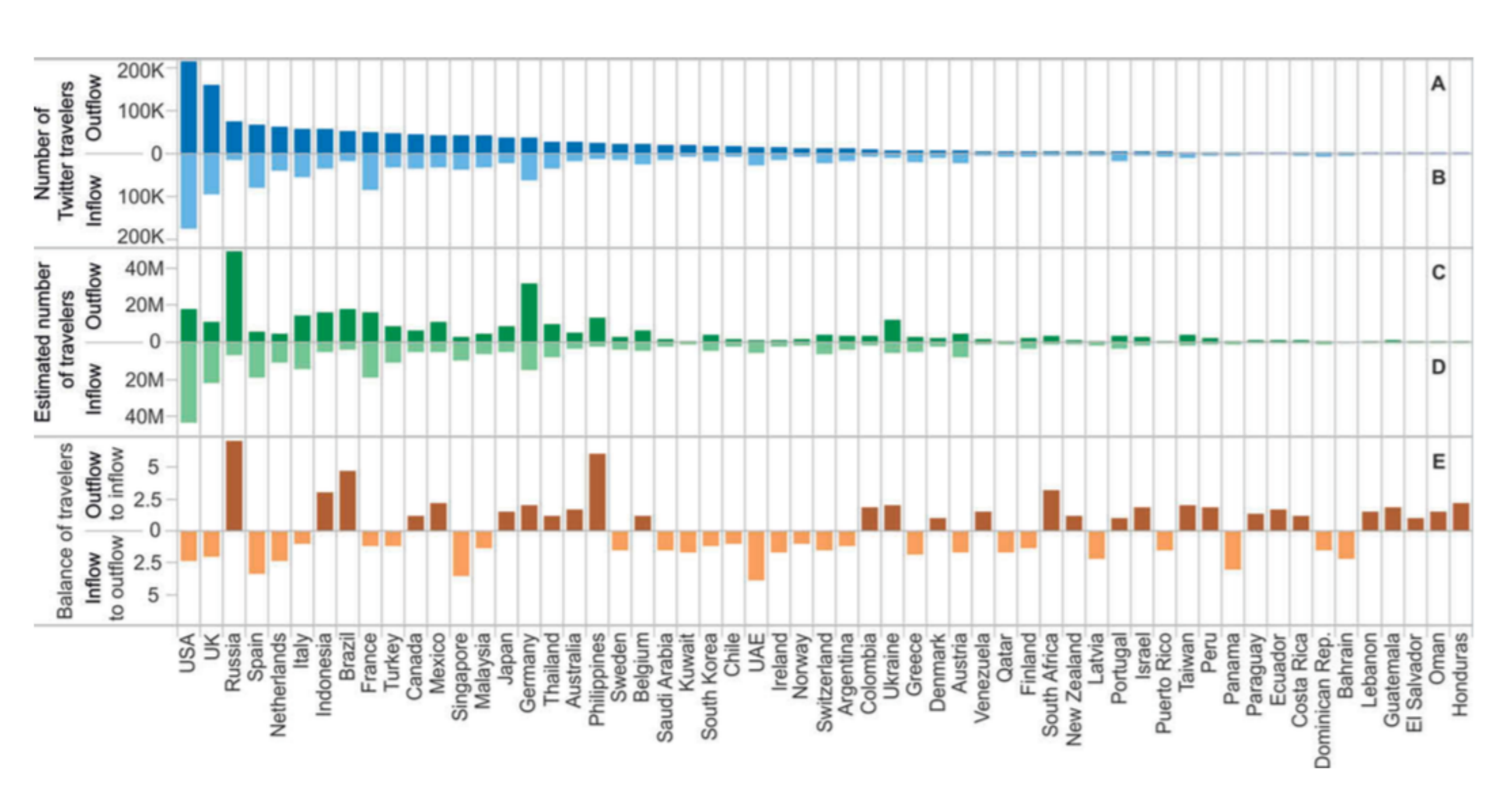}
\caption{Country-specific analyses of travel based on Twitter users and on the estimated total number of travelers. Panel A shows the number of Twitter users residing in a country and traveling to another while panel B shows the number of users visiting this particular country. Panels C and D represent the number of Twitter travelers normalized by the  extent of Twitter usage in their home country. Finally  panel E represents the yearly ratio between the estimated inflow and outflow of travelers, revealing which countries were the origin or destination of international travel. Figure from~\cite{hawelka_2014_geo}.}
\label{fig:2014_Hawelka_CGIS_Fig5}
\end{figure}

Another valuable source of location information are Online Social Network (OSN) or Location-Based Social Network (LBSN) services, that attract hundreds of millions of users worldwide. Since the introduction of GPS and wifi chips in smartphones, social network providers have been able to collect valuable data on both the social connections as well as precise geographical locations of their users.
Indeed, services such as Twitter, Facebook, Foursquare and Flickr collect geotagged data every time a user enables localization for the content being posted (e.g., checking-in at a restaurant with friends); this is associated with geographic coordinates, a time stamp, and additional information relative to the location itself or the content published by the user.
The mobility profiles of the users can be obtained from the list and number of locations other than their home which they have visited over the period of the study (see \figurename~\ref{fig:2014_Hawelka_CGIS_Fig5} for an example). These profiles can range from intra-urban routine trips to worldwide travel~\cite{noulas_2012_tale, hawelka_2014_geo}. 
From the geographical locations, by inferring a user's home location, their radius of gyration can be determined using the geographical distance between the home location and reported locations.
The jump length distribution can be determined by defining a jump as the geographical distance between consecutive reports for a single user and frequency of reports can be used to assess temporal patterns in human mobility~\cite{hawelka_2014_geo}. Other measures such as visitation frequency and predictability are also obtainable~\cite{jurdak_2015_understanding}. \\

Geotagged data can also be used to assess social interactions between individuals. Scellato \et\cite{scellato_2011_socio} analyzed three online location-based social networks: Foursquare, Brightkite, and Gowalla, finding their exhibition of common properties found in most other real world complex networks, such as fat tailed degree distributions, high clustering coefficient, and short average path length between nodes, confirming the small-world nature of LBSN's~\cite{Newman_book}. Furthermore, the probability of a social link  between two users was found to exhibit a distance dependence, decreasing with geographical separation between the users.
Compared to CDR's, data collected from OSN's and LBSN's, have the advantage of having more contextual information associated with the geographical positions and the users, enabling the study of mobility in a broader context. For example, data collected from Twitter users has been applied to topics such as social networks~\cite{java_2007_why, huberman_2008_social, kwak_2010_what}, evolution of moods~\cite{pak_2010_twitter, bollen_2011_modeling, golder_2011_diurnal}, and crisis management~\cite{sakaki_2010_earthquake,maceachren_2011_senseplace2, thom_2012_spatiotemporal} among other features.

Yet, as in all other data sources, limitations persist. Besides requiring extensive clean up (such as removing users with suspiciously high activity, unreasonably fast movement between two consecutive check-ins or incomplete data \cite{hawelka_2014_geo}), mobility and interaction data originating from online social networks needs further validation in order to be considered truly representative of the general population. Sloan \et\cite{sloan_2015_tweets} found that Twitter data originating from the UK has significant demographic differences compared to the wider population data from UK census. Furthermore, there appear to be demographic differences between the users who enable geotagging and those who do not (only $\approx 3\%$ of the users were found to have geotagging enabled).

\section{Metrics, Physics and Scales}
\label{sec:scaling}

In this section, we will discuss some of the fundamental metrics used to characterize mobility as well as the associated spatio-temporal scales at which they are relevant. We will then move onto a discussion of some of the physics associated with mobility, including the relations between distance, time and velocity. We end the section with a discussion of energy arguments and interpolation of spatio-temporal scales through the lens of multimodality. 

\subsection{General Metrics} \label{sec:metrics}

\subsubsection{Jump Lengths}

A key factor in modeling human mobility is the distance an individual travels in a given time period. Measures of distance are often dependent on the source of data being used, and the terms flight length, jump length, displacement, and trip refer to different distance measures that may be extracted from data. For example, early measures of mobility leveraged information about the spatial trajectories of bank-notes (one can think of this as an aggregation of many individual trips as a given banknote necessarily changes many hands). Conventionally, the distance between two instances of the appearance of a banknote in the measured data is termed a \emph{jump length}~\cite{brockmann_2006_scaling}. Later, higher resolution measurements of movement were provided by Call Detail Records (CDRs). Here one could reliably measure the location (and corresponding displacement) of \emph{individuals} based on placement of mobile phone towers that are pinged when one makes a call. Associated with the displacement between two successive calls is a time interval~\cite{gonzalez_2008_understanding}, that provides an estimate of the stay of an individual in a location (waiting-times), although it is not possible to detect the position of the user between two consecutive calls. Even higher resolution spatio-temporal data available from Global Positioning System (GPS) allows one to make more stringent definitions of displacement. In addition to the standard definition of jump-lengths~\cite{shin_2008_levy}, one can now define a displacement between \emph{stops}, i.e. the displacement of individuals between two locations, given that they spent a \emph{minimum amount of time} per location~\cite{bazzani_2010_statistical}. 
Indeed, the use of the terms ``stop'', ``trip'' or ``displacement'' reflects human behavioral tendencies that motivate people to go from point A to point B. It is therefore important to verify that a ``stop'' corresponds to an actual behavior and that it is not artificially generated by oversampled or undersampled signals~\cite{turchin_1998_quantitative}. 

Regardless of how one chooses to constrain it, the jump length, typically denoted as $\Delta r$, is defined as the euclidean distance between $r(t)$ and $r(t+dt)$ corresponding to locations  recorded at intervals $t$ and $t+dt$. Of particular interest is the distribution of $\Delta r$ within a population; how likely is it that a random member of the population will travel a distance $r$ from their origin location in a time $dt$? In order to measure this, when modeling human mobility it is common to consider the probability distribution function (PDF) of jump lengths, $P(\Delta r)$. This may be defined as the probability of finding a displacement $\Delta r$ in a short time step $dt$. 

\begin{figure}[t!]
\centering
\includegraphics[width=0.8\textwidth]{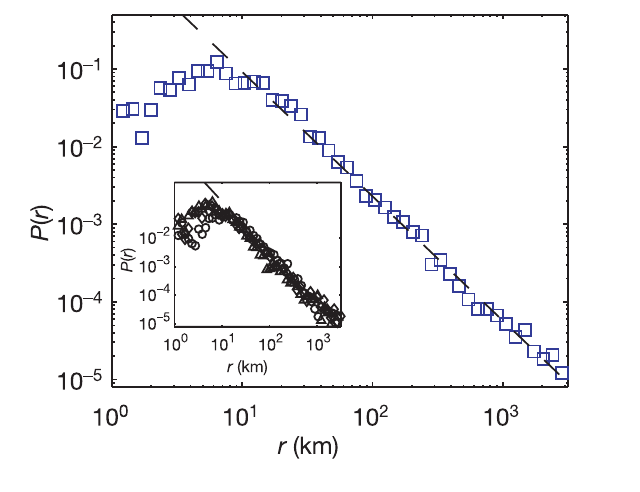}
\caption{The short-time dispersal kernel of bank notes. The measured probability density function $P(r)$ of traversing a distance $r$ in less than $T = 4$ days is depicted in blue symbols. It is computed from an ensemble of 20,540 short-time displacements. The dashed black line indicates a power law $P(r)\sim r^{-(1+ \beta)}$ with an exponent of $\beta \sim 0.59$. The inset shows $P(r)$ for three classes of initial entry locations (black triangles for metropolitan areas, diamonds for cities of intermediate size, circles for small towns). Their decay is consistent with the measured exponent $\beta = 0.59$ (dashed line). Figure from~\cite{brockmann_2006_scaling}.}
\label{fig:scales_brockmann_1c}
\end{figure}

Brockmann \et~\cite{brockmann_2006_scaling} determined $P(\Delta r)$ for the trajectories of dollar bills and the distribution of jump lengths was observed to follow a power law, $P(\Delta r) \sim \Delta r^{-(1 + \beta)}$, with exponent $\beta = 0.59 \pm 0.02$, independent of the size (in population) of the entry point of a bank note (see Fig.~\ref{fig:scales_brockmann_1c}).
The power-law behavior of $P(\Delta r)$ has also been observed in the trajectories of mobile phone users ~\cite{gonzalez_2008_understanding, song_2010_modelling}.  
The empirical distribution of displacements obtained from mobile phone data analyzed in \cite{gonzalez_2008_understanding} is well approximated by a truncated power-law distribution: 
\begin{equation}\label{eq:pdr}
P(\Delta r) = (\Delta r - \Delta r_0)^{-(1 + \beta)} \exp{(- \Delta r / \kappa)}
\end{equation}
where $r_0$ and $\kappa$ represent cutoffs at small and large values of $\Delta r$. 
The value of the observed scaling exponent in Eq.~(\ref{eq:pdr}) is $\beta = 0.75 \pm 0.15$, 
not far from $\beta = 0.59 \pm 0.02$ obtained from bank note dispersal, suggesting that the two distributions may capture the same fundamental mechanism driving human mobility patterns. 
Zhao \et ~\cite{zhao_2015_explaining} measured the distribution of jump lengths using two GPS data sets that, along with location and time stamps, also included transportation mode. From this, the jump length $\Delta r$ was determined as the longest straight line distance between two locations without a change of direction. Wait times, $\Delta t$, were taken to be the time spent in a particular transportation mode. A single trip may be made up of several flights, each of which have a corresponding transportation mode taken directly from the dataset. The transportation modes were grouped into four categories; Walk/Run, Car/Bus/Taxi, Subway/Train and Bike.
The distribution of jump lengths is found to be log-normal if each transportation mode is considered individually, whereas $P(\Delta r)$ for all modes combined follows a truncated power-law with $\beta = 0.55$ and $0.39$  for the two datasets. The distribution of wait times, $P(\Delta t)$, was found to be exponential, corresponding to a large number of Walk/Run flights that connect the use of other transportation modes within a single trip and therefore have only a short duration. This result provides further insight into studies using on bank note and mobile phone data ~\cite{brockmann_2006_scaling, gonzalez_2008_understanding, song_2010_modelling} in that the power-law behavior of the distribution of jump lengths is the result of combining several distinct log-normal distributions, corresponding to the different transportation modes available, with exponentially distributed wait times. 
GPS devices installed in cars transmit the location of the car to a high degree of accuracy every few seconds when the car is in motion. The analysis of such data from over 150,000 cars in Italy over a period of one month~\cite{pappalardo_2013_understanding} found that the probability density function, $P(\Delta r)$ had two different regimes; an exponential distribution up to a characteristic distance of $\sim$20 km corresponding to inter-city travels, and a power-law with $\beta=1.53$ corresponding to intra-city movement.  Here, $\Delta r$ corresponds to the true length of a single trip. Trips are distinguished by records that are greater than 20 mins apart; as the GPS device only transmits when the car is moving, using a threshold of 20 mins allows for small stops, such as waiting at traffic lights or re-fueling, to be accounted for and included within a single trip.
The exponent, $\beta$, is significantly higher than those obtained for bank note dispersal~\cite{brockmann_2006_scaling} and mobile phone data~\cite{gonzalez_2008_understanding, song_2010_modelling}. The authors suggest this is due to the limitation in the GPS data caused by restrictions on the size of the area covered (500km in length), and the physical limitation on the distance an individual will drive in a single trip.

\subsubsection{Mean Square Displacement (MSD)}\label{sec:msd}

Research into the distribution of jump lengths suggests that individual trajectories can be described by L\'evy flights, a family of models associated with random walks (described in detail in \sectionname~\ref{sec:rw}). In this context, a common measure of the potential exploration of area by an individual is the Mean Square Displacement (MSD):
\begin{equation}
\mathrm{MSD}(t) = \langle (\mathbf{r}(t) - \mathbf{r}_0)^2 \rangle \equiv \langle \mathbf{\Delta r}(t)^2 \rangle.
\end{equation}
Here $\mathbf{r}_0$ is a vector marking the origin of the individual relative to some reference point, or in other words, the location at which a particular trajectory starts, while $\mathbf{r}(t)$ measures the subsequent position of the individual at time $t$. 
The scaling of MSD with time provides a measure of the type of diffusion of individuals relative to their starting point in a trip.
The MSD(t) has the unit of an area and it corresponds to the average squared distance from the origin after a time $t$. 
In general, if the individual mobility trajectories follow a so-called Continuous Time Random Walk, (Cf. \sectionname~\ref{sec:ctrw}), the MSD follows the form $\langle \Delta \mathbf{r}(t)^2 \rangle \sim t^{\nu}$ with $\nu = 2 \alpha /\beta$ where $\alpha$ and $\beta$ are the exponents of the waiting-time, i.e. the time interval between two consecutive jumps, and jump length PDFs~\cite{brockmann_2006_scaling}. 
In particular, for a random walk (ordinary diffusion) $\nu = 1$.
Yet a CTRW type process is not entirely a realistic representation of how humans actually move. In particular, a random walker will tend to drift away from the origin of its trajectory rather rapidly, the longer the elapsed time. 
Whereas, as one would imagine, empirical measurements indicate that people have a tendency (on average) to return home on a daily basis~\cite{gonzalez_2008_understanding}. Furthermore, the means of transportation constrain the maximum jump length, therefore it has been observed that the PDF's $P(\Delta t)$ and $P(\Delta r)$ have exponential cutoffs; this implies that their variances are finite and hence in the long-time limit the scaling of the MSD will asymptotically converge to that of Brownian motion, $\mathrm{MSD(t)} \sim t^{1/2}$.

\begin{figure}[t!]
\centering
\includegraphics[width=0.8\textwidth]{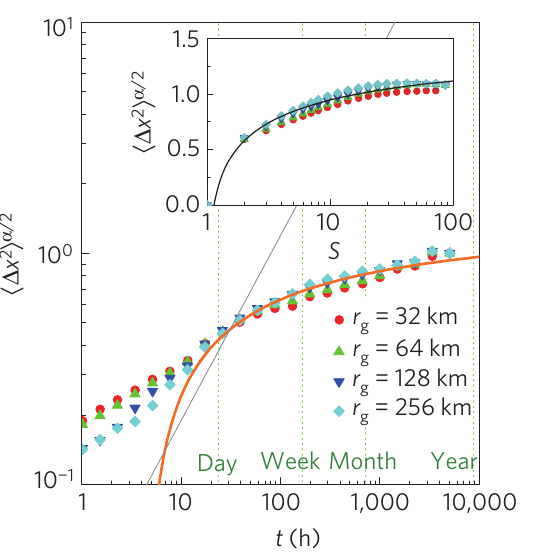}
\caption{MSD versus time for groups of individuals with different radii of gyration. The gray line represents the analytic prediction of CTRW (Cf. \sectionname~\ref{sec:ctrw}) whereas the orange line represents the analytic prediction of the asymptotic behavior $MSD(t) \sim (\log \log(t))^{2/\beta}$, which more accurately reflects the movement pattern of humans. Figure from~\cite{song_2010_modelling}.}
 \label{fig:scales_song_2}
\end{figure}

To account for this discrepancy, refinements were made in~\cite{song_2010_modelling}, where analysis was restricted to high-resolution data with a combination of mobile phone records that provided locations each time a call was made, along with location data recorded by mobile services at hourly intervals. The newer measurements revealed that human mobility apparently follows an ultra-slow diffusive process characterized by a slower than logarithmic growth of the MSD with time, $\mathrm{MSD}(t) \sim (\log \log(t))^{2/\beta}$ (see \figurename~\ref{fig:scales_song_2}). 

Yet, the picture is typically more complex. Indeed, a common problem arising when calculating the MSD, is that the location of an individual's origin is often ambiguous~\cite{shin_2008_levy}. One method of overcoming this issue is to take the average of MSD values measured by varying the origin among all locations that an individual visits~\cite{dimilla_1993_maximal, maruyama_2003_truncated}. Such an approach was taken in~\cite{shin_2008_levy} where they analyzed the GPS traces of individuals in a number of different location, including a university campus (KAIST), New York City, Disney World, and a US state fair. The averaging procedure applied to the MSD revealed two types of temporal behavior; up to a certain time ($\sim 30$ minutes) the participants moved super-diffusively (as expected in a CTRW process), whereas for periods greater than $30$ minutes, the individuals moved sub-diffusively. 
It is here that the notion of scales becomes important. The authors explain this dichotomous behavior by showing that at short time scales, the distribution of jump lengths follow a power-law distribution, whereas at longer time scales, jump lengths are more homogeneous, following a Gaussian distribution. While the combination of power law waiting times and jump lengths lead to super-diffusive behavior at shorter spatio-temporal scales, the Gaussian distribution of jump lengths (while maintaining a power law behavior for the waiting times) leads to mobility that appears sub-diffusive~\cite{vazquez_1999_diffusion} at larger spatio-temporal scales. 

%

\subsubsection{Radius of Gyration}\label{sec:rg}
The sub-diffusive behavior of humans at certain scales suggests that they tend to move a \emph{characteristic distance} away from their starting locations. This distance can be quantified by the so-called, radius of gyration, $r_g$, defined as the root mean square distance of a set of points from a given axis. Traditional applications of this measure are rooted in physics (where it is related to the mass moment of inertia) and engineering (distribution of cross sectional area in a column). A general formulation of $r_g$ is given by:
\begin{equation}\label{eq:rg_def}
 r_{g} = \sqrt{\frac{1}{N} \sum_{i=1}^{N}(\mathbf{r}_i - \mathbf{r}_0)^2}
\end{equation}
where $\mathbf{r}_i$ are the coordinates of the $N$ individual points (or $N$ measurements of location) and $\mathbf{r}_0$ is the position vector of the center of mass of the set of points, $\mathbf{r}_{cm} = \sum_{i=1}^{N} \mathbf{r}_i / N$. When applied to human mobility, the radius of gyration can be used to characterize the typical distance of an individual from the center of mass of their trajectory.
The radius of gyration is not equivalent to the square root of the Mean Squared Displacement, defined in Eq.~\ref{eq:rg_def}, although the two quantities are related because the former is a central second moment while the latter is a second moment about the origin. 
\begin{figure}[ht!]
\centering
\includegraphics[width=0.7\textwidth]{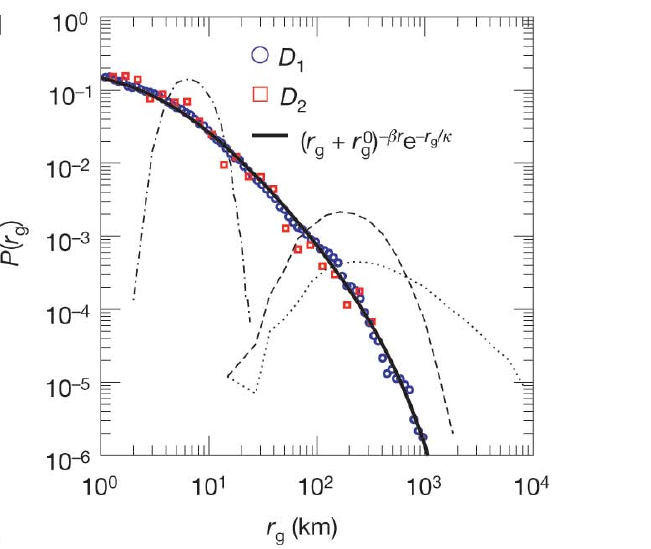}
\caption{The distribution $P(r_g)$ of the radius of gyration measured for two sets of mobile phone users labeled $D_1$ and $D_2$, where $r_g(T)$ was measured after 6 months of observation. The solid line represents a truncated power-law fit. The dotted, dashed, and dot-dashed curves show $P(r_g)$ obtained from Random Walk, L\'evy flight, and truncated L\'evy flight models respectively. Figure from~\cite{gonzalez_2008_understanding}.}
\label{fig:scales_gonzalez_1}
\end{figure}

Gonzalez \et\cite{gonzalez_2008_understanding} used mobile phone call records to determine the radius of gyration for two sets of users; one set with high temporal resolution (position recorded every two hours over one week), and one set with lower resolution on a longer (6 month) scale. The radius of gyration was calculated for each user, $a$, up to a time $t$  using the recorded positions, $\mathbf{r}_i^{(a)}$, $i=1...,N^{(a)}$. The value of $\mathbf{r}_0$ was taken as the center of mass of each users trajectory; $\mathbf{r}_{cm}^{a} = \sum_{i=1}^{N^{(a)}} \mathbf{r}_i^{(a)}/ N^{(a)}$.
The distribution of $r_g$, $P(r_g)$, was found to follow a truncated power law of the form:
\begin{equation}\label{eq:prg}
P(r_g) = (r_g - r_g^{(0)})^{-(1+\beta_r)} \exp{(- r_g / \kappa_r)}
\end{equation}
where $r^{(0)}_0$ represents the minimum radius of gyration cutoff due to the spatial sensitivity of the data, and $\kappa_r$ represents an upper cutoff mostly due to the finite size of the study area (see Fig. \ref{fig:scales_gonzalez_1}).

\begin{figure}[t!]
\centering
\includegraphics[width=0.8\textwidth]{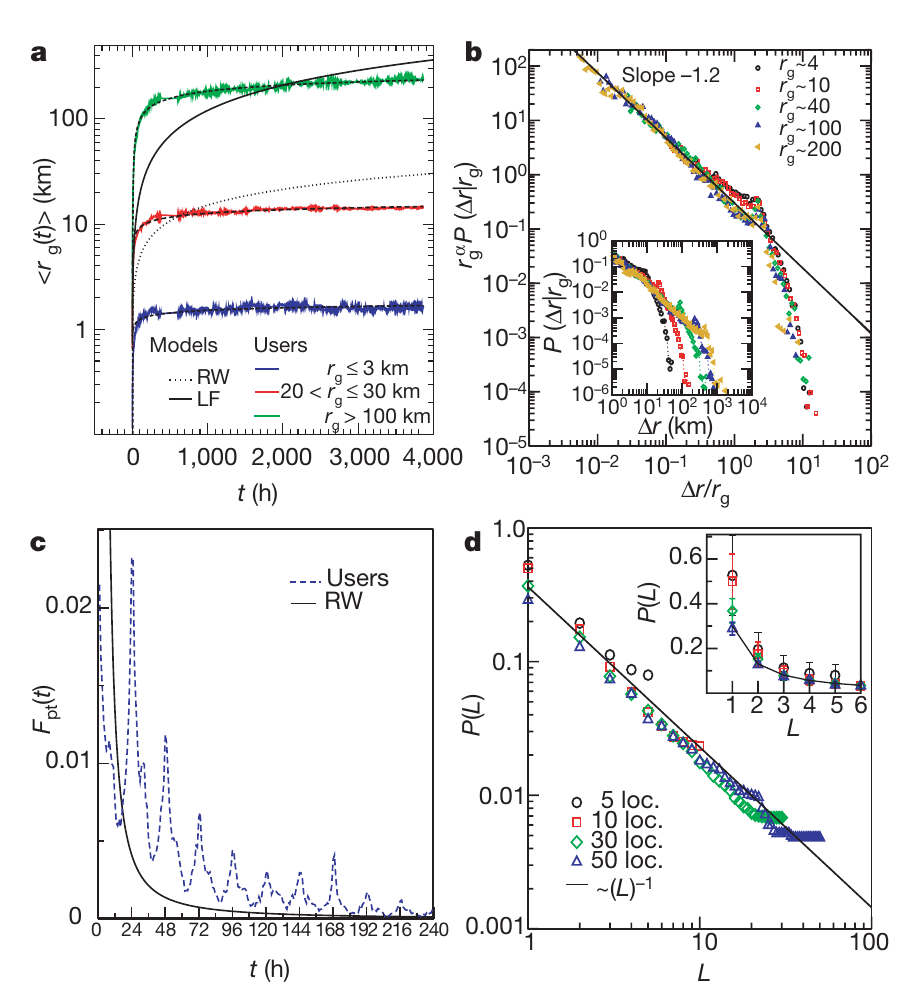}
\caption{(a) Radius of gyration $r_g(t)$ versus time for mobile phone users separated into three groups according to their final $r_g(T)$, where $T=6$ months. 
The dashed curves correspond to a logarithmic fit of the form $A + B \ln(t)$, where $A$ and $B$ are time-independent coefficients that depend on $r_g$. 
(b) Probability density function of individual travel distances $P(\Delta r | r_g)$ for users with $r_g =$ 4, 10, 40, 100 and 200 km. {\it Inset} Each group displays a different $P(\Delta r | r_g)$ distribution. {\it Main} After rescaling the distance and the distribution with $r_g$, the different curves collapse to a power law (solid line).
(c) Return probability distribution, $F_{pt}(t)$. The prominent peaks capture the tendency of humans to return regularly to the locations they visited before, in contrast with the smooth asymptotic behavior $\sim 1/(t ln(t)^2)$ (solid line) predicted for random walks. 
(d) A Zipf plot showing the frequency of visiting different locations. The symbols correspond to users that have been observed to visit $n_L = 5, 10, 30,$ and $50$ different locations. Denoting with $L$ the rank of the location listed in the order of the visit frequency, the data are well approximated by $R(L) \sim L^{-1}$. The inset is the same plot in linear scale, illustrating that $40\%$ of the time individuals are found at their first two preferred locations; bars indicate the standard error. Figure from~\cite{gonzalez_2008_understanding}.}
\label{fig:scales_gonzalez_2}
\end{figure}

Indeed, the measured value of the scaling exponent $\beta_r = 0.65 \pm 0.15$ indicates a significant degree of heterogeneity in the travel habits of the observed population. Measuring the conditional jump length distribution, $P(\Delta r | r_g)$, revealed that users with small $r_g$ travel mostly over small distances, whereas those with large $r_g$ tend to display a combination of many small and a few larger jump sizes. Once one accounts for this heterogeneity in travel habits by rescaling the distribution with respect to $r_g$, it leads to a collapse of the data onto a single curve thus,
\begin{equation}\label{eq:rgcoll}
P(\Delta r| r_g )  \sim \Delta r^{-(1+\beta_c)} F(\Delta r / r_g) \, .
\end{equation}
Here $\beta_c = 0.2 \pm 0.1$ is a ``universal'' scaling exponent, and $F(x)$ is a scaling function that is constant for $x<1$ and rapidly decreasing for $x \gg 1$. 

It is important to note that the distributions described above are not independent but are related by the equality $P(\Delta r) = \int_{r_0}^{\infty} P(\Delta r | r_g) P(r_g) d r_g$, where $P(\Delta r$) is the jump-length distribution introduced previously. If $P(\Delta r)$ has a power-law scaling exponent $\beta$, then we have $\beta = \beta_c + \beta_r$, which seems to be in good agreement with empirically measured values. 
This form of scaling corresponds to a type of random walk called a L\'evy flight (\sectionname~\ref{sec:levy}), and the results suggest that this may be the behavior of individuals up to their associated characteristic distance $r_g$ (and beyond which saturation effects take over). 

\begin{figure}[t!]
\centering
\includegraphics[width=0.6\textwidth]{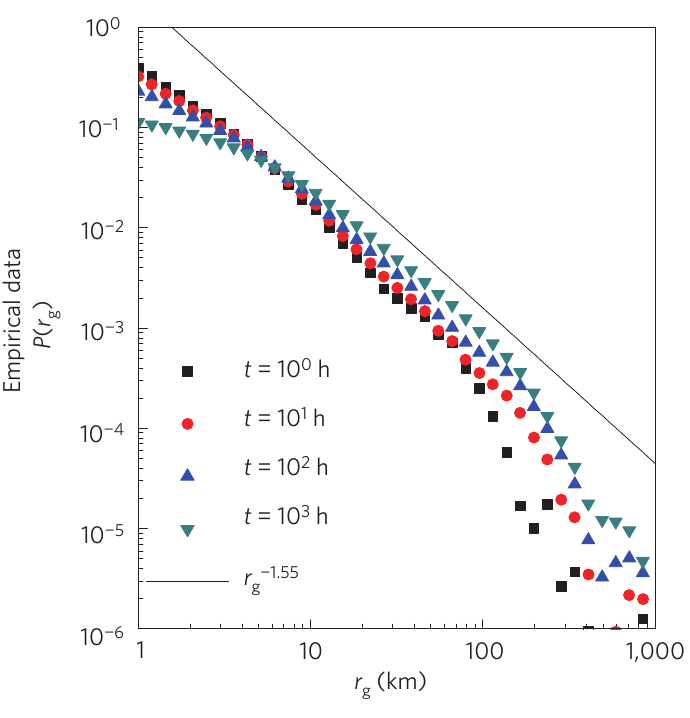}
\caption{The distribution $P(r_g)$ of the radius of gyration $r_g$ for mobile-phone users at different moments of time. The straight line, shown as a guide to the eye, represents a power-law decay with the exponent $1+\alpha \approx 1.55$. Figure from~\cite{song_2010_modelling}.}
\label{fig:scales_song_1}
\end{figure}

A way to uncover the saturation effect is to examine the time evolution of the radius of gyration, i.e. $r_g(t)$. 
In ~\cite{gonzalez_2008_understanding}, it was found that the average $r_g$ of mobile phone users displays a logarithmic increase with time, $\langle r_g(t) \rangle \sim A + B \ln t$ (see \figurename~\ref{fig:scales_gonzalez_2}), in contrast to a pure random walk where the scaling is of the form $\langle r_g(t) \rangle \sim t^{\theta}$. The observed saturation in $r_g$ can be attributed to the regularity in travel patterns of the individuals, and specifically to the high probability for an individual to return to a few highly-frequented locations. As pointed out before, the true picture is a mixture of behaviors. Supporting the conclusions made in~\cite{shin_2008_levy} while measuring the $\mathrm{MSD}$, Zhao \et\cite{zhao_2008_empirical} argue that humans engage in a mixture of superdiffusive and subdiffusive behaviors (depending on temporal scale), and the latter specifically can be explained as a consequence of the saturation of $r_g$. This was later reproduced by Song \et\cite{song_2010_modelling} and demonstrated to be the consequence of the fat-tailed distribution of the jump lengths (\figurename~\ref{fig:scales_song_1}).

The radius of gyration is a measure to characterize the typical distance traveled by an individual and it depends both on the mutual distance of the locations visited and on the time spent (or the total number of visits) in each location. 
However, the radius of gyration does not allow us to quantify the relevance of each location in determining an individual's characteristic mobility. Indeed, an individual who spends a majority of time in their most visited locations, e.g. home and work, will have a large $r_g$ if these two locations happen to be quite far from each other. Conversely, even if these most visited locations are close to each other, large values of  $r_g$ may be reported if the individual happens to visit a number of distant locations. 

\begin{figure}[t!]
\centering
\includegraphics[width=0.8\textwidth]{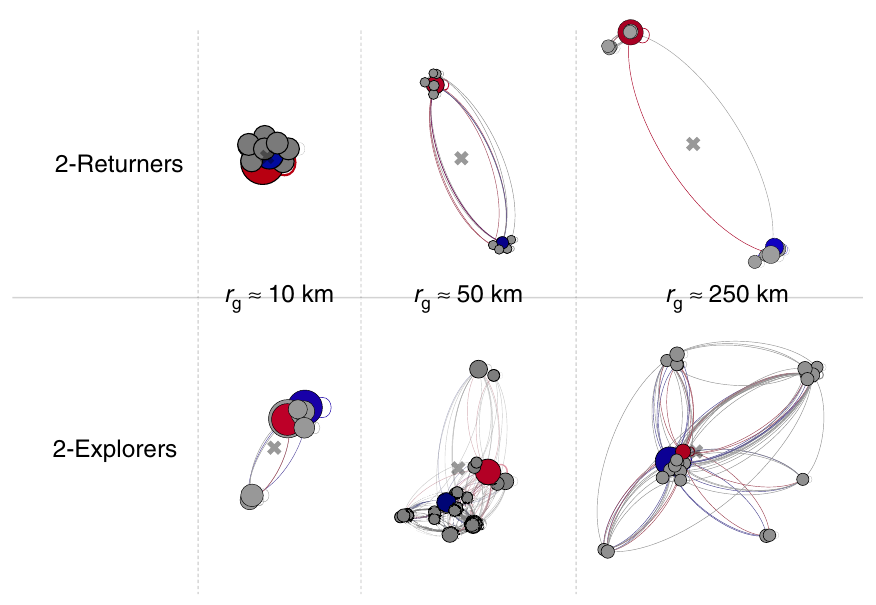}
\caption{The mobility networks of returners and explorers for $k = 2$. Nodes (circles) indicate the geographic locations visited by the individual, and each link denotes a travel observed between two locations. When the total $r_g$ is small, the two most important locations (red and blue) are close to each other for both two-explorers and two-returners. As the total radius increases the behavior of two-returners and two-explorers starts to differ; for returners, the two most important locations move away from each other; for explorers, they stay close and other clusters of locations emerge far from the center of mass (the gray cross). Figure from~\cite{pappalardo_2015_returners}.}
\label{fig:scales_pappalardo_2015}
\end{figure}

To disentangle these effects, one can study the influence of frequency of location visits on the characteristic distance traveled, by studying the $k$-radius of gyration, $r_g^{(k)}$, defined as the radius of gyration computed over an individual's $k$ most frequently visited locations $L_1, \dots, L_k$ (see next section) ~\cite{pappalardo_2015_returners}. One can express this modified form as
\begin{equation}
\label{eq:k_radius}
r_g^{(k)} = \sqrt{{\frac{1}{N_k}} \sum_{j = 1}^{k} n_j (\mathbf{r}_j - \mathbf{r}_{cm}^{\ (k)})^2},
\end{equation}
where $N_k$ is the sum of the visits to the $k$ most frequented locations, $\vec{r}_{cm}^{\ (k)}$ is the center of mass computed on those  locations and $n_j$ is the number of visits to the $j$-th most visited location. As an example, if $r_g^{(2)} \simeq r_g$, then the characteristic traveled distance is dominated by the two most frequented locations, whereas if  $r_g^{(2)} \ll r_g$ than the two most frequented locations do not offer an accurate characterization of the individual's travel pattern, requiring us to consider more locations.

The analysis of GSM and the GPS data revealed that there exist two distinct classes of individuals, {\em returners} and {\em explorers}. 
The characteristic distance traveled by $k$-returners is dominated by their recurrent movement between a few preferred locations, and their radius of gyration is well approximated by their $k$-radius of gyration for $k \geq 2$.
In contrast, $k$-explorers have a tendency to wander between a varying number of different locations and their $k$-radius of gyration is very small compared to their overall $r_g$ (see Fig.~\ref{fig:scales_pappalardo_2015}).

\subsubsection{Most Frequented Locations and Motifs}
When considering patterns in human mobility, particularly movements within a single day or week, it is essential to distinguish between locations based upon their importance. As already mentioned, people have a tendency to return home on a daily basis and therefore most daily and weekly trajectories will start and finish at the same location.
One method of quantifying the importance of a location is the use of ranks; the most visited location (likely home or work) would have rank 1, a school or local shop may have rank 2 or 3, etc. 
In ~\cite{gonzalez_2008_understanding}, the rank of a location was determined for each individual mobile user by the number of times their position was recorded in the vicinity of the cell tower covering that location. It was found that visitation frequency follows a Zipf law, that is the probability of finding a user at a location of rank $L$ is approximately $P(L) \sim 1/L$. 

Another method of distinguishing between locations is to construct each individual's mobility pattern as a network. Schneider \et\cite{schneider_2013_unravelling} used data from both mobile phone users and travel survey respondents to construct weekday mobility networks for each individual. These profiles consisted of nodes to represent locations visited and directed edges to represent trips between the locations. Every daily trip network started and ended at the home location of each user, which was determined as the location they were most often recorded at between 3:00am and 3:30am. For mobile phone users, only days where calls were made, and therefore location was recorded, during 8 or more 30-min time slots were included. Daily networks were constructed for weekdays only in order to identify patterns in mobility during a typical day.


It was found that $\sim$ 90\% of the recorded trips made by all users can be described with just 17 daily networks. These 17 trip patterns can be described as motifs; a sub-network within a complex network ~\cite{alon_2007_network}. In this case, a daily network was considered to be a motif if it occurred more than 0.5\% in the datasets. The result that the daily patterns of human mobility can be constructed by just 17 trip networks suggests that these motifs represent the underlying regularities that exist in our daily movements and are therefore useful for the accurate modeling and simulation of humans' mobility patterns.

\subsubsection{Origin-Destination Matrices}\label{sec:odmatrix}

The Origin-Destination (OD) matrix {\bf T} is the standard object in aggregated mobility studies and transport planning. It provides an estimate of the number of individuals traveling between locations in a given area, over a given period of time. More precisely, an OD matrix is a $n\times m$ matrix where $n$ is the number of different ``Origin'' zones, $m$ is the number of ``Destination'' zones, and $T_{ij}$ is the number of people traveling from zone $i$ to
zone $j$. More commonly, an area under study is partitioned to an equal number of origin and destination points and therefore, $n=m$. The size $n$ of the OD matrix then depends on the spatial scale/resolution at which the data has been collected. Traditionally, zones are administrative units, whose size may vary from census and electoral units to entire municipalities, departments or states, depending on the question that motivated the development of the OD matrix. Obviously, the maximal spatial resolution of an OD matrix depends on the data source.

An OD matrix can be empirically derived, being estimated from
household or roadside travel surveys, traffic counts, and more
recently from individual digital footprints (see below). It can
also be the output of a model, like in the classic 4-step transport
model widely used in urban planning\footnote{1-Generate travel demands and offers in each spatial unit; 2-Distribute trips in space (generally thanks to a gravity model (see \sectionname~\ref{sec:models}); 3-Evaluate the model choice; and 4-Assign trips to routes}. OD
matrices and the 4-steps framework have been used in transport
planning since at least the middle of the $20^{\text{th}}$ century, and there exists an enormous literature that covers its origins, uses, models, synthetic indicators, and so on~\cite{ortuzar_2011_modeling}.

OD matrices are of specific importance to models of aggregate flow, that is models of human mobility at the population level, rather than individual level. Examples of such models are gravity and intervening opportunities models (\sectionname~\ref{sec:models}). 
In empirical OD matrices, $T_{ij}$ indicates the number of travelers from $i$ to $j$ measured during observation time window, whereas if the OD matrix is a model's estimate or prediction $T_{ij}$ usually indicates the expected (average) number of travelers between the two locations. The OD matrix diagonal elements are usually 0 to indicate that people only move between distinct locations. In some studies the diagonal elements are larger than zero and indicate the number of non-traveling individuals in the origin location (or those who travel within the origin location). 
The fraction of travelers from the origin $i$ to all other locations can be calculated as $p_{ij} = T_{ij} / T_i$,  where $T_i = \sum_j T_{ij}$. In this case, the entry $p_{ij}$ may represent the probability of an individual located at $i$ to select location $j$ as their destination over all other possible locations.




Until the 2000's, the traditional approaches to estimate OD matrices consisted in relying on travel surveys or counting. The drawback of such approaches are related to their cost to setup, frequency of updates, and that they cover a limited sample of individuals (a few hundreds or thousands of households)~\cite{iqbal_2014_development}. Consequently, to circumvent these difficulties, a large body of literature has dealt with extracting OD matrices from recently available, individual digital footprints \cite{white_2002_extracting, caceres_2007_deriving, isaacman_2010_tale, calabrese_2011_estimating, jiang_2013_review, iqbal_2014_development, lenormand_2014_cross, alexander_2015_origin, toole_2015_path}. 
Early work on this topic has already been summarized in a dedicated review a few years ago~\cite{caceres_2008_review}. Here we don't go deeply into the details of the numerous methods that have been proposed, and give only the general picture for the case of average daily journey-to-work commuting. 

Starting from time-sequences of successive locations, the goal is to identify important places (``anchor points'') for each individual, such as the residence and workplace. The heuristic underlying method is straightforward: due to the circadian regular daily rhythms of human activity, one can make the reasonable assumption that for most individuals, the most frequent location of mobile devices will be very near the residence during non-working hours of weekdays, and during weekends. Similarly, during working hours on weekdays, it is likely that for most individuals their devices will be located just next to their workplace or location of primary activity (which include students, unemployed, retired, and employed individuals with a non-fixed workplace)~\cite{tizzoni_2014_use, lenormand_2014_cross, alexander_2015_origin, toole_2015_path}. In some cases validation of methodology was conducted by asking a small group of volunteers to complete a survey, whose results were then compared to the key locations extracted from the recorded activity of their mobile devices \cite{isaacman_2010_tale}.

Alexander \et\cite{alexander_2015_origin} proposed a method for constructing an OD matrix from mobile phone records. Such a method allows for data at the individual level to be aggregated in a way that lends itself to analysis at the population level. Implementing the proposed method, in order to infer a user's origin and destination, clustered locations are extracted from the mobile phone record. By clustering locations in close proximity to one another into a single origin or destination, noise from the data, such as inexact triangulation, is eliminated thus allowing for a more accurate analysis. The clustered locations are then assigned a type: work, home, or other. The assignment takes into account the time of day a user is observed there, the duration of their stay, and the day of the week.
By the nature of mobile phone records, location is only recorded when a user makes or receives a call, therefore the arrival times and duration of stay at a given location may not reflect the true times. To correct for this, surveys on trips may be used to assign true arrival times and duration from a probability distribution derived from the survey data. Then, trips can be constructed for every user between each location at which they are observed. In order to ensure that the final OD matrix is representative, users that do not satisfy certain conditions, such as on average of at least one visit to the home location per day, may be removed from the data set. The daily trips are then combined for all users and locations allowing for the aggregate flow between all possible locations to be determined and an origin destination matrix to be constructed.

Lenormand \et\cite{lenormand_2014_cross} compared the journey-to-work OD matrices extracted from three different data sources (census, mobile phone CDR data, and Twitter data) in Madrid and Barcelona, two of the largest European urban areas. Projecting residential and workplace locations identified in each dataset on regular square grids, in order to have a common spatial resolution for each dataset, they measured the correlation between the OD matrices estimated through each data source, as a function of the level of spatial aggregation. They found good correlation for square grids of size $2km$ ($\rho > 0.9$), and very good correlation between the various sources when aggregating the individual data sources (Twitter, CDR data) on the map of administrative units (municipalities) used by the travel surveys ($\rho > 0.99$) (here $\rho$ is the Pearson correlation coefficient). Using a slightly different method that aggregates nearby cell towers when identifying important locations, Alexander \et\cite{alexander_2015_origin} estimated OD matrices of daily average commuting trips from CDR data in the Boston area, and then performed a sensitivity analysis of the correlation between this OD and the OD issued from transport surveys at different levels of spatial aggregation. Tizzoni \et\cite{tizzoni_2014_use} also compared OD estimated from census and mobile phone data at different scales of spatial aggregation. They found lower correlation values, but applied a simpler method to estimate the OD from mobile phone data, when the aforementioned papers applied more stringent filters to remove outliers, fusion phone antennas close to one another, etc. Several algorithms in the literature are reviewed in \cite{toole_2015_path}, and further successive refinements and improvements in the methodology have been made since~\cite{jiang_2013_review,iqbal_2014_development,c_2015_analyzing,alexander_2015_origin}.

\subsection{Physics of mobility}
\subsubsection{Distance, travel time, and effective speed}

\begin{figure}[t!]
  \centering
\includegraphics[width=.8\textwidth]{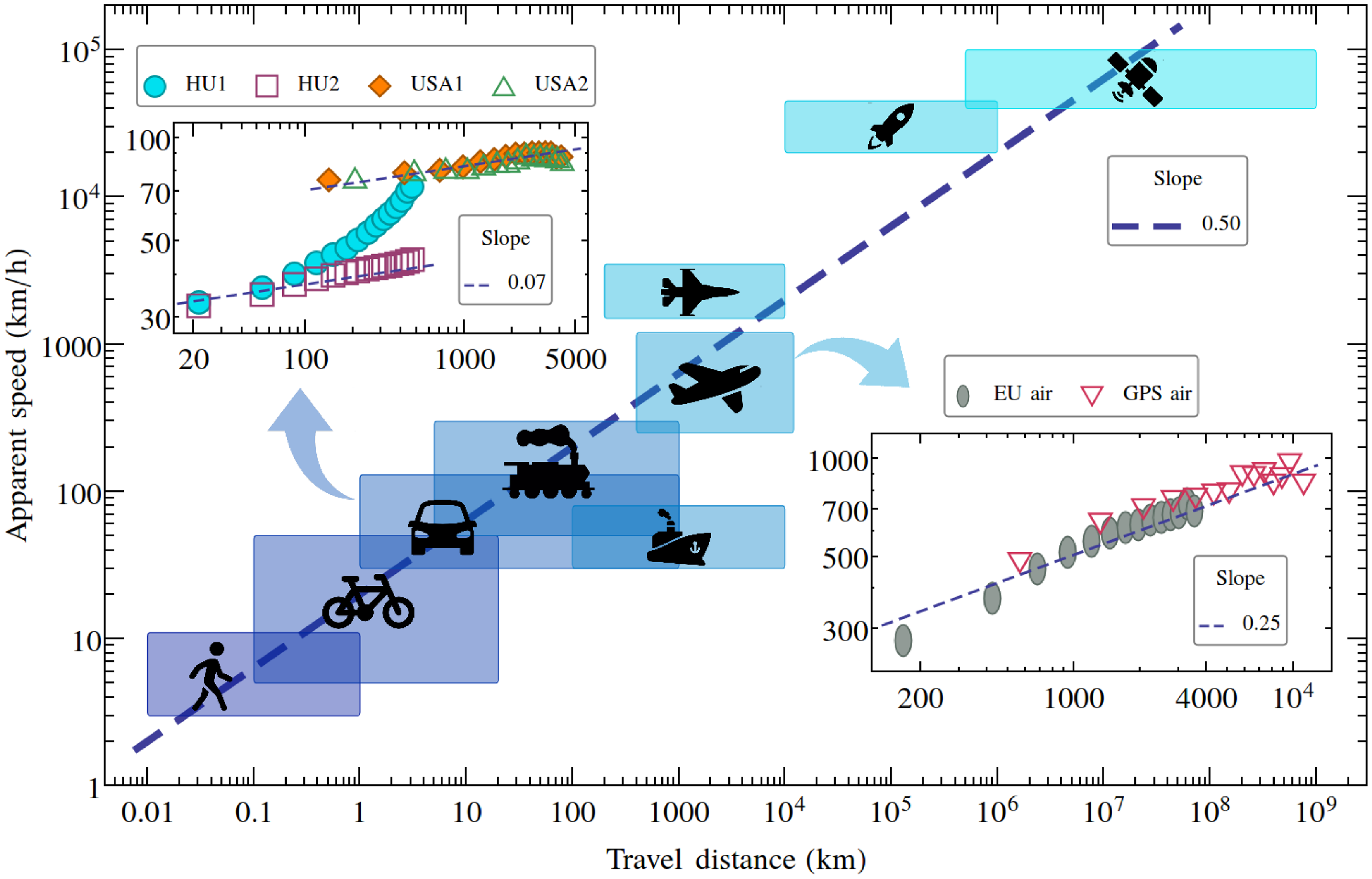}
  \caption{Apparent speed versus travel distance. The boxes represent the intervals for the different transportation modes. In the insets, the average result for cars (top left) and for plane travel (bottom right) are shown. The dashed lines represent power law fits. Figure from \cite{varga_2016_further}.}
\label{fig:scales_varga}
\end{figure}

In essence, the metrics described above, deal with the concepts of distances traveled and the time elapsed within and between journeys. Another important feature, which has not been discussed, is the notion of \emph{speed}. To understand the relation between these quantities, it is important to note that mobility occurs over a large variety of distances. Framing it in the context of urban agglomerations, intra-urban mobility occurs over distances typically in the range of 1-10 kms, while inter-urban displacements occur generally over a wider range of the order of 100 kms (or more for large countries such as the US), and inter-country and intercontinental trips cover distances typically of the order of 1000 kms. While to first order, one may think of time spent on a given trip being roughly proportional to the distance traveled, one has to take into account the mode of transportation, which itself depends on the distance. For short-range travel, slow transportation modes with many stops are utilized (such as walking or public transportation), and for longer distances, one typically takes fast trains or planes with comparatively fewer stops. Thus the relation between distance, travel-times, and therefore speed, is fairly nuanced, as can be seen in \figurename~\ref{fig:scales_varga}, where we show the relation between \emph{apparent} speed (\ie the geodesic  distance divided by the travel time) and the travel distance~\cite{varga_2016_further}.

In general, a trip can consist of multiple connections and may in fact be multimodal (multiple components of travel) with the corresponding waiting-time distributions associated with the modes of transportation~\cite{gallotti_2014_anatomy}. Also, it is observed that the apparent speed $\overline{v}$ increases with travel distance according to the following power-law functional form~\cite{varga_2016_further} 
\begin{equation}
\overline{v}\sim r^\beta,
\end{equation}
where $\beta\approx 0.5$. This particular dependence is primarily due to a combination of the hierarchical structure of transportation systems~\cite{gallotti_2016_stochastic} and the fact that waiting-times (parking, take-off, landing, etc) decrease in proportion to  trip distance. 

\begin{figure}[t!]
  \centering
\includegraphics[width=.8\textwidth]{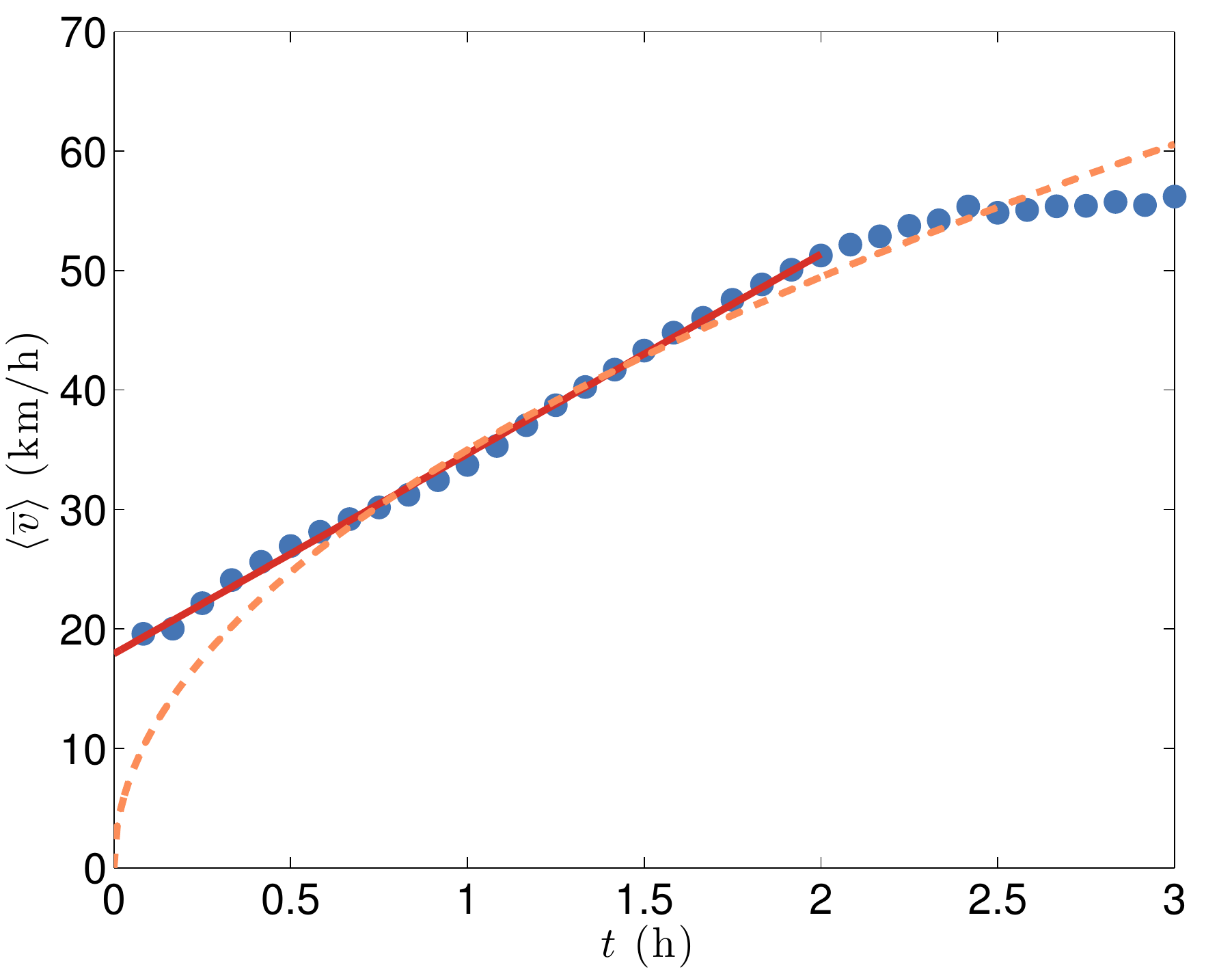}
  \caption{Empirical average speed versus the duration of the trip obtained from GPS data for cars (blue dots). The red solid line represents a fit to the form seen in Eq.~\eqref{eq:at} with $v_0 = 17.9$ km h$^{-1}$ and $a = 16.7$ km h$^{-2}$.  Observe the saturation at $t > 2$ hours due to the finite number of layers in the transportation hierarchy. The orange dashed lines represents the best fit to $\langle \mid \overline{v} \mid \rangle \propto \sqrt{t}$ which corresponds to a Brownian acceleration model. Figure from \cite{gallotti_2016_stochastic}.}
\label{fig:scales_gallotti}
\end{figure}

For example, it has been observed for public transportation as well as personal cars~\cite{gallotti_2016_stochastic} that a plot of the average trip velocity as a function of the trip duration reveals an effective acceleration (\figurename~\ref{fig:scales_gallotti}) of the form 
\begin{equation}
\langle\overline{v}\rangle = v_0+at.
\label{eq:at}
\end{equation}
This effective acceleration is a direct consequence of the hierarchical organization of roads: the longer the trip, the more likely that higher velocity roads such as highways are used, thus trips can be decomposed into a ascending cascade to faster roads, followed by a descending cascade when approaching the target location. \equationname~\eqref{eq:at} implies that the distance evolves with trip duration as 
\begin{equation}
r\simeq v_0t+\frac{1}{2}at^2,
\label{eq:dt2}
\end{equation}
from which the relation $\langle\overline{v}\rangle \sim \sqrt{r}$ directly follows~\cite{varga_2016_further}.

\begin{figure}[t!]
  \centering
\includegraphics[angle=-90,width=.85\textwidth]{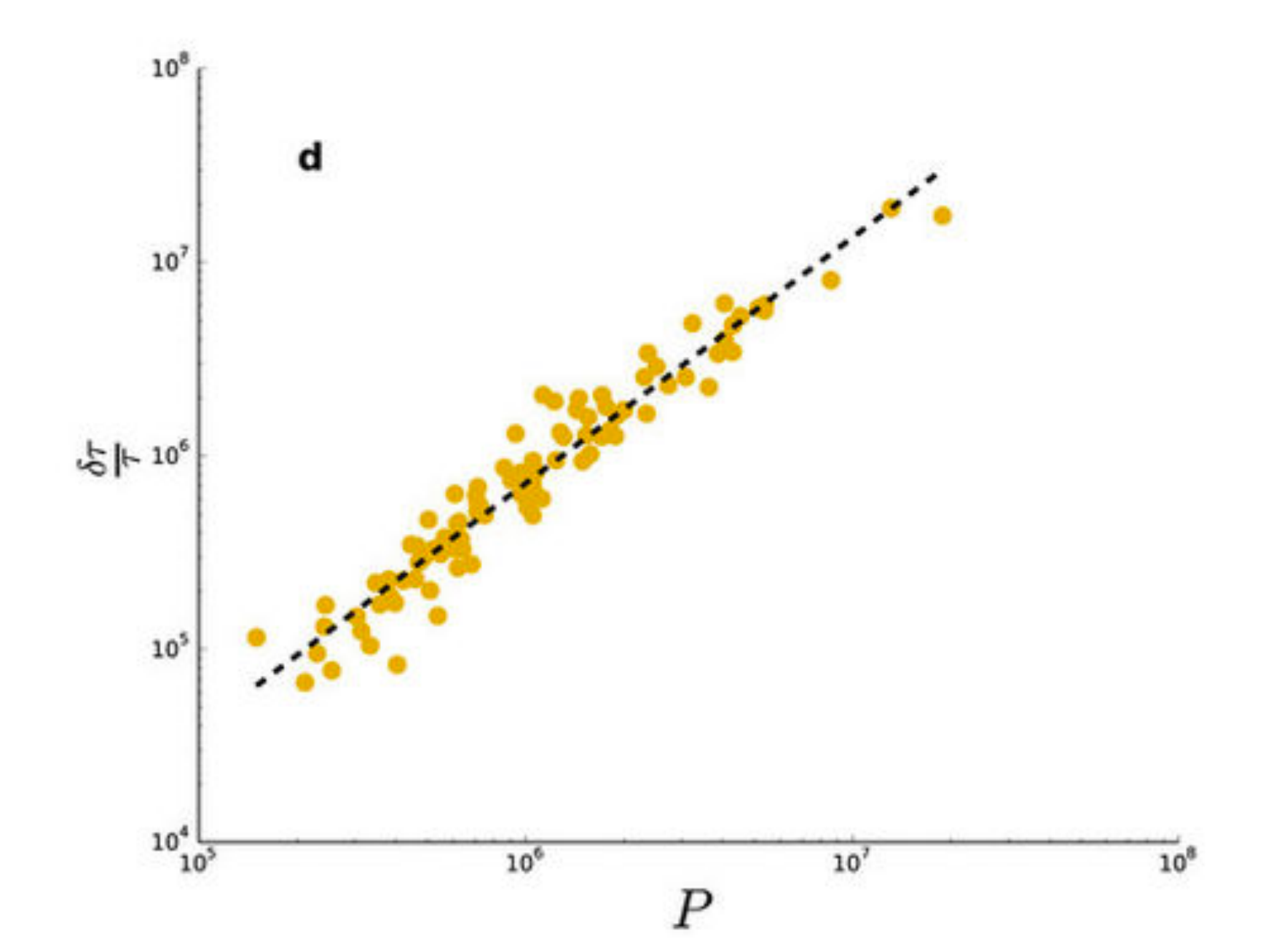}
  \caption{Variation of the total delay due to congestion with population for 97 urbanized areas in the US. A power law fit gives an exponent  of $1.270 \pm 0.067$. Figure from \cite{louf_2014_how}.} 
\label{fig:louf_congestion}
\end{figure}

\subsubsection{Travel time budget}
In the context of intra-urban mobility, Marchetti~\cite{Marchetti_1994}, based on ideas developed by Zahavi~\cite{Zahavi_1977}, proposed the idea of a travel time budget of about one hour per day irrespective of location. This implies that with improvements in transportation technology and corresponding increase in speed, a greater amount of distance is covered within the budgeted time, thus allowing for urban sprawl. This assumption can be reformulated as the \emph{rational locator hypothesis} \cite{levinson_2005_rational} which posits that individuals maintain approximately a constant journey-to-work travel times by adjusting their home and workplace. This was revisited in \cite{levinson_2005_rational} on data for travel times in Washington DC for 1968, 1988, 1994, and Twin Cities for 1990 and 2000. The results show that there is a strong dependence on geography: for the Washington DC greater urban area, travel times are relatively stable, while the results for Twin Cities show a marked increase of the commute time.
This has also been confirmed in another study \cite{louf_2014_how} that discussed the impact of congestion on mobility patterns. In particular, it has been measured for US cities that typical travel delay due to congestion increases with the population as 
\begin{equation}
\Delta\tau\sim P^{1+\delta}
\end{equation}
where $\delta\approx 0.3$, indicating that average commuting time increases with population, due to the (nonlinear) congestion effects, in sharp contrast with the travel time budget hypothesis (See \figurename~\ref{fig:louf_congestion}).

%

\subsubsection{Energy arguments}

\begin{figure}[t!]
  \centering
\includegraphics[angle=-90,width=.85\textwidth]{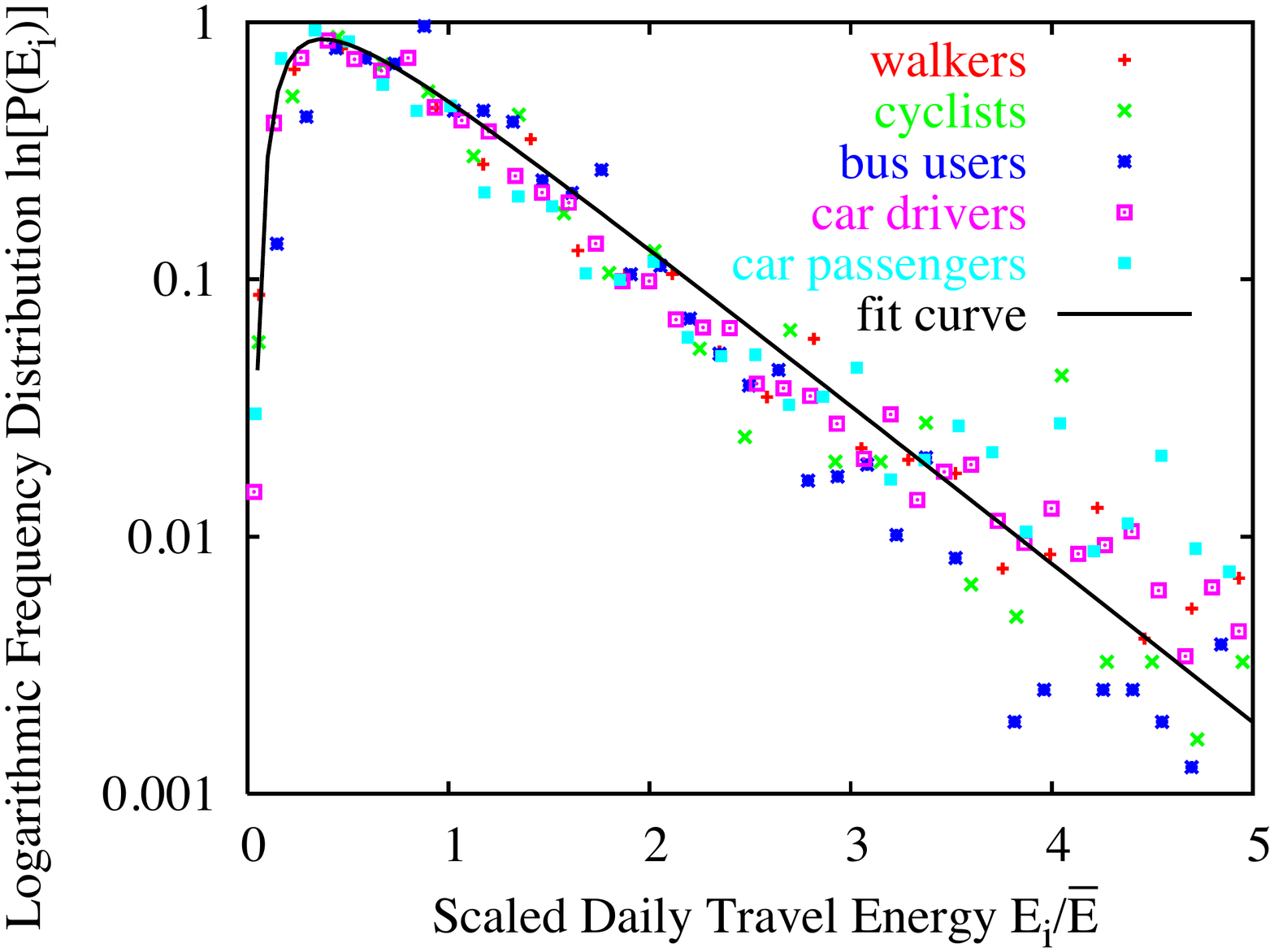}
  \caption{Rescaled travel time distribution for different transport modes (linear-log). Points represent different travel modes and the solid line is a fit to the rescaled energy distribution~\eqref{eq:energyrescaled}. Figure from \cite{kolbl_2003_energy}.}
\label{fig:helbing2}
\end{figure}

Mobility, ultimately, is about energy. Indeed, moving an object of a certain mass at a certain speed requires a given amount of energy. Consequently, one would expect to apply energy concepts to understand human travel behavior \cite{kolbl_2003_energy}. First, it has been observed that the average travel times for different modes of transportation are inversely proportional to the energy consumption rates measured for the respective physical activities. Second, when daily travel-time distributions of different transport modes such as walking, cycling, bus, or car travel are appropriately scaled, they turn out to have a universal functional relationship (see \figurename~\ref{fig:helbing2}). 

This corresponds to a canonical-like energy distribution (with exceptions for short trips) hinting at a law of constant average energy consumption related to daily traveling. The argument, first proposed by Kolbl \et\cite{kolbl_2003_energy},  goes as follows.
If we define the energy $E_i$ spent per transportation mode $i$, the average energy consumption per day $\overline{E}$ is constant and independent of the mode of transportation. The corresponding 
entropy is given by
\begin{equation}
S=-\int P(E_i)\ln P(E_i)\mathrm{d}E_i,
\end{equation}
and the constraints on the energy distribution read
\begin{equation}
\int P(E_i)\mathrm{d}E_i =1; \quad \int E_iP(E_i) =\overline{E}\mathrm{d}E_i,
\end{equation}
which leads to the canonical distribution
\begin{equation}
P(E_i)\sim \mathrm{e}^{-\beta E_i}.
\end{equation}
However, the data indicates that the probability to expend small energy is vanishing, and this was accounted for by introducing a cutoff of the form $e^{-\alpha\overline{E}/E_i}$. This cutoff term is reminiscent of the so-called ``Simonson effect''~\cite{hettinger_1989}, where short trips are not very likely to be taken with a given mode if the energy spent is much less than $\alpha \overline{E}$. The complete energy distribution is then given by
\begin{equation}
P(E_i)\sim \mathrm{e}^{-\alpha\overline{E}/E_i-\beta E_i}.
\label{eq:energyrescaled}
\end{equation}
This finding, highlights the importance of physical concepts in understanding mobility and more generally social phenomena. In particular, it contains only physical variables such as travel times and energies, that are both measurable, in contrast with utilities introduced in classical choice modeling that describe preferences that are not measurable.  Yet, as the authors point out, there are several issues that need to be accounted for, including measurement errors as well as multimodal trips that combine different transportation modes. 

\subsection{Interpolation of scales: the importance of multimodality}
\label{sec:multimodal}

As discussed previously, mobility occurs over multiple spatio-temporal scales (multimodal) and thus a comprehensive picture of human mobility requires an accounting of the effects of multimodality. Indeed, the transition between different modes of transport necessarily implies a temporal cost that increases in proportion to the complexity of a trip (number of modes, or indeed level of mode). For example, in the UK, a rough estimate shows that roughly $23\%$ of travel time is accounted for by connections made between modes in trips \cite{gallotti_2014_anatomy}.

\begin{figure}[t!]
\centering
\includegraphics[angle=0, width=0.8\textwidth]{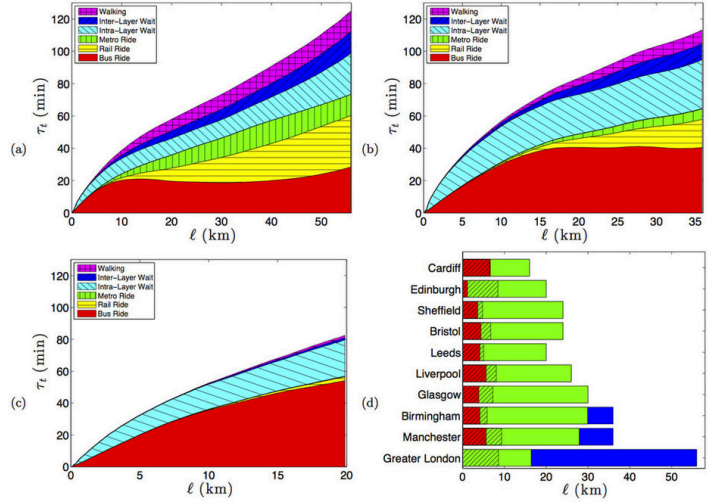}
\caption{The Anatomy of the transportation networks of selected cities in the UK. (a) London, (b) Manchester, (c) Edinburgh: total travel time in function of trip length separated by mode of travel. (d) Different colors represent different regimes versus the trip length. (i) Red: the trips are mostly done on the bus layer and display waiting times larger than riding times (a regime not present in London). (ii) Green: riding times exceed waiting times and most of the distance is covered in the bus layer. (iii) Blue: riding times exceed waiting times and most of the distance is covered in the metro and rail layers. Figure from~\cite{gallotti_2014_anatomy}.}
\label{fig:anatomy}
\end{figure}

As showed in \figurename~\ref{fig:anatomy}, the average structure of travel time varies with trip length $\ell$ and between cities. With the exception of London, short trips are composed of longer waiting times than riding times (i.e more time is spent being stationary than on the move), where the waiting times are mostly intra-layer in nature, due to bus-bus interchanges. If the transportation network is particularly multi-modal, inter-layer waiting times and walking times start to play a significant role for longer values of $\ell$. 

The effect of multiple scales on movement can be unpacked by defining two types of trajectories. One is the (ideal) quickest path that connects two points and neglects delays due to inter-modal connections (walking and waiting times). For this ideal path to exist, one would need perfect synchronization between modes (bus schedules aligning perfectly with train schedules for example). The other (more realistic) type of trajectories are  so-called \emph{time-respecting} paths that are the quickest paths accounting for inter-modal effects such as arrival, departure and connection times. 

\begin{figure}[t!]
 \centering
 \includegraphics[angle=0, width=0.7\textwidth]{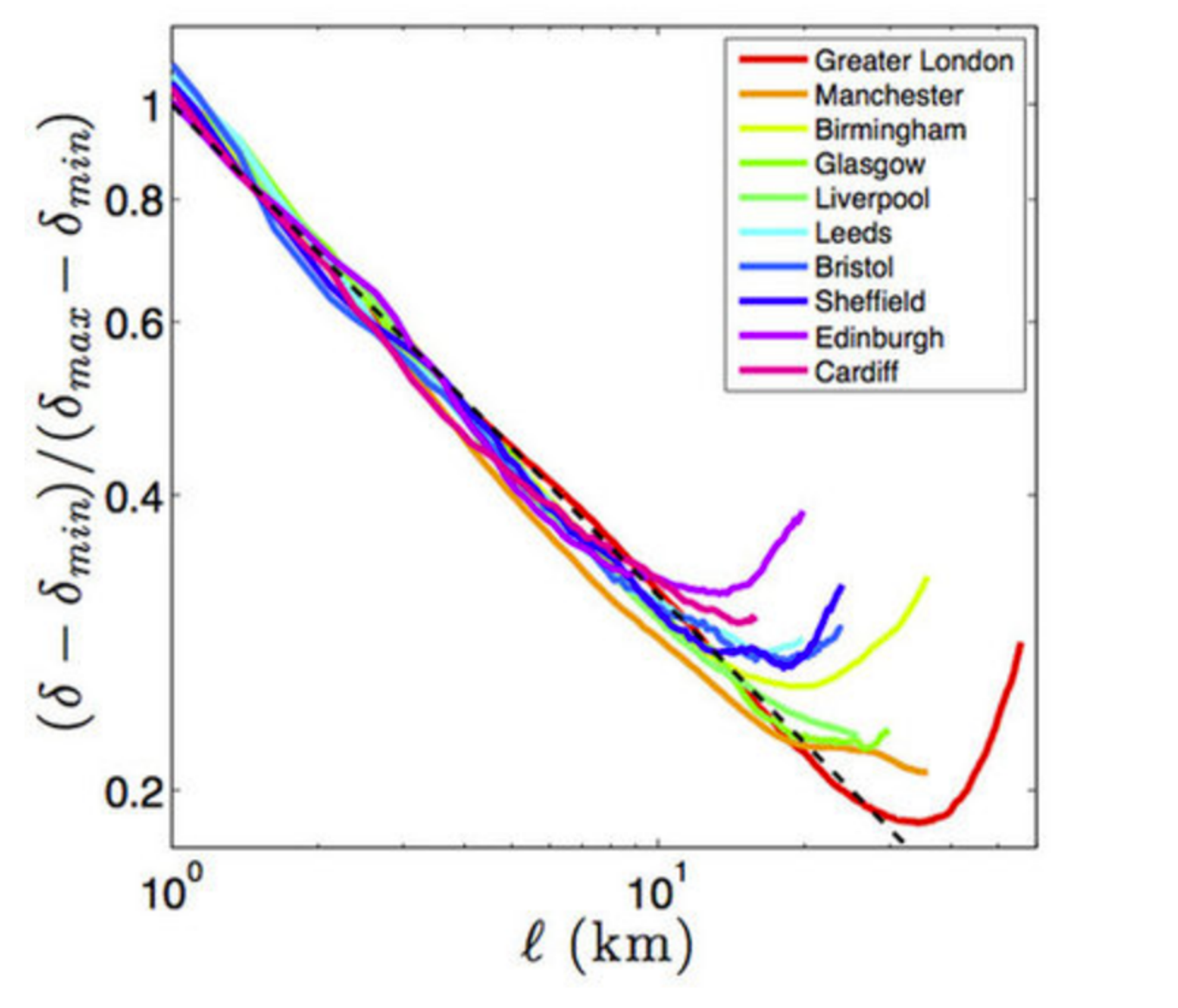}
 \caption{Dependence of the synchronization inefficiency $\delta$ on path length $l$ for cities in the UK. All cities appear to collapse onto a curve described by Eq.~\eqref{eq:deltafun}. Figure from~\cite{gallotti_2014_anatomy}.}
 \label{fig:anatomyb}
 \end{figure}

For OD pairs $i$ and $j$, denoting the ideal path time as $\tau_m(i,j)$ and the corresponding time-respecting path as $\tau_t(i,j)$, the delay due to the lack of synchronization can be measured by
\begin{equation}
\delta(i,j)=\frac{\tau_t(i,j)}{\tau_m(i,j)}-1.
\end{equation}
Averaging this ratio over all OD pairs in a city, one obtains a characteristic delay $\overline{\delta}$. As an example, for cities in the UK the variation of $\overline{\delta}$ with trip length $l$ is unimodal, reaching its maximum $\delta_{max}$ for short trips and then decreasing with distance $\ell$ according to 
\begin{equation}
\overline{\delta} \approx \delta_{min} + \frac{\delta_{max}-\delta_{min}}{\ell^\nu},
\label{eq:deltafun}
\end{equation}
where $\nu\approx 0.5$. This surprising collapse (see \figurename~\ref{fig:anatomyb}) suggests an underlying process describing the accumulation of waiting and walking times along time-respecting paths.

It is clear for a given trip that the number of modes and the frequency of their use play an important role in determining $\overline{\delta}$. To measure their effect a natural quantity is the average number of stop events per unit time 
\begin{equation}
\Omega = \frac{\sum_\alpha C_\alpha}{\Delta t_{\alpha}},
\end{equation}
where $C_\alpha$ is the number of stop events in mode $\alpha$ and $\Delta t_{\alpha}$ the duration spent in that mode.

The quantity $\Omega$ can be thought of as a measure of the efficiency of transportation modes in terms of synchronization. The connection between  $\bar\delta$ and $\Omega$ appears to follow a power law \cite{gallotti_2014_anatomy}
\begin{equation}
\bar\delta \sim  \Omega^{-\mu},
\label{eq:deltaomega}
\end{equation}
where $\mu\approx 0.3\pm 0.1$.  The trend is due to the fact that larger values of $\Omega$ implies larger frequency and thus a better synchronization between modes. The observed small value of $\mu$ is however bad news in terms of efficiency: Decreasing inefficiency by a factor of 2 requires a 10-fold increase in $\Omega$ meaning essentially that the trip would need to be unimodal.


Thus as the results show, for large cities with multi-modal transportation system, the temporal cost due to interchange between layers plays a key role in the statistics of trips. Indeed, a challenge is posed due to the entanglement between  the temporal and multilayer aspects of the system. 




\section{General Mobility Models}
\label{sec:models}


Models of human mobility can be aimed at reproducing individual mobility patterns or general population flows. 
In both cases one must necessarily take into account the characteristic spatial and temporal scales of the mobility process, which can vary from hundreds of meters to thousands of kilometers and from hours to years.
For this reason, each of these cases have been tackled with distinct modeling frameworks. 
Individual mobility is subject to a certain level of uncertainty associated with free will and arbitrariness in the actions of individuals, leading to a degree of stochasticity in trip patterns. Consequently, minimal models borrow concepts and methods from random walks and Brownian motion. 
However, several studies highlighted that individual trajectories are far from random, possessing a high degree of regularity and predictability, which can be exploited to predict an individual's future whereabouts and to construct realistic generative models of individual mobility. 
At the level of population flows, models describe the aggregate mobility of many individuals and aim to reproduce Origin-Destination (OD) matrices by estimating the average number of travelers between any two spatial regions (e.g. municipalities) per unit time (i.e. daily in the case of commuting flows, yearly in the case of migration flows).
Most modeling approaches derive the mobility flows as a function of a few relevant variables of the regions considered, such as mutual distances, population levels, GDP per capita, etc. 
The next sections review the state-of-the-art mobility models starting from individual-like approaches, moving on to flows at the population level and concluding by the interpolation between these scales  (inter-modality).

\subsection{Individual-Level (Random walks)}
\label{sec:rw}

A random walk is mathematically defined as a path formed by successive discrete random steps although they can be described in the continuum limit as well  (Sec.~\ref{sec:ctrw}). The simplest version, however, deals with spatial displacements $\Delta X_{i}$ that are taken at discrete times $t_i$. If $x_0 = 0$ corresponds to the initial position of the walker at time $t_i=0$ (we can arbitrarily fix the origin in the initial position of the walker), then the position after $N$ steps is given by the random variable
\begin{equation}
X(t_N) = \sum_{i=1}^{N} \Delta X_{i},
\end{equation}  
where each displacement $\Delta X_{i}$ is a random variable extracted from a probability distribution $f(\Delta x )$,   
and draws are assumed to be statistically independent. The probability distribution $f(\Delta x)$ can be used to determine the probability density function (PDF), $P(x,t)$ for the process to be at position $x$ at time $t$ which completely characterizes the nature of the walk and the related spatial and temporal measures. For example, the mean square displacement (\sectionname ~\ref{sec:msd}) corresponds to the second moment $\mathrm{MSD}(t) = \langle X(t)^2 \rangle$, 
and the $n$-th moment is obtained from
\begin{equation}
\langle X(t)^n \rangle = \int_{-\infty}^{\infty} x^n P(x,t) d x, \label{individual:moment}
\end{equation}
where brackets indicate ensemble averages over multiple realizations of walks.

Of particular interest when analyzing models of individual mobility is the scaling of the square root of the mean squared displacement (RMSD), $R(t) = \sqrt{\mathrm{MSD}(t)}$, with time $t$; a relationship that characterizes the speed of displacement from the origin with time. The scaling of the MSD can be used to categorize the type of diffusive motion of the random walker. Ordinary Brownian motion (Sec.~\ref{sec:brownian}) has a MSD that scales linearly with time (for any spatial dimension) therefore on average we expect that after a time $t$ the distance of the walker from the origin is proportional to the square root of the elapsed time; 
$R(t) \sim t^{1/2}$. Taking this definition of diffusion, random walks that have displacement growing at a slower rate than $t^{1/2}$ are said to be {\it sub-diffusive}. In contrast, if displacement grows at a rate faster than $t^{1/2}$, the random walk is classed as {\it super-diffusive}. The models of random walks that can lead to these different behaviors are outlined below. 

As an example, the most basic form of a random walk is the discrete symmetric random walk in 1D, for which $\Delta X_{i} = \pm 1$ with equal probability. For this simple case after an elapsed time time $t = N$ (where $N$ is the number of steps taken by the walker), the first two moments are $\langle X_N \rangle$ = 0 and $\langle X_{N}^{2} \rangle = N$. 

\subsubsection{Brownian Motion}
\label{sec:brownian}

Brownian motion is a class of random walk originally developed to describe the motion of a particle suspended in a fluid (liquid or gas) \cite{einstein_1905_movement}. Such a particle undergoes many rapid instantaneous collisions with much smaller particles in the medium, resulting in a trajectory characterized by a series of irregular and random displacements. Mathematically, a 1-dimensional Brownian motion is a random walk in the space of real numbers $\mathbb{R}$ with independent and normally distributed increments where the probability to observe a displacement of magnitude $X$ from the origin location after a time $t$ is Gaussian distributed with mean zero and variance proportional to $t$. Brownian motion can be defined as a limit of the discrete symmetric random walk. Let us assume that the particle can take steps of length $1/\sqrt{k}$ to its left or right with equal probability, and that after time $t$ the particle has taken $N=t\,k$ steps. For a given $k$, the displacement of the particle at time $t$ is given by
\begin{equation}
X_{k}(t) = \frac{1}{\sqrt{k}} \sum_{i=1}^{tk} \Delta X_{i},
\end{equation}
and taking the limit $k \to \infty$ leads to Brownian motion as a consequence of the Central Limit Theorem (CLT). 
In fact, in the large $k$ limit $X_{k}(t)$ tends to $X(t)$, whose PDF is 
\begin{equation}
P(x,t) = \frac{1}{\sqrt{2 \, \pi\, t \sigma^2}} e ^{\frac{-(x-\mu k t)^2}{2\, t \sigma^2}},\label{eq:3}
\end{equation}
where $\mu = \langle \Delta X \rangle$ and $\sigma^2 = \langle \Delta X^2 \rangle$ are the mean and variance of the random walk displacements.
For this case, the first moment is zero and $\sigma^2=1$, so the MSD is simply $t$ and the scaling of the RMSD is $R(t) \sim t^{1/2}$, corresponding to ordinary diffusion. 
Using a similar limit, it can be shown that a $d$-dimensional random walk converges to a $d$-dimensional Brownian motion, for which the probability that the walker is found at distance $x$ from the initial position is 
$P(x,t) = (4 \pi \, t )^{-d/2} e ^{\frac{-x^2}{4\, t}}$,
and the RMSD always scales as the square root of $t$, $R(t) \sim t^{1/2}$.

\subsubsection{L\'evy Flight}
\label{sec:levy}

Unlike Brownian motion, there is a class of random walks called L\'evy Flights, for which one cannot use the CLT. A L\'evy Flight is composed of a series of small displacements, interspersed occasionally by a very large displacement. It is formally defined as the sum of independent identically distributed random variables whose PDF for a single jump has a divergent second moment due to a long-tailed distribution of the form
\begin{equation}
f(\Delta x) \sim \frac{1}{\Delta x^{1+\beta}}, \label{eq:7}
\end{equation} 
with $0 < \beta < 2$. 
If the displacement after $k$ steps of size $1 / k^{1/\beta}$ is defined as $X_k(t)$, then the random variable $Z_N$ defined as the re-scaled sum of $N = t\, k$ independent random variables distributed as \equationname~(\ref{eq:7}), takes the form
\begin{equation} \label{eq:zy}
Z_N = \frac{1}{(t \, k)^{1/\beta}}\sum_{i=1}^{t\, k}\Delta X_i = \frac{1}{t^{1/\beta}}X_k(t).
\end{equation}
The rescaled variable satisfies a generalization of the CLT, namely the L\'evy-Khintchin theorem which states that the PDF of $Z_N$ in the limit $N \to \infty$ is a so called $\alpha$-stable (L\'evy) distribution. These distributions do not have a closed formed in real space; instead the characteristic function can be written in Fourier space, and for this particular example, the tail would show the same power law behavior as \equationname~(\ref{eq:7}). Changing variables from $Z_N$ to $X_k$ via \equationname~(\ref{eq:zy}) we obtain
\begin{equation}
Z_N \to f(z); \qquad
X_k(t) \to \frac{1}{t^{1/\beta}}f\left(\frac{x}{t^{1/\beta}}\right). 
\end{equation}
The MSD is equal to the second moment
\begin{equation}
\langle X(t)^2 \rangle \sim \int_{0}^{\infty} x^2 \frac{1}{t^{1/\beta}}f\left(\frac{x}{t^{1/\beta}}\right) d x = t^{2/\beta} \int_{0}^{\infty} y^2 f\left(y\right) d y
\label{levy:MSD}
\end{equation}
and hence RMSD for a L\'evy flight scales super-diffusively: $R(t) \sim t^{1/\beta}$.

The jumps in L\'evy flights can be considerable, but occurs within the same time step as a short one (which is unrealistic), and thus L\'evy flights can be seen as only a rough approximation to actual human trajectories.

\subsubsection{CTRW}
\label{sec:ctrw}

The random walk models discussed thus far have been discrete in time. In each time interval, an instance of a jump is dictated by the corresponding jump-length distribution. A continuous time random walk (CTRW) is a random walk in which the number of jumps made in a time interval $d t$ is also a random variable or equivalently, the time elapsed between jumps (wait-time $\Delta T$) is also a random variable. 

If the PDF of jump-lengths is $f(\Delta x)$ and that for wait-times is $\phi(\Delta t)$, and these are independent, the CTRW consists of pairwise random and independent events with $\Delta X$ and $\Delta T$ drawn from the joint PDF, $P(\Delta x, \Delta t)  = f(\Delta x)\phi(\Delta t)$, which denotes the probability that a jump of length $\Delta x$ is taken after a time $\Delta t$. According to this model, after $N$ steps, the total displacement, $X_N$, and the total elapsed time $T_N$ are given by:
\begin{equation}
X_N = \sum_{i=1}^{N} \Delta X_i; \qquad
T_N = \sum_{i=1}^{N} \Delta T_i. 
\end{equation}\\
The PDF of the process, $P(x,t)$ can be Fourier-Laplace transformed to give:
\begin{equation}
W(k,u) = \frac{1-\tilde{\phi}(u)}{u\, (1-\tilde{\phi}(u)\, \tilde{f}(k))} \label{eq:ctrw:2}
\end{equation}
where $\tilde{\phi}(u)$ and $\tilde{f}(k)$ are the Laplace and Fourier transforms of $\phi(\Delta t)$ and $f(\Delta x)$ respectively. Taking the inverse transform we get
\begin{equation}
P(x,t) = \frac{1}{2\, \pi}\frac{1}{2\, \pi \, i} \int du \int dk \, e^{u\, t-i\, k\, x}\, W(k,u). \label{eq:ctrw:3}
\end{equation}
This expression for $P(x,t)$ can be analyzed according to the asymptotic behavior of distributions $\phi(\Delta t)$ and $f(\Delta x)$. In particular four types of models are obtained, depending on whether none, either, or both distributions have heavy tails. The properties of these models and their application to human mobility are discussed below.

\begin{figure}[t!]
\centering
\includegraphics[width=0.8\textwidth]{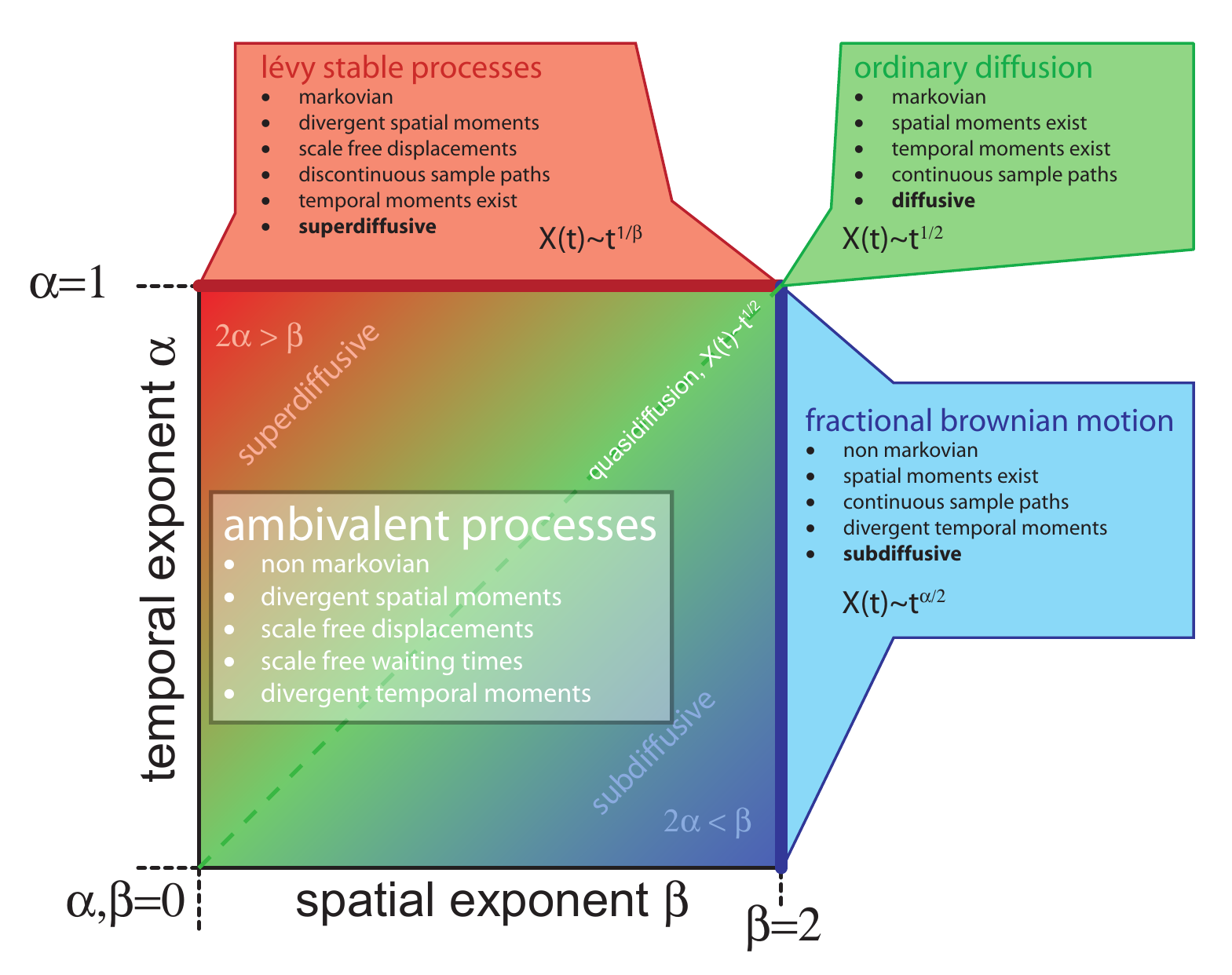}
\caption{ Schematic of the different (asymptotic) classes of CTRW defined in the text, as a
function of the waiting-time and jump-length exponents $ 0 < \alpha < 1$ and $0 < \beta < 2$. L\'evy flights, fractional Brownian motion
as well as ordinary diffusion are limiting cases of the more general class of ambivalent processes. Figure from~\cite{brockmann_2006_scaling}.}
\label{fig:brockmann_2006_scaling_1_2S2}
\end{figure}

\noindent {\bf Ordinary Diffusion:} If the expectation value of $\phi(\Delta t)$ and the variance of $f(\Delta x)$ are both finite, i.e $1 - \tilde{\phi}(u) \sim \tau \, u$ and $\tilde{f}(k) \sim 1 - (\sigma \, k)^2$ then from \equationname~(\ref{eq:ctrw:2}) and \equationname~(\ref{eq:ctrw:3}), asymptotically:
\begin{equation}
P(x,t) \sim \frac{1}{\sqrt{t}}e^{-x^2/D\, t}
\end{equation}
where $D$ is a diffusion constant. Therefore a CTRW with well defined jump-length and waiting-time distributions, is asymptotically (with time) equivalent to Brownian Motion.\\ 

\noindent {\bf L\'evy Flights:} If $f(\Delta x) \sim \Delta x^{-(1+\beta)}$ ($0<\beta<2$), and $\phi(\Delta t)$ has a finite variance, a L\'evy flight is recovered. Following the same procedure as Sect.~\ref{sec:levy}, the PDF of the process is given by
\begin{equation}
P(x,t) \propto \frac{1}{t^{1/\beta}}G(x/t^{1/\beta}) ,
\end{equation}
where $G(x/t^{1/\beta})$ is a scaling (limiting) function. Therefore if the distribution of jump lengths is a power-law and the wait-times are well defined, $R(t) \sim t^{1/\beta}$ and a continuous time random walker will follow a super-diffusive path, equivalent to a L\'evy Flight.\\

\noindent {\bf Fractional Brownian Motion:} Conversely, if jump-lengths with a finite variance are combined with a power-law distribution of wait-times
$\phi(\Delta t) \sim \Delta t^{-(1+ \alpha)}$, ($0 < \alpha < 2$), the PDF is
\begin{equation}
P(x,t) \sim \frac{1}{t^{\alpha/2}}H(x/t^{\alpha/2}) ,
\end{equation}
with $H$ being a non-Gaussian limiting function. In this case, the effect of the distribution of waiting times is to slow down the random walk. Here $R(t) \sim t^{\alpha/2}$, consequently the walk is sub-diffusive for $\alpha <1 $ and super-diffusive for $1 < \alpha <2$.\\

\noindent {\bf Ambivalent Processes:} The fourth variant occurs when both $f(\Delta x)$ and $\phi(\Delta t)$ are heavy-tailed. In this case, $R(t) \sim t^{\alpha/\beta}$, and hence the nature of the diffusive behavior is fully specified by $\alpha$ and $\beta$. For $\beta < 2\, \alpha$, the CTRW is super-diffusive and for $\beta > 2\, \alpha$, it is sub-diffusive. If $\beta = 2\, \alpha$, the random walk converges to ordinary diffusion/Brownian motion, despite the diverging moments of the respective distributions. A schematic of these limiting cases is provided in \figurename~\ref{fig:brockmann_2006_scaling_1_2S2}. Given the sensitivity of the models to the parameters of the wait times and jump lengths, the importance of their accurate measurement from data cannot be over-stated. In any event, the ambivalent process model is most often used for describing the mobility of individuals. Analysis of various data sources (GPS, CDRs, Dollar bills) has found that both jump length distributions and the distribution of wait times display power-law behavior~\cite{brockmann_2006_scaling, gonzalez_2008_understanding, zhao_2008_empirical, song_2010_modelling}. The parameter ranges measured from data correspond to $\alpha$ estimated from empirical data range from $0.42 \leq \alpha \leq 0.8$~\cite{zhao_2008_empirical,song_2010_modelling} and values of $0.31 \leq \beta \leq 0.75$ ~\cite{zhao_2008_empirical,gonzalez_2008_understanding}.

\subsubsection{Preferential Return}
\label{sec:epr}

An important aspect of human behavior, missing from the models discussed thus far, is the tendency of individuals to return to one or more locations on a daily basis (so-called preferential return). In a CTRW process, the number of distinct sites $S$ visited by a random walker in time $t$ is given by
\begin{equation}
S(t) \sim t^{\mu} ,
\label{individual:17}
\end{equation}
with $\mu = \alpha$ for a CTRW and $\mu =1$ for a L\'evy Flight~\cite{gillis_1970_expected}. The analysis of CDR's (\sectionname~\ref{sec:cdr}), revealed a power-law distribution of jump-size with exponent $\alpha = 0.8$, while $\mu$ was independently measured to be $\mu = 0.6 \pm 0.02$, considerably less than the theoretical prediction. Indeed for random walks, the probability of an individual to visit any distinct location becomes asymptotically uniform, whereas  analysis of visitation patterns~\cite{gonzalez_2008_understanding} from data suggests that the rank-frequency visitation frequency follows a Zipf's law: 
\begin{equation}
f_{k} \sim k^{-\zeta}, 
\label{individual:18}
\end{equation}
where $k$ corresponds to the rank of location according to frequency of visit. This implies the distribution of visitation frequencies follows $P(f) \sim f^{-(1+1/\zeta)}$.

Furthermore the scaling of the MSD in the CTRW model  suggests that an individual will asymptotically drift away from the origin (home) in contrast for what is known from daily experience. To account for the human trait of returning to locations,  Song \et~\cite{song_2010_modelling} include two extensions to the CTRW model: \emph{exploration} and \emph{preferential return}. 

\begin{figure}[t!]
\centering
\includegraphics[width=0.8\textwidth]{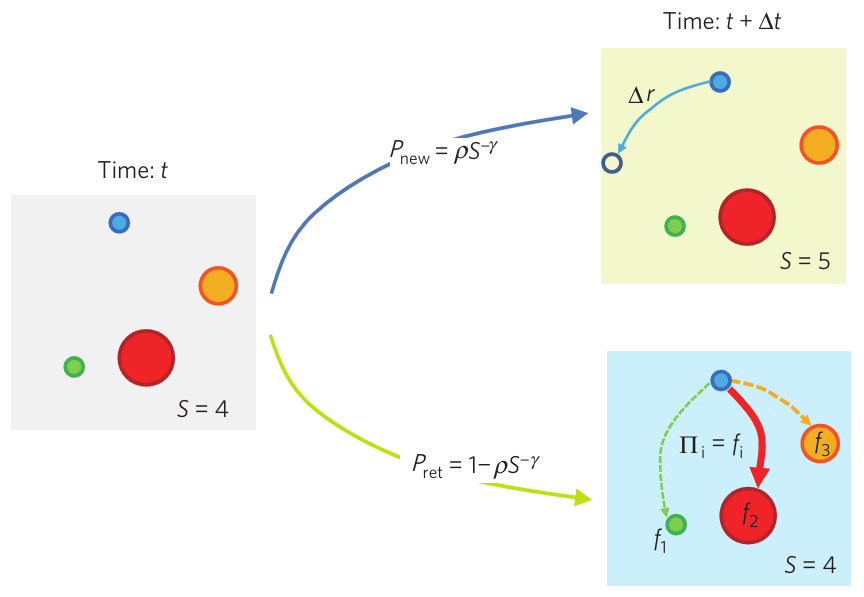}
\caption{Schematic description of the Exploration and Preferential Return model. Starting at time $t$ from the configuration shown in the left panel, indicating that the user visited previously $S = 4$ locations with frequency $f_i$ that is proportional to the size of circles drawn at each location, at time $t + \Delta t$ (with $\Delta t$ drawn from the $P(\Delta t)$ fat-tailed distribution) the user can either visit a new location at distance $\Delta r$ from their present location, where $\Delta r$ is chosen from the $P(\Delta r)$ fat-tailed distribution (exploration; upper panel), or return to a previously visited location with probability $P_{ret} = 1- \rho S^{-\gamma}$, where the next location will be chosen with probability $\Pi_i = f_i$ (preferential return; lower panel). Figure from~\cite{song_2010_modelling}.}
\label{fig:song_2010_2}
\end{figure}

Exploration is defined as the probability for an individual to move to a \emph{previously unvisited} location according to
\begin{equation}
\label{eq:pnew}
P_{new} = \rho \, S^{-\gamma}, 
\end{equation}
where $\rho$ and $\gamma$ are parameters of the model and $S$ increases by one for each new visit.
Preferential return on the other hand is the probability to return to a previously visited location and is thus complementary to exploration: $P_{ret} = 1-\rho \, S^{-\gamma}$ (see \figurename~\ref{fig:song_2010_2} for a schematic). 

The probability to visit a previous location $i$, denoted $\Pi_i$, is determined by the number of previous visits to $i$, i.e., $\Pi_i = f_i$ where $f_i$ is the visitation frequency. The parameters of the model are restricted to $0<\rho<1$ and $\gamma \ge 0$. Considering the total number of steps, $N$, an individual makes in time $t$, and noting that $dS/dN =P_{new}$, the number of distinct locations visited in time $t$ is given by $S \sim N^{1/(1+\gamma)}$. For a power-law waiting time distribution $\phi(\Delta t)$, we have $t\sim n^{1/\alpha}$ and hence the exponent in \equationname~(\ref{individual:17})is
\begin{equation}
\mu = \frac{\alpha}{1+ \gamma},
\end{equation}
thus leading to slower rate of exploration than predicted by CTRW processes. 

Furthermore, the number of visits to location $i$ at step $N$, $m_i(N)$, increases according to 
\begin{equation}
\frac{d m_i}{d N} = \Pi_i \, (1-P_{new}) \, ,
\end{equation}
where $\Pi_i = f_i = m_i / \sum_i m_i(N)$. As $S(t) \to \infty$, for $\gamma > 0$, $P_{new} \to 0$ and therefore $d m_i/d N = m_i/\sum_i m_i(N)$. Noting that the sum of visits over all locations is equivalent to the number of steps taken, $\sum_i m_i(N) = N$, the expression $m_i(N) = N/N_i$ is obtained, where $N_i$ denotes the first jump to location $i$. It is clear that the likelihood of visiting a location increases with number of earlier visits, and as such, the rank $k_i$ of location $i$ follows the relation $k_i = S(n_i) \sim n_i^{1/(1+ \gamma)}$. Combined with the fact that $f_i$ is proportional to $m_i(N)$ this results in the relation $f_k \sim k^{-\zeta}$ with $\zeta = 1 + \gamma$, thereby accounting for the relation in \equationname~(\ref{individual:18}).

 The MSD is related to $S$ via
\begin{equation}
\mathrm{MSD} ^{\beta/2} \sim \log \Big(\frac{1-S^{1-\zeta}}{\zeta -1}\Big) + c ,
\end{equation}
where $c$ is a constant. Given the measured range of the parameters, this leads to three asymptotic regimes: $\mathrm{MSD} \sim (\log t)^{2/\beta}$ for $\zeta<1$; $\mathrm{MSD} \sim (\log\log t)^{2/\beta}$ for $\zeta=1$;  and $\mathrm{MSD}  \to X_{max}$ for $\zeta > 1$ where $X_{max}$ denotes the saturation point of the MSD.  Measured values suggest that $\zeta = 1.2 \pm 0.1$ indicating a saturation of movement at long times~\cite{song_2010_modelling}. This regime corresponds to an individual's motion being dominated by their most visited location and is more in line with expected human behavior.


\subsubsection{Recency}
\label{sec:recency}
The concept of recency was introduced to solve discrepancies that emerges under the standard preferential return assumptions. Specifically, the fact that the earlier a location is discovered, the more visits it will receive, leading to a cumulative advantage precluding people from changing preference of location (unlike what is observed in data). Barbosa \et \cite{barbosa_2015_effect} proposed a model in which the exploration phase in human movements also considers recently-visited locations and not solely frequently-visited locations. Two rank variables $K_{f}$ and $K_{s}$ are defined to characterize, respectively, the \emph{frequency} and \emph{recency} of a given location in the context of an individual's trajectories. More precisely, the rank variables can be described as:
\begin{itemize}
	\item[$K_s$] is the recency-based rank. A location with $K_s = 1$ at time $t$ means that it was the previous visited location. $K_s = 2$ means that such location was the second-most-recent location visited up to time t and so on.
	\item[$K_f$] is the frequency-based rank. A location with $K_f =1$ at time $t$ means that it was the most visited location up to that point in time. Similarly, a location with $K_f = 2$ is the second-most-visited location up to time $t$, and so on.
\end{itemize} 

The model can be described as follows: first, a population of $N$ agents is initialized and scattered randomly over a discrete lattice with $L\times L$ cells, each one representing a possible location. The initial position of each agent is accounted as its first visit. At each time step agents can visit a new location with probability similar to the preferential return model, Eq.~\eqref{eq:pnew}. 

  
The return phase happens with probability $P_{ret}$ analogous to \equationname~\eqref{eq:pnew}, with the caveat that return jumps selects location $i$ from  frequently visited ones with probability $\alpha$ and recently visited locations with probability $1 - \alpha$ thus:
\begin{equation} \label{eq:recency}
\begin{array}{ll}
P^s_{ret} = (1 - \alpha) \, P_{ret} & \Pi_i \propto k_s(i)^{-\nu} ,\\
P^f_{ret} = \alpha \, P_{ret} & \Pi_i \propto k_f(i)^{-1-\gamma} ,
\end{array}
\end{equation}
where $k_s(i)$ is the recency-based rank and $k_f(i)$ is the frequency-based rank of the location $i$. When $\alpha = 1$ the preferential return model is recovered. The empirical measurements for the two datasets considered had $\alpha = 0.1$, and $\nu = 1.6$. $\gamma = 0.6$ was kept the same as~\cite{song_2010_modelling}.

\begin{figure}[t!]
\centering
\includegraphics[width=0.9\textwidth]{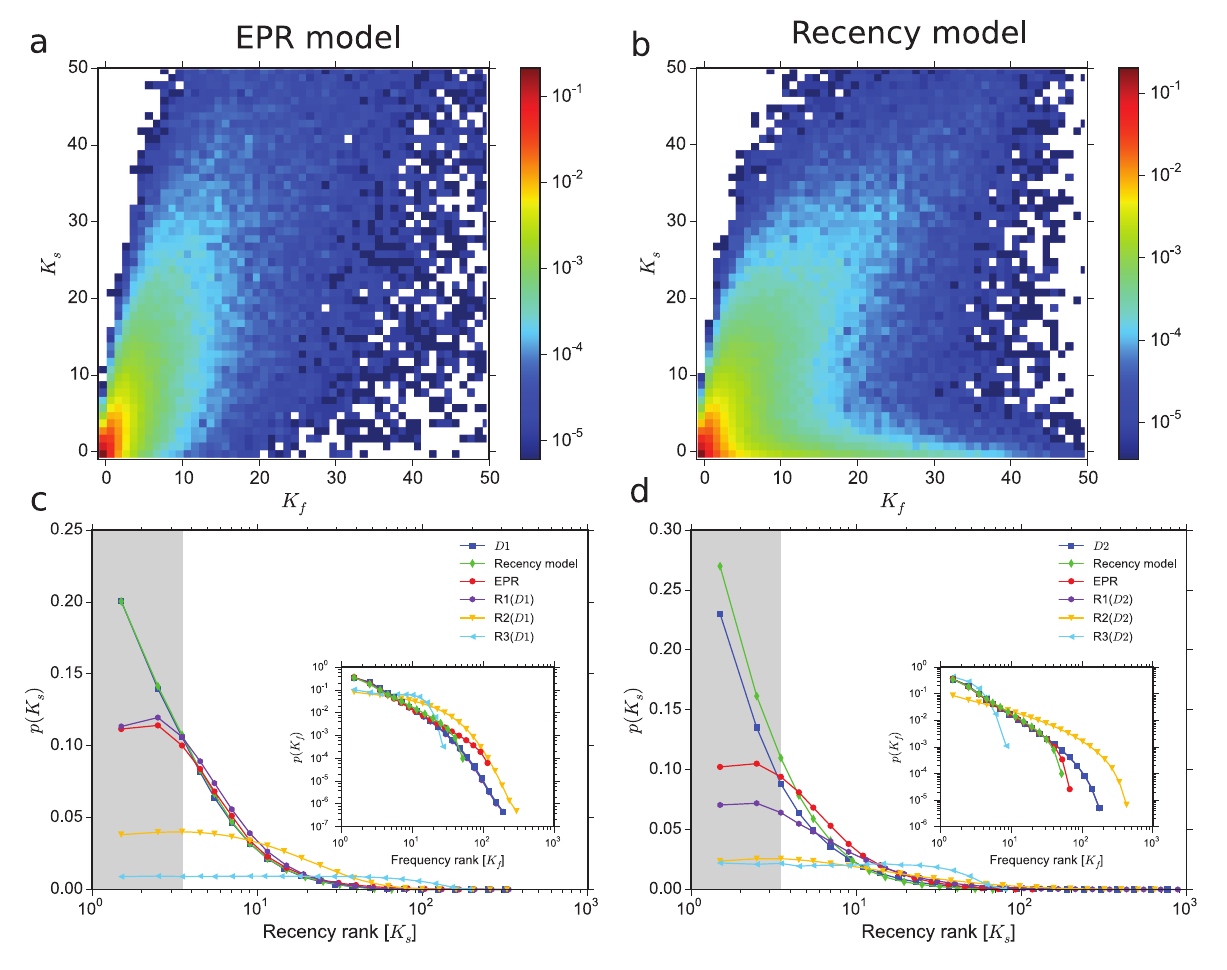}
\caption{Comparison between the Preferential Return (EPR) model and the recency-based (RM) model. (a) The analysis of the return ranks generated by the EPR model shows that it reproduces a pattern similar to the one observed from the empirical analysis.
(b) Probability of return to recently-visited locations (i.e.,low $K_s$). (c) Distribution of the frequency ranks, the preferential return mechanisms (labelled EPR) exhibits a power-law distribution. The activation of the recency mechanism does not affect the frequency rank distribution. (d) $K_s$ distribution, the EPR mechanism does not capture the power-law behavior observed on the empirical data. Figure from~\cite{barbosa_2015_effect}.}
\label{fig:barbosa_2015_6}
\end{figure}

As seen from \figurename~\ref{fig:barbosa_2015_6} the preferential return model does not capture the broader distribution of $p(K_{f},K_{s})$ for recently-visited locations, an effect captured by the recency-based refinements. The primary differences can be seen for the distribution of the recency rank $K_s$ seen in log-linear scale.

\subsubsection{Social-based models}
\label{sec:social}

It is natural to assume that two individuals who have a social interaction, such as friends, family or colleagues, do not always move independently \cite{axhausen_2005_social,carrasco_2006_exploring,dugunji_2005_discrete}. Occasionally, they will share full trips, destinations or origins. The trips can be also synchronized if the objective is to meet somewhere or go back home after a meeting. Furthermore and closing the loop, the social network of an individual typically reflects the geography of their life with tighter connections with people spatially closer or at least in clusters related to the places in which the person has previously resided. These correlations have been observed in several publications. For example, a relation between distance and ``online friendships" was described in 2005 by Liben-Nowell \et \cite{liben-nowell_2005_geographic} and was later confirmed using surveys \cite{carrasco_2008_how,vandenberg_2013_path}, social networks \cite{carrasco_2008_collecting,carrasco_2008_agency} and  mobile phone records \cite{lambiotte_2008_geographical,krings_2009_urban,phithakkitnukoon_2012_socio}. Indeed this feature formed the basis of a generative model describing the behavior of population level aggregate economic indicators across urban systems~\cite{Pan_2013_UrbanScaling}. Possible methods for exploiting this intuitive observation towards improving forecasting of individual movement has been explored in \cite{de_2013_interdependence}.
For example, nonlinear time series prediction methods based on the delay embedding theorem by Takens~\cite{Takens_1981_Delay} can be used to forecast an individual's future mobility given a detailed history of past movements and the assumption of a certain degree of determinism in mobility patterns.
Furthermore, improvements in prediction accuracy by about one or two orders of magnitude has been observed when including the time series of past movements of members of the social network (individual's acquaintances)  

Conversely, online~\cite{eagle_2009_inferring,crandall_2010_inferring,picornell_2015_exploring} social links can be inferred from co-occurrences of individuals in space and time. In particular, the probability that two users have a friendship link on the Flickr social network grows significantly with the number of distinct geographical locations that they both visited within a given time threshold, where the spatial and temporal information are extracted from the geo-tagged photos they published. 
The probability of being friends is higher if the size of the regions and the temporal range between the two observations decrease. 
This result can be explained by simple mobility models exploiting the fact that friends are more likely to spend time together in the same place, often live close to each other and that the jump-size distribution decays as a power law. Similarly, the location of a person may be predicted by those of their near contacts~\cite{backstrom_2010_find}. Two individuals with similar mobility patterns are generally in close proximity in the who-calls-whom social network of mobile phone users \cite{wang_2011_human}. 
Indeed, strong correlations exist between various classical topological measures of the proximity of two mobile phone users in the social network, such as Adamic-Adar, Jaccard and Katz, and various measures of spatio-temporal proximity, such as co-location rates and spatial cosine similarity. 
Combining co-location information with information on the relative positions of users in the social network it is possible to predict the formation of new social ties with higher precision than using only information derived from the proximity in the social network. 

The idea of interconnecting mobility and social interactions has been considered in different theoretical contexts. Detailed models have been proposed in the area of transportation to take into account the relation between social network and transport demand. This includes microsimulation of transport systems \cite{axhausen_2005_social,carrasco_2006_exploring,dugunji_2005_discrete} in which agents may have communication between them and have common objectives \cite{paez_2007_social}. Social groups and their sizes can impact transport demand  \cite{molin_2007_social}, daily schedules, social relations and face-to-face interactions \cite{arentze_2008_social,carrasco_2009_social,hackney_2011_coupled,ronald_2012_modeling,sharmeen_2014_dynamics}. A simpler framework has been introduced from a Physics perspective, as for example formulated by~\cite{gonzalez_2006_system}. The system is composed of $N$ agents within a radius $r$ moving in a 2D square of lateral size $L$ and periodic boundary conditions. The starting position of the agents and their directions of movement are randomly selected. All the agents have the same initial speed $v_0$ and they travel in straight lines until they ``crash". Every crash produces a new social tie between the pair of agents involved. After every crash, the velocities of the two agents are updated randomly selecting new directions of movement and increasing their speeds according to the expression:
\begin{equation}
|\vec{v}_i (t)| = v_o + c\, k_i(t), 
\end{equation}
where $\vec{v}_i (t)$ is the velocity of agent $i$ at time $t$ immediately after the crash, $c$ is a constant and $k_i(t)$ is the number of connections of $i$ in the social network. The number of agents $N$ is constant, but after a certain period of time old agents are removed and substituted with new agents with speed $v_o$ and no social links, allowing the system to reach a stationary state. The characteristics of the emerging social network are analyzed. The degree distribution depends on the constant $c$ and on the lifetime of the agents in the system. For small $c$, the distribution is well approximated by a Poissonian distribution, but deviates from this shape with increasing $c$ and average degree $\langle k \rangle$. Beyond the distribution, the model can reproduce other properties of real social networks such as the degree-degree correlations, the size of the large connected cluster and the abundance of cliques.

\begin{figure}[t!]
\centering
\includegraphics[width=0.6\textwidth]{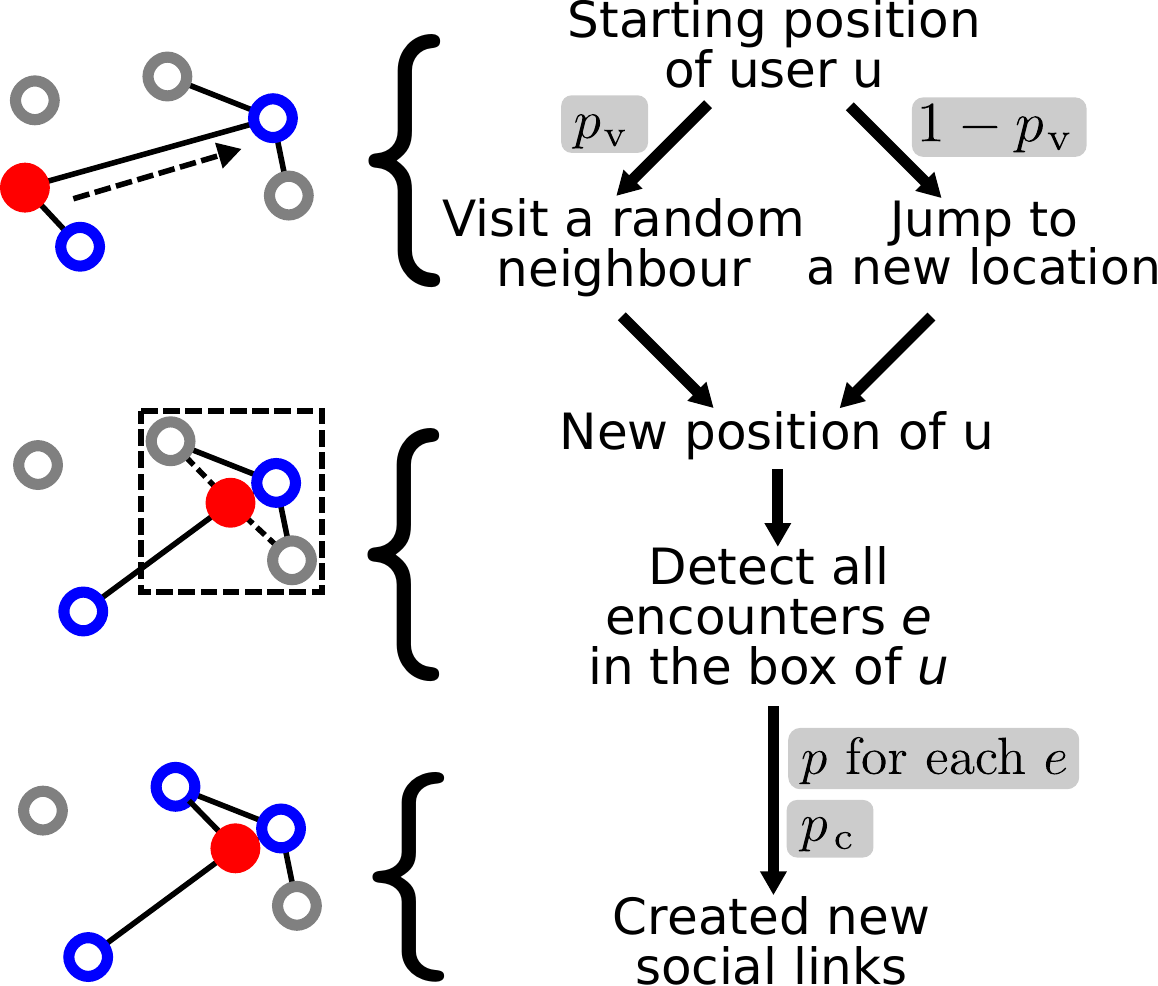}
\caption{Sketch showing the main ingredients of Grabowicz's model. The agent's update of position and network is marked in red, while its contacts are in blue. The model has two main steps: one in which the mobility is determined in terms of either visiting a friend with probability $p_v$ or a L\'evy-like jump otherwise. After this, a new social link may be established with  probability $p$ in the neighborhood of the new position or at random with $p_c$. Figure from~\cite{grabowicz_2014_entangling}.}
\label{fig:grabowicz_2014_entangling}
\end{figure}

More realistic mobility models in the context of generated social networks were explored in~\cite{grabowicz_2014_entangling,toole_2015_coupling}. In particular the former proposed a model inspired by the methods described in Sec.~\ref{sec:epr} with important variations in the main ingredients (See \figurename~\ref{fig:grabowicz_2014_entangling}). The initial conditions were set such that the agents are placed according to measured population density in the area considered. Then at every step, each agent decides to visit a social contact, with probability $p_v$ or, otherwise moves according to a L\'evy-like flight with probability $1-p_v$, with the destination chosen in proportion to the population density in that area. This precludes agents from moving to unphysical locations such as water-bodies, geographic barriers and the like. After movement, the agent generates a new directed social link with probability $p$ in the neighborhood of her present position and with probability $p_c$ with a random agent. The parameter $p_c$ represents social relations that emerge online, inspired by the observations made by~\cite{liben-nowell_2005_geographic}. Note that the  jump distribution can be calibrated from measurements made in~\cite{song_2010_modelling}, and the tie formation parameter $p$ can be sampled at different scales from measurements made in~\cite{crandall_2010_inferring}. 
This leaves only two free parameters in the model: the probability of visiting friends $p_v$ and that of creating new links regardless of the distance $p_c$. These parameters were adjusted to reproduce an error function containing topological and  geographical properties of a set of online social networks such as Twitter, Gowalla and Brightkite. With these parameters, the model reproduces the degree distributions found in~\cite{gonzalez_2006_system} as well as the spatial dependence of link distribution, the probability of reciprocal ties, the overlap of the social environment, the density of triangles and the disparity of the triangles in terms of distances of their vertices. The model is also amenable to a  mean-field analytical treatment that demonstrates the necessity of accounting for social effects on mobility (visiting friends for example) to obtain realistic reproductions of the behavior seen in  social networks. 

The variant proposed by~\cite{toole_2015_coupling} instead focuses on behavior seen at shorter time-scales such as intra-day mobility, for which period the social network can be approximated as static. Like the preferential return model, at each step an agent decides to return to a previously visited location with probability $1-\rho\, S^\gamma$ or explore a new one with $\rho\, S^\gamma$, but there is an additional ``social pressure'' component. This occurs with probability $\alpha$ where one of the agent's contacts are selected at random with probability proportional to the co-similarity in the location visiting profiles. Consequently, the new location is chosen from the list of the contact's previously visited locations. In the absence of social pressure, (probability $1-\alpha$) the behavior is the same as Sec.~\ref{sec:epr}. The introduction of the social pressure component does a better job at reproducing the re-visitation profiles measured in CDR data than models without this component.  

\subsection{Population-Level}
\label{sec:poplev}

Mobility information gathered at the individual level can be aggregated to study the flows of individuals traveling from one region to another at different spatio-temporal scales. These flows can be organized in the framework of Origin-Destination (OD) matrices (see \sectionname~\ref{sec:odmatrix}). Such a format, with all possible combinations of origins and destinations for trips, is easily transformed into a directed weighted network in which nodes denote locations (for example counties or municipalities) and link weights correspond to the flow of travelers between the two locations. As discussed earlier in the review, OD matrices can be empirically estimated from transportation surveys, traffic counts or individuals' geolocated ICT data. An OD matrix, or its corresponding network, provides useful information on the travel demand between the origin and destination areas, representing a valuable asset widely studied and used in Geography, Transportation research and Urban Planning. Therefore, being able to obtain accurate estimation of OD matrices is crucial for both modeling and applications, and this problem has attracted the interest of researchers and decision makers for decades.

Note that individual mobility patterns are straightforwardly aggregated into flows, however the inverse problem, i.e the disaggregation of flows, is typically not possible. Instead, one has to resort to determine dependences between mobility flows and a limited set of ``static" attributes of the locations that would allow predictions on how changes in these attributes can potentially influence future travel demand. To this end, various spatial interaction models have been proposed to predict flows of individuals based on a small number of key local attributes. Considering a region of interest divided into $n$ locations, the purpose of these models is to estimate the number of trips $T_{ij}$ from location $i$ to location $j$ from the socio-economic characteristics of the populations of $i$ and $j$, and their spatial distribution. Models of spatial flows have been traditionally developed starting from the principle of entropy maximization subjected to various constraints. In the strongest version, the constraints are the number of people leaving and entering each location, but in softer ones there can be other proxy variables to represent the demand and attraction of the trips' origins and destinations such as population levels, the finite amount of resources for travels, utility functions to describe individual choices over competing alternatives, etc \cite{wilson_1967_statistical,mcfadden_1974_measurement,benakiva_1985_discrete,sagarra_2013_statistical,sagarra_2015_role}.

\begin{figure}[t!]
\centering
\includegraphics[width=0.8\textwidth]{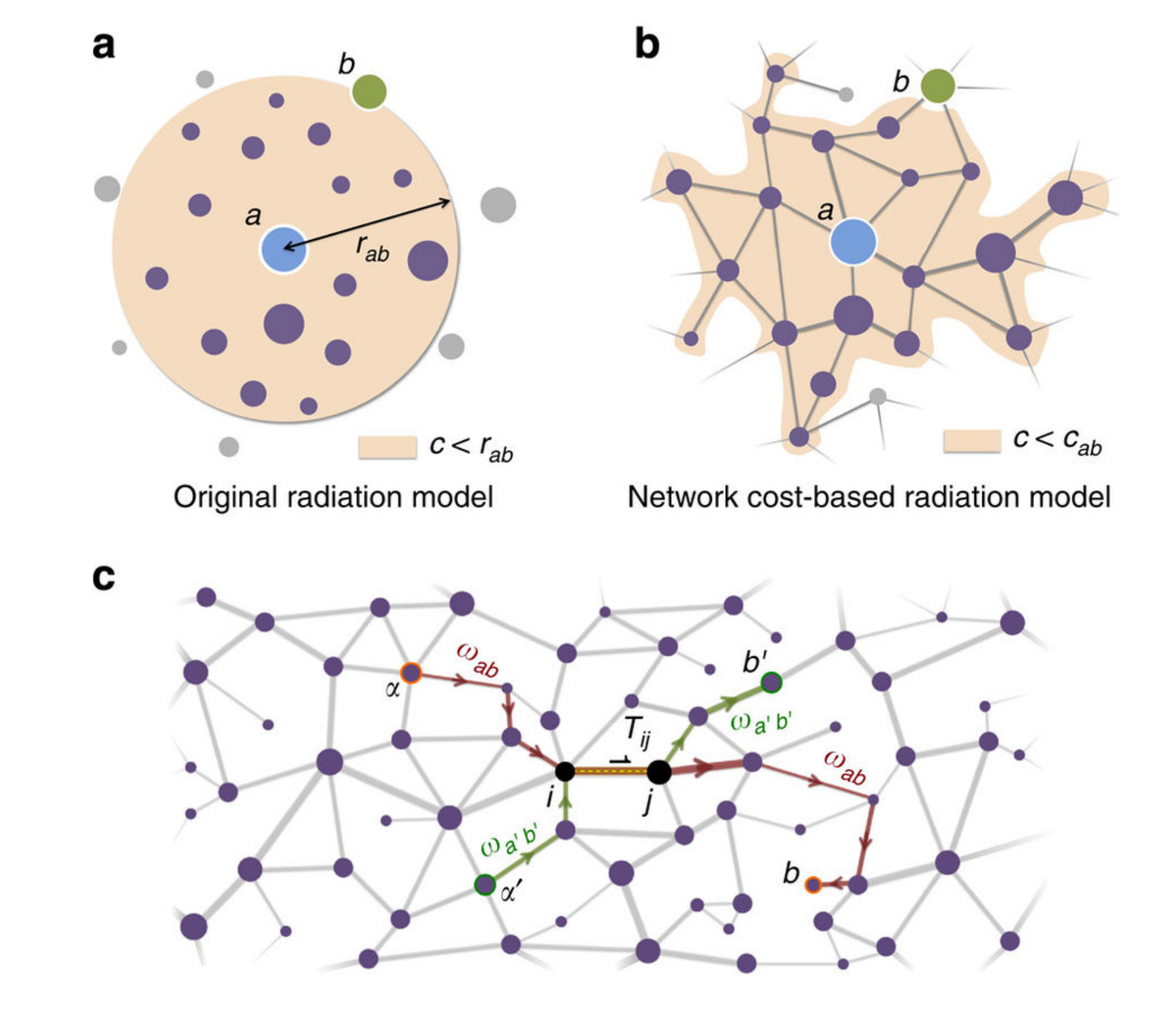}
\caption{Differences between distance-based and intervening opportunity models. (a) The radiation model uses distance as a search criterion. (b) The cost-based radiation model uses network travel cost as a search criterion, which usually has a heterogeneous distribution. (c) The flow $T_{ij}$ through edge $(i, j)$ is the sum of contributions from all those mobility fluxes $\phi_{ab}$ whose minimal cost paths $\omega_{ab}$ contain (i, j). Figure from~\cite{ren_2014_predicting}.}
\label{fig:ren_2014_predicting}
\end{figure}

In the simplest version, the objective is to infer flows from the product of two types of variables: one type that depends on an attribute of each single location (e.g. the population), and the other type that depends on a quantity relating a pair of locations (e.g. the distance or travel time). 
The differences between the various models consist of the choice of variables considered, and the specific functional forms in which these variables enter.
Over the years, two main schools of thought have emerged. 
The first one assumes that the number of trips between two locations is 
a decreasing function of 
their distance, giving rise to the so-called gravity models \cite{carey_1858_principles,zipf_1946_p1}; the second variant postulates that the number of intervening opportunities, defined as the number of potential destinations between two locations, determines the mobility flow between them, and models that share this assumption are called Intervening Opportunities models \cite{stouffer_1940_intervening}. 
An illustrative example that highlights the fundamental differences between distance and intervening opportunities is shown in \figurename~\ref{fig:ren_2014_predicting}. In addition to providing the mathematical framework to model human mobility flows \cite{barthelemy_2011_spatial,thiemann_2010_structure,jung_2008_gravity}, these models have found successful applications in estimating other spatial flows, including cargo shipping volume \cite{kaluza_2010_complex} and social interactions from inter-city phone calls \cite{krings_2009_urban,p._2011_uncovering}.

\subsubsection{Gravity models}
\label{sec:gravity}

Preceded conceptually by the work of H.C. Carey in 1858 about land use \cite{carey_1858_principles} and the retailing models of W. J. Reilly in 1931 \cite{reilly_1931_law}, George K. Zipf proposed in 1946 an equation to calculate mobility flows inspired by Newton's law of gravitation \cite{zipf_1946_p1}\footnote{Although the model is not stated as a \emph{gravity} model, Zipf draws a parallel between his model and a two dimensional ``gravitation" equation.}. In his work, Zipf highlights the importance of the distance for human migration patterns where the magnitude $T_{ij}$ of a migratory flow between two communities $i$ and $j$ can be approximated by 
\begin{equation}
T_{ij} \propto \frac{P_{i} \, P_{j}}{r_{ij}} ,
\end{equation}
where $P_{i}$ and $P_{j}$ are the respective populations and $r_{ij}$ the distance between $i$ and $j$.

\begin{figure}[t!]
  \centering
 \includegraphics[width=0.8\textwidth]{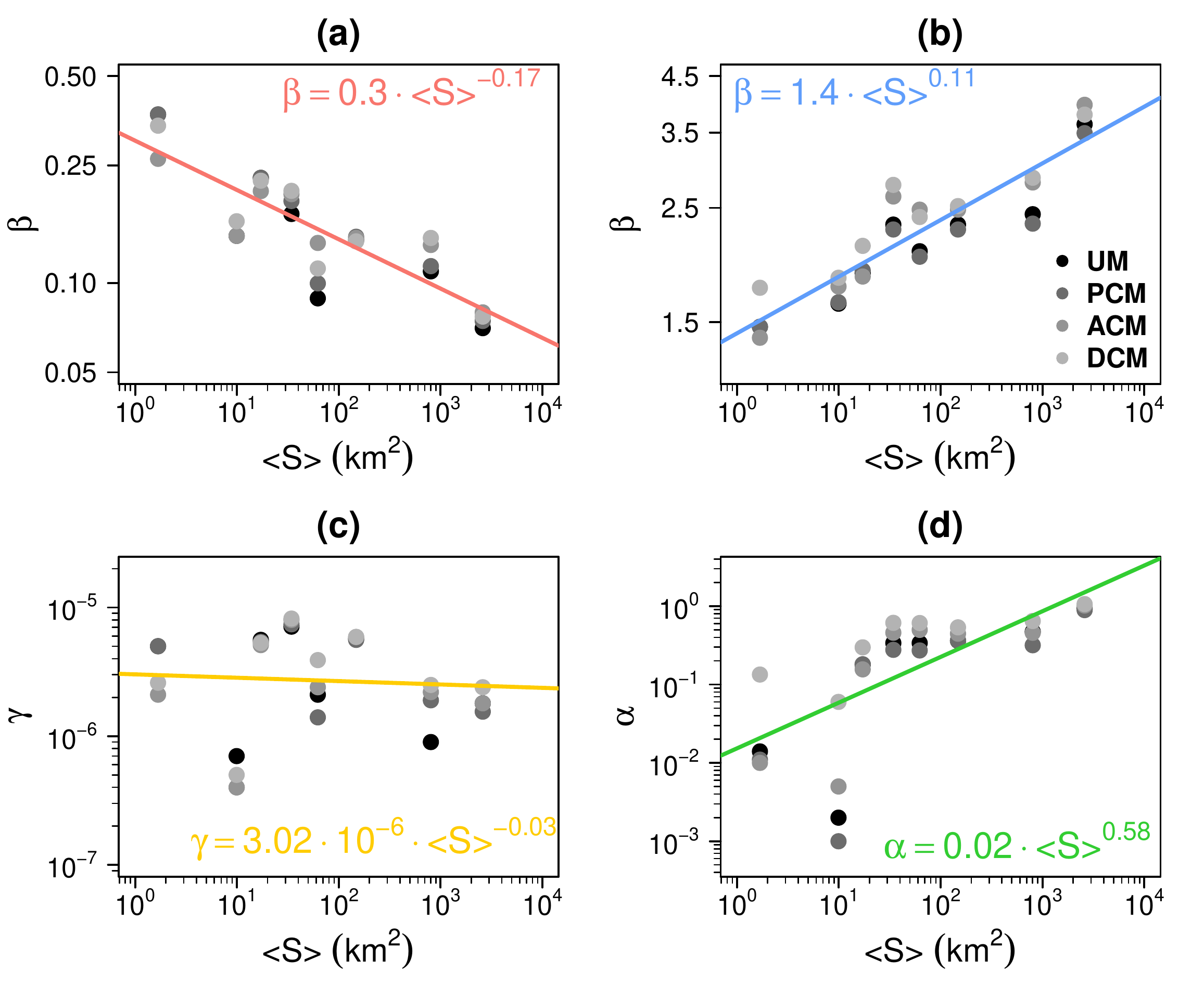}
  \caption{The distance exponent $\beta$ (\equationname~\eqref{eq:fij}) as a function of average unit surface area. (a) Normalized gravity laws with an exponential distance decay function. (b) Normalized gravity laws with a power distance decay function. (c) Schneider's intervening opportunities law. (d) Extended radiation law. Figure from \cite{lenormand_2015_systematic}.}
\label{fig:systematic}
\end{figure}

The basic assumptions of this model are that the number of trips leaving $i$ is proportional to its population,  the attractivity of $j$ is also proportional to $P_j$, and finally, that there is a cost effect in terms of distance traveled. These ideas can be generalized assuming a relation of the type:
\begin{equation}
T_{ij} = K m_i m_j f(r_{ij}) ,
\label{eq:grav}
\end{equation}
where $K$ is a constant, the masses $m_i$ and $m_j$ relate to the number of trips leaving $i$ or the ones attracted by $j$, and $f(r_{ij})$, called a ``friction factor'' or ``deterrence function", is a decreasing function of distance. As in the original version of the model, the masses usually are functions of population (not necessarily linear); common functional forms used in the literature are  $m_i = {P_i}^\alpha$ or $m_j = {P_j}^\gamma$~\cite{ortuzar_2011_modeling}. Unlike the original version, however, other variables, such as gdp-per-capita, may factor in the definition of the masses~\cite{mccullogh_1989_generalized,li_2011_validation}. The distance function $f(r_{ij})$ is commonly modeled with a power-law or an exponential form, although more complicated functions can be considered, such as a combination of the two, 
\begin{equation}
f(r_{ij}) = \alpha \, r_{ij}^{-\beta} \, e^{- \gamma r_{ij}}.
\label{eq:fij}
\end{equation} 
Indeed, the optimal form of the function  may change according to the purpose of the trips, the spatial granularity of the locations, and the transportation mode \cite{barthelemy_2011_spatial}. For example, in the case of commuting flows, the value of the exponent is highly correlated with the scale~\cite{lenormand_2012_universal,lenormand_2015_systematic} as can be seen in \figurename~\ref{fig:systematic}. In other  applications, the distance may not be the appropriate variable to quantify the cost of travel between two locations, and other variables such as travel time or economic (i.e. monetary) cost of a trip may offer better characterizations.

The ability to estimate, even if as a crude approximation, trip-flows, and consequently, traffic demand between two different locations as a function of their local properties, has made the gravity model widely popular in transport planning \cite{erlander_1990_gravity,ortuzar_2011_modeling}, in studies of geography \cite{wilson_1970_urban} and spatial economics \cite{karemera_2000_gravity,patuelli_2007_network}. It has been also used in situations where the knowledge of mobility flows is essential but the empirical data is not available as in the case of the characterization and modeling of epidemic spreading patterns~\cite{xia_2004_measles,viboud_2006_synchrony,balcan_2009_multiscale,balcan_2010_modeling,li_2011_validation}. 
Despite its widespread use and historical popularity, one must keep in mind that the gravity model is a gross simplification of travel flows, and in many cases, falls far short of capturing actual empirical observations~ \cite{simini_2012_universal,masucci_2013_gravity,lenormand_2015_systematic}.  Furthermore, the model requires the estimation of a number of free parameters, making it rather sensitive to fluctuations or incompleteness in data~\cite{jung_2008_gravity,simini_2012_universal}.\\ 

\noindent {\bf Constrained gravity models.} Some of the limitations apparent in the gravity model, can be resolved via certain constrained versions. For example, one may hold the number of people originating from a location $i$ to be a known quantity $O_i$, and the gravity model is then used to estimate the destination, constituting a so-called \emph{singly-constrained} gravity model of the form,
\begin{equation}
T_{ij} = K_i O_i m_j  f(r_{ij}) =  O_i  \frac{m_j f(r_{ij})}{\sum_k m_k \, f(r_{ik})}.
\label{eq:scgm}
\end{equation}
As can be seen, in this formulation, the proportionality constants $K_i$ depend on the location of the origin and its distance to the other places considered.
One can go further and fix now also the total number of travelers \emph{arriving} at a destination $j$ as $D_j = \sum_i T_{ij}$, leading to a {\it doubly-constrained gravity model}. For each Origin-Destination pair, the flow is calculated as
\begin{equation}
T_{ij}= K_i  O_i  L_j D_j \, f(r_{ij}) ,
\label{eq:dcgm}
\end{equation}
where there are now two flavors of proportionality constants,
\begin{equation}
K_i =  \left[ \sum_j L_j \, D_j \, f(r_{ij}) \right]^{-1}, \qquad
L_j =  \left[ \sum_i K_i \, O_i \, f(r_{ij}) \right]^{-1}, 
\end{equation}\label{eq:KL}
usually calibrated with an \textit{Iterative Proportional Fitting} procedure \cite{deming_1940_least}. 

The use of singly-, doubly- or non-constrained models depends on the amount of information available and on the pursued objective. If the aim is to approximate the mobility flows and  transport demand from indirect socio-economic variables of different geographical areas, then one employs non-constrained models. On the other hand, if out-going or in-going flows are empirically measured quantities,  and the objective is to estimate the elements of the OD matrix $T_{ij}$, then one employs constrained models.\\

\noindent {\bf Maximum entropy derivation of gravity models.} 
While the gravity model may seem rather \emph{ad hoc} in terms of the form of \equationname~\eqref{eq:grav}, an argument in favor of this functional form was provided by Alan Wilson \cite{wilson_1970_entropy} in the framework of classical transportation theory combined with entropy maximization. Essentially, the argument posits that in the absence of detailed information, out of all possible variants of OD matrices, the most probable ones are those that can be obtained with the highest number of trip configurations under opportune global and local constraint satisfaction. The objective is thus to find the set of flows $\{T_{ij}\}$ that maximizes the number of possible configurations of trips associated with it, respecting the possible constraints. 

Let $\Omega(\{T_{ij}\})$ be the number of distinct arrangements of individuals (configurations) that give rise to the set of flows $\{T_{ij}\}$, 
corresponding to the number of ways in which $T_{11}$ individuals can be selected from the total number of travelers $T = \sum_{ij} T_{ij}$; $T_{12}$ from the remaining $T - T_{11}$ and so on and so forth.   
Then we have that 
\begin{equation} 
		\Omega(\{T_{ij}\}) = 
\frac{T !}{T_{11}! \, (T - T_{11})!} \frac{(T - T_{11})!}{T_{12}! \, (T - T_{11} - T_{12})!} \dots = 
\frac{T !}{\prod_{ij} T_{ij}!} .
\label{eq:omega}
\end{equation}  
One then determines the maximum of $\Omega$ using Lagrange multipliers, subject to the constraints: 
\begin{equation}
 \label{eq:constraints}
  \left\{ 
    \begin{array}{l}  
        \sum_j T_{ij}=O_i\\					
				$\,$ \\	
				\sum_i T_{ij}=D_j  \\ 
        $\,$ \\	
			\sum_{ij} T_{ij}\, C_{ij}=C, \\ 
    \end{array} 
	\right.
\end{equation}
where $C_{ij}$ is the cost of travel from location $i$ to location $j$. The first two constraints ensure that trips originating and terminating in every location are equal to their observed values, whereas the third constraint fixes a total cost for all trips, $C$. 
In the limit of a large number of trips, $T$, the configuration that maximizes $\Omega$ is by far the most probable, and hence dominates over all other configurations. 
The resulting doubly-constrained gravity model takes the form
\begin{equation}
    T_{ij}=K_i O_i  L_j D_j \, e^{-\beta \, C_{ij}} ,
		\label{eq:doubly}
\end{equation}  
where the values of $K_i$ and $L_j$ are set in order to fulfill the first two constraints in \equationname~\eqref{eq:constraints}.
The Lagrange multiplier $\beta$ appearing in \equationname~\eqref{eq:doubly} controls the effect of cost on flows, and its value is empirically calibrated. It is worth noting that one can introduce a power law distance decay $f(r_{ij}) = r_{ij}^{- \beta}$ by considering a cost function of the form $C_{ij} \propto ln(r_{ij})$. These arguments have been further developed in a series of references~\cite{sagarra_2013_statistical,sagarra_2015_role} that also includes the distinguishable nature of the travelers and therefore necessarily modifying the statistics.\\

\noindent {\bf Using gravity models to estimate mobility flows.} In order to calibrate and apply gravity models to estimate mobility flows within a given region the following procedure is usually adopted. 
\begin{enumerate}
\item First, depending on the objectives, a flavor of gravity model is selected from either the unconstrained version, \equationname~\eqref{eq:grav}, or one of the constrained versions, i.e  \equationname~\eqref{eq:scgm}, or \equationname\eqref{eq:dcgm}.%

\item Second, the set of independent variables, population size, gdp or gdp-per-capita, distance, etc, as well as their relation with the local outflow, the attractiveness and the travel cost must be established. Although the choice of functions are somewhat arbitrary, common forms are power laws for the origin and destination populations, and exponential or power laws for the distance dependence. 
These particular functional forms are chosen to enable a fast and accurate calibration of the model, as it ensures that the logarithm of the flow depends linearly on some functions of the populations and the distance, allowing researchers to apply linear regression methods to determine the parameter values. 
\item Third, the parameter values are selected in order to maximize the fit between the flows estimated by the gravity model and the empirical flows observed in the region of interest. 
The best fit values of the parameters are determined using an optimization algorithm that either minimizes some error function between the model's estimates and the observed data \cite{cha_2007_comprehensive}, or maximizes the likelihood function of the observed data given the model's parameters \cite{flowerdew_1982_method}. 
Generalized Linear Models (GLM) \cite{nelder_1972_generalized} are a generalization of linear regression that are usually applied to fit the parameters of globally and singly constrained gravity models. 
GLM methods are more adapt than Ordinary Linear Regression (OLM) as it allows for the use of a wider and more realistic range of  probabilistic models to capture fluctuations in flow estimates. 
\end{enumerate}

Both the model training and parameter calibration steps can be employed either on the entire dataset or on subsamples, particularly if the aim is to validate the performance of the model in relation to the remaining set. 

\subsubsection{Intervening opportunities models}
\label{sec:Intervening}

Along with the gravity model, one of the first attempts to provide a conceptual and formal model of human mobility was introduced in 1940 by Stouffer \cite{stouffer_1940_intervening}. Contradicting a long tradition in the social sciences -- that distance is the central factor in determining mobility -- Stouffer proposed a conceptual framework in which distance and mobility are not directly related. Instead, Stouffer suggested that what plays the key role in determining migration is the number of \emph{intervening opportunities} or the cumulative number of \emph{opportunities} between the origin and the destination. In the paper, Stouffer does not provide a precise definition for ``opportunities'', leaving it to be defined depending on the social phenomena under investigation.

The basic idea behind the intervening opportunities (IO) model is that the decision to make a trip is not explicitly related to the distance between origin and destination, but to the relative accessibility of opportunities for satisfying the objective of the trip. An opportunity is a destination that a trip-maker considers as a possible termination point for their journey, and an intervening opportunity is a location that is closer to the trip maker than the final destination but is rejected by the trip maker. The law of intervening opportunities as proposed by Stouffer in 1940 states ``The number of persons going a given distance is directly proportional to the number of opportunities at that distance and inversely proportional to the number of intervening opportunities''.
Stouffer used this model to estimate migration patterns between services and residences \cite{stouffer_1940_intervening}. 
The traditional form of the intervening opportunities model is usually given by Schneider's version of Stouffer's original model \cite{schneider_1959_gravity}. 
Schneider's hypothesis states that ``The probability that a trip ends in a given location is equal to the probability that this location offers an acceptable opportunity times the probability that an acceptable opportunity in another location closer to the origin of the trips has not been chosen''. 
More formally, 
the flow $T_{ij}$ from the origin location $i$ to the $j$-th location ranked by travel cost from $i$ is given by:
\begin{equation}
  T_{ij}=O_i \frac{e^{-L \, V_{ij-1}}-e^{-L\, V_{ij}}}{1-e^{-L \,V_{in}}} .
  \label{IO}
\end{equation}
Here, $O_i$ is the total number of trips originating from $i$ and the second term represents the probability that one of these trips ends in location $j$. The denominator is a normalization factor ensuring that the probabilities sum to 1 (i.e. $\sum_j T_{ij}=O_i$). This probability depends on $V_{ij}$, the cumulative number of opportunities up to the $j$-th location ranked by travel cost from the origin location $i$ (and $n$ is the total number of locations in the region considered). 
Usually, the population, $m_j$, or the total number of arrivals, $D_j=\sum_i T_{ij}$, are assumed to be proportional to the number of ``real opportunities'' in location $j$. 
The value of the parameter $L$ can be seen as the constant probability of accepting an opportunity destination. As in the case of the gravity model, the value of $L$ is adjusted in order to obtain simulated flows as close as possible to observed data. 

Several studies and variants of intervening opportunities models have been developed on this concept~\cite{heanus_1966_comparative, ruiter_1967_toward, haynes_1973_intermetropolitan, wilson_1970_urban, fik_1990_spatial, akwawua_2001_development}. The gravity and the intervening opportunities models have been compared several times during the second half of the twentieth century, showing that generally both models perform comparably~\cite{david_1961_comparison, pyers_1966_evaluation, lawson_1967_comparison, zhao_2001_refinement}. In fact one can cast the intervening opportunities model as a special variant of the gravity model with the friction factor $f(r_{ij})$ replaced by a function of the number of opportunities between the two locations, $f(V_{ij}) = e^{-L \, V_{ij-1}}$ \cite{eash_1984_development, zhao_2001_refinement}. Consequently, the intervening opportunities model can also be derived from entropy maximization methods as discussed earlier in the context of the constrained gravity model.  Indeed, some authors  have proposed hybrid gravity-opportunities models taking into account both the effect of distance and the number of opportunities between locations~\cite{willis_1986_flexible, goncalves_1993_development}.

However, unlike the gravity approach and despite relative good performances, the intervening opportunities model has suffered a growing loss of popularity. This is mainly due to the lack of research effort into the implementation and calibration of the model, attributable to the fact that the theoretical and practical advantages of opportunities models over the gravity ones are not overwhelming \cite{ortuzar_2011_modeling}.

\begin{figure}[t!]
\centering
\includegraphics[width=0.7\textwidth]{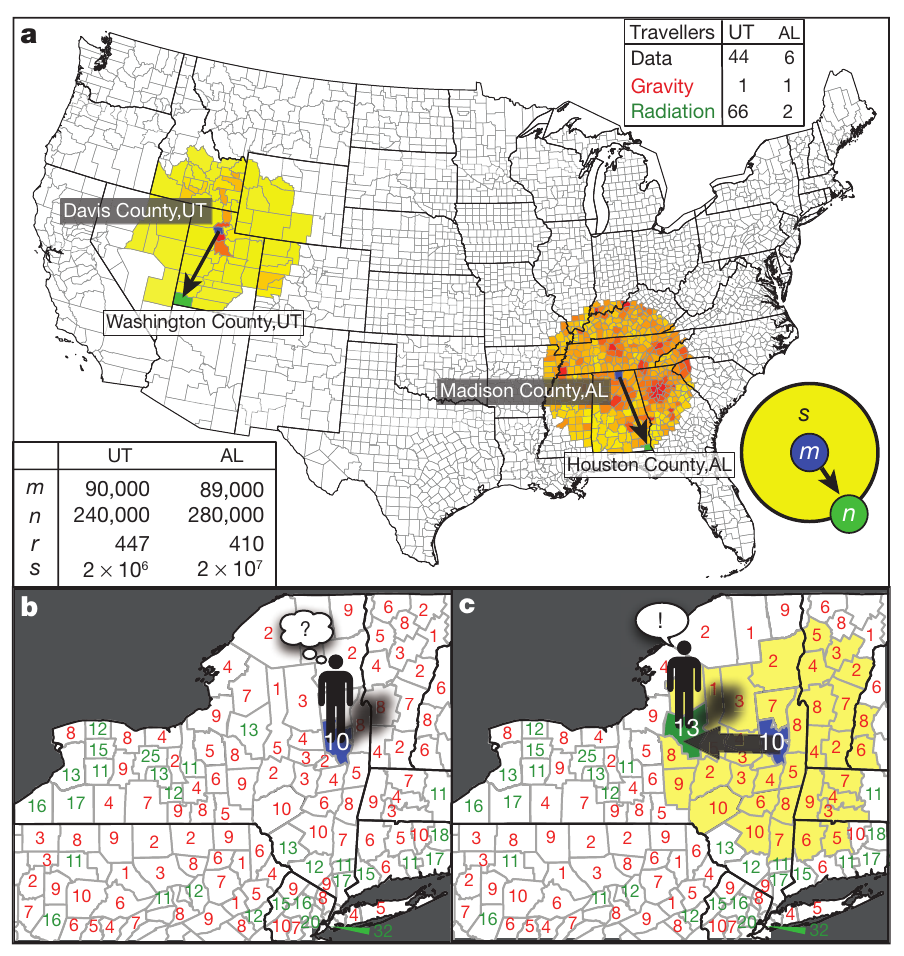}
\caption{ Schematic of the the radiation model. (a) Commuting flows in two pairs of counties,
one in Utah (UT) and the other in Alabama (AL), with similar
origin (m, blue) and destination (n, green) populations and comparable distance $r$ between
them (see bottom left table). Number of travelers in the data, as predicted by the gravity model and finally for the radiation model shown as upper right inset. The definition of
the radiation model: (b) An individual (e.g. living in Saratoga County, NY) applies for
jobs in all counties and collects potential employment offers. The number of job
opportunities in each county is chosen to be proportional to the resident
population. Each offerÕs attractiveness (benefit) is represented by a random variable
with distribution $p(z)$, the numbers placed in each county representing the best offer
among the jobs in that area. Each county is marked in green (red) if its best offer
is better (lower) than the best offer in the home county. (c) An individual
accepts the closest job that offers better benefits than his home county.  Figure from~\cite{simini_2012_universal}. 
\label{fig:fig_simini_2012_universal_1}}
\end{figure}

\subsubsection{The radiation model}
\label{sec:radiation}

The concept of intervening opportunities and Schneider's hypothesis has recently triggered a renewed interest thanks to the recently proposed radiation model \cite{simini_2012_universal}. The radiation model assumes that the choice of a traveler's destination consists of two steps. 
First, each opportunity in every location is assigned a fitness represented by a number, $z$, chosen from some distribution $p(z)$, whose value represents the quality of the opportunity for the traveler. 
Second, the traveler ranks all opportunities according to their distances from the origin location and chooses the closest opportunity with a fitness higher than the traveler's fitness threshold, which is another random number extracted from the fitness distribution $p(z)$ (see \figurename~\ref{fig:fig_simini_2012_universal_1}). 
As a result, the average number of travelers from location $i$ to location $j$, $T_{ij}$, takes the form:
\begin{equation}
   T_{ij}=O_i \, \frac{1}{1-\frac{m_i}{M}} \frac{m_i \, m_j}{(m_i+s_{ij})\, (m_i+m_j+s_{ij})} .
	\label{eq:rad}
\end{equation}
Here again, the destination of the $O_i$ trips originating in $i$ is sampled from a distribution of probabilities that a trip originating in $i$ ends in location $j$. This probability depends on the number of opportunities at the origin $m_i$, at the destination $m_j$ and the number of opportunities $s_{ij}$ in a circle of radius $r_{ij}$ centered in $i$ (excluding the source and destination). This conditional probability needs to be normalized so that the probability that a trip originating in the region of interest ends in this region is equal to $1$. In case of a finite system it is possible to show that this is equal to $1-\frac{m_i}{M}$ where $M = \sum_i m_i$ is the total number of opportunities \cite{masucci_2013_gravity}. In the original version of the radiation model, the number of opportunities is approximated by the population, but the total inflows $D_j$ to each destination can also be used \cite{lenormand_2012_universal, masucci_2013_gravity, lenormand_2015_systematic}. The great advantage of the radiation model compared with other spatial interaction models is the absence of a parameter to calibrate with observed data.  In particular, the flows defined in \equationname~\eqref{eq:rad} are independent of the fitness distribution $p(z)$.
However, this advantage represents also a limitation since the model does not seem to be very robust to changes in the spatial scale \cite{lenormand_2012_universal, masucci_2013_gravity, liang_2013_unraveling, lenormand_2015_systematic}. To overcome this drawback, a radiation model with opportunities' selection \cite{simini_2013_human} and an extended radiation model \cite{yang_2014_limits} have been proposed. In this extended version, the conditional probability to perform a trip between two locations according to the spatial distribution of opportunities is derived under the survival analysis framework introducing a parameter $\alpha$ to control the effect of the number of opportunities between the source and the destination on the location choice. The addition of this scaling parameter seems to greatly improve the performance of the model, and similar to the gravity model, its value can be inferred from the scale of the study region according to the homogeneity of the opportunities' spatial distribution \cite{yang_2014_limits,lenormand_2015_systematic}.

\begin{figure}[t!]
\centering
\includegraphics[width=0.8\textwidth]{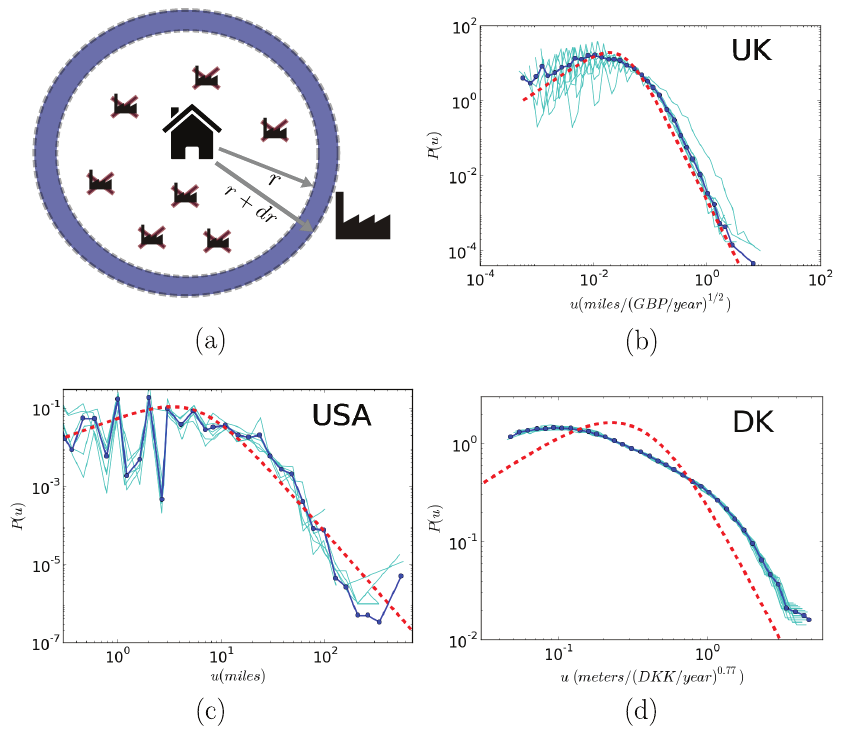}
\caption{ The closest opportunity model (a) Sketch with the main ingredients of the model: the residence place of the agents and the opportunities with their correspondent quality until the last one at distance $r$ is selected. In (b), (c) and (d) rescaled distributions of commuting distances. In blue, the empirical data for several years in the three countries, in dark blue the averaged empirical distributions, and superimposed in red the model fits using a single parameter. Figure from~\cite{carra_2016_modeling}. 
\label{fig:carra_2016_modeling_2}}
\end{figure}


An interpretation of the radiation model 
in the context of job search and, consequently, for the formation of commuting flow networks has been given in \cite{carra_2016_modeling}. 
The basic components are individuals residing within a demarcated geographical area, who are seeking employment. Job locations are uniformly distributed in space and characterized by a fitness parameter $z$, which itself is drawn from a distribution that includes aspects such as wage and worker-convenience. Individuals have a certain tolerance level, $z^{\ast}$, and will search in a progressively increasing radius from their residence, with the search terminating at the first instance of the condition $z >  z^{\ast}$. Using extreme value statistics, it is possible to determine the distance distribution, $P(r)$, between residence and job places, which is independent of the distribution of $z$ (for non-pathological cases), and was calculated to be of the form 
\begin{equation}
P(r) = \frac{2\, \rho\, \pi\, r}{(1+ \rho\, \pi \, r^2)^2} ,
\end{equation}
where $\rho$ is the density of jobs. The model satisfactorily reproduces some of the empirical features of the travel distance distributions in commuting data as can be seen in \figurename~\ref{fig:carra_2016_modeling_2}.

\subsubsection{Comparison between models}
\label{sec:comparison}

The models described so far provide theoretical expectations for the OD matrices in terms of the flow values $T_{ij}$ between geographical regions $i$ and $j$. Before employing these for practical applications, it is essential to validate the expected (calculated) flows against empirical evidence, that may be limited in the sense of being restricted in space compared to regions of interest, or indeed to shorter time windows than the temporal period of interest. By evaluating the performance of each model, one selects 
the one that most closely matches the empirical data (if available). The accuracy, in general, will depend on the spatio-temporal scale as well as on available (meta) information and the extent of missing data. It is crucial that comparison between models is done on an equal footing, i.e using the same extent of available data and in equal spatio-temporal resolutions. As discussed, the model output is the value of $T_{ij}$, subject to constraints, which range from \equationname~\eqref{eq:grav}, where only the total sum of the flows is imposed via the parameter $K$, the singly constrained framework models based on intervening opportunities or the closest opportunity case, which is production constrained, in the sense that outflows $O_i$ are fixed \emph{ a priori} and must be preserved for every location, and finally to the doubly-constrained models, where both the outflows $O_i$  and the inflows $D_i$ in every region $i$ are preserved. 

\begin{figure}[t!]
\centering
\includegraphics[width=0.7\textwidth]{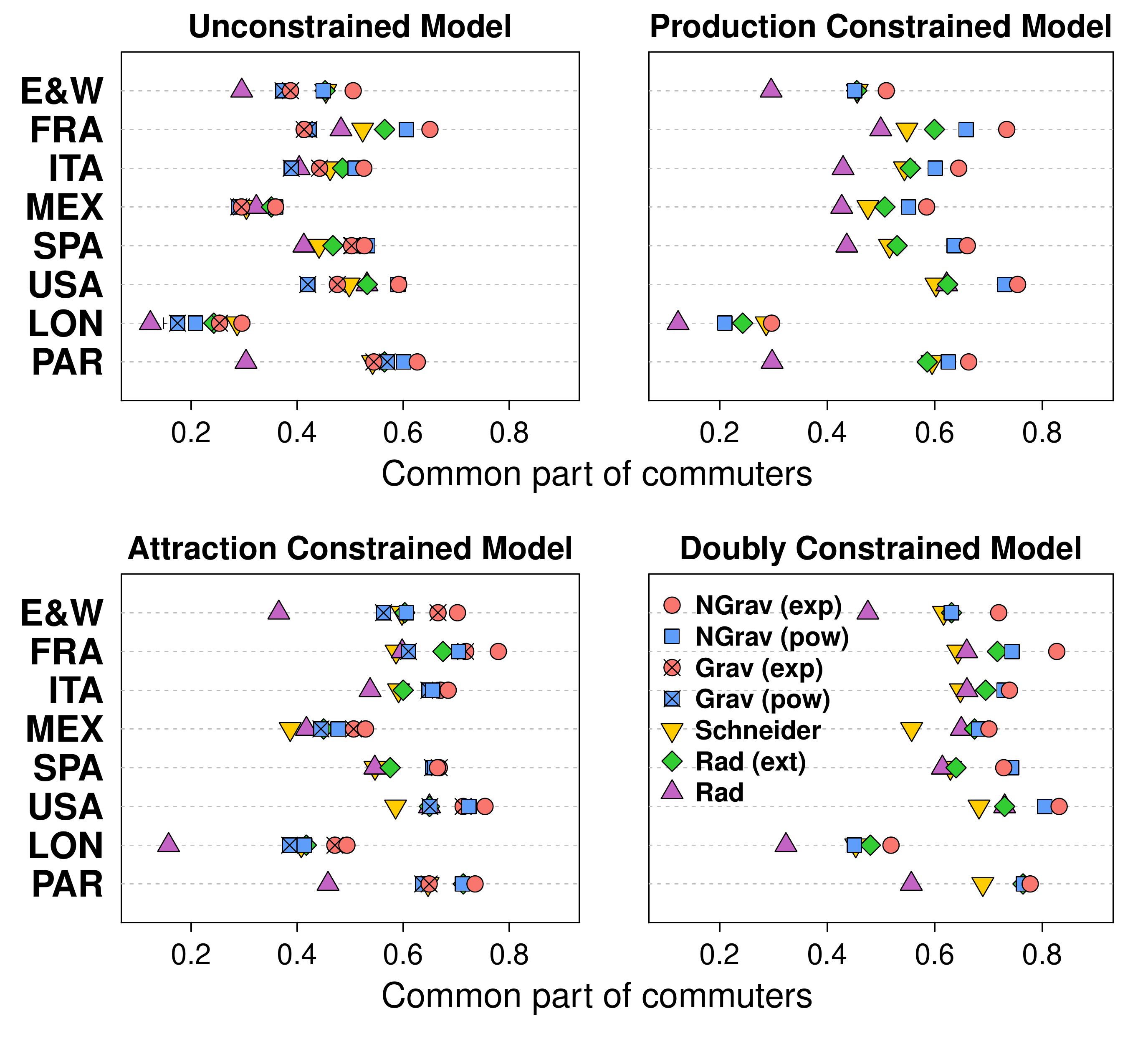}
\caption{Performance comparison of the different models described in the text using the CPC metric defined in \equationname~\eqref{eq:cpc} applied to census commuting data from England and Wales (E\&A), France (FRA), Italy (ITA), Mexico (MEX), Spain (SPA), USA. Additional census data at the city-scale was gathered from and London (LON) and Paris (PAR). The different symbols and colors represent the different flavors of the model. The short-forms exp (exponential) and pow (power-law) refer to the functional forms of distance dependence. Figure from~\cite{lenormand_2015_systematic}. 
\label{fig:lenormand_2015_systematic_3}}
\end{figure}

In general, all of the described models share some common assumptions:
\begin{enumerate}
\item The trip distribution $P_{ij}$ that generate the flows $T_{ij}$ is independent of the trip production $O_i$.
\item In both the unconstrained and singly-constrained models, choice of travel destinations are statistically independent (i.e there are no memory effects).  
\item Flows are estimated as a product of variables related to opportunities and distance (i.e. variables are "separable").
\end{enumerate}
Under these assumptions, it is possible to draw statistical laws that govern the distribution of travels $P_{ij}$, and from thereon, build models with varying levels of constraints. This is important to adapt the models to the available information and to fairly compare between them.

One such comparison was carried out in~\cite{lenormand_2015_systematic}, where commuting data was obtained from the census offices of England and Wales (E\&A), France (FRA), Italy (ITA), Mexico (MEX), Spain (SPA), USA and then in shorter spatial scales, from the cities of London (LON) and Paris (PAR). Several variants of the gravity, intervening opportunities and radiation models were considered with multiple levels of constraints. The predicted flows $T_{ij}$ were compared with the empirical data using two related metrics: The \emph{common part of commuters} ($CPC$) for all location pairs $i,j$ with positive flows both in the empirical data set and in the model prediction, defined as
\begin{equation}
CPC = \frac{\sum_{i,j} min (T^m_{ij}, T^e_{ij})}{T} = 1 - \frac{1}{2} \, \frac{\sum_{i,j} |T^m_{ij}-T^e_{ij}|}{T}.
\label{eq:cpc}
\end{equation}      
Here $T = \sum_{i,j} T^e_{ij}$ is the total number of commuters, $T^m_{ij}$ the model prediction for the flow, and $T^e_{ij}$ the empirical value. The $CPC$ is one, if the agreement is perfect and zero if there is no overlap between data and model. The ability of the models to
recover the topological structure of the original network was assessed through the second metric, termed the \emph{common part of links}
(CPL) and defined as

\begin{equation}
CPL = \frac{2 \, \sum_{i,j} 1_{T^m_{i,j} > 0} \, 1_{T^e_{i,j}}}{\sum_{i,j} 1_{T^m_{i,j} > 0} + \sum_{i,j} 1_{T^e_{i,j} > 0}} ,
\end{equation}
where $1_{T_{i,j} > 0}$ is the Heaviside function. The CPL measures the proportion of links in common between
the simulated and the observed networks, it is null if there is no link in common and one if both networks are topologically
equivalent. The results in terms of the $CPC$ are shown in Fig. \ref{fig:lenormand_2015_systematic_3}. For this choice of datasets,  the gravity model is seen to moderately outperform the other models. One must note of course, that this might change across different datasets.

\subsection{Intermodality}
\label{sec:intermodal}

\begin{figure}[t!]
\centering
\includegraphics[width=\textwidth]{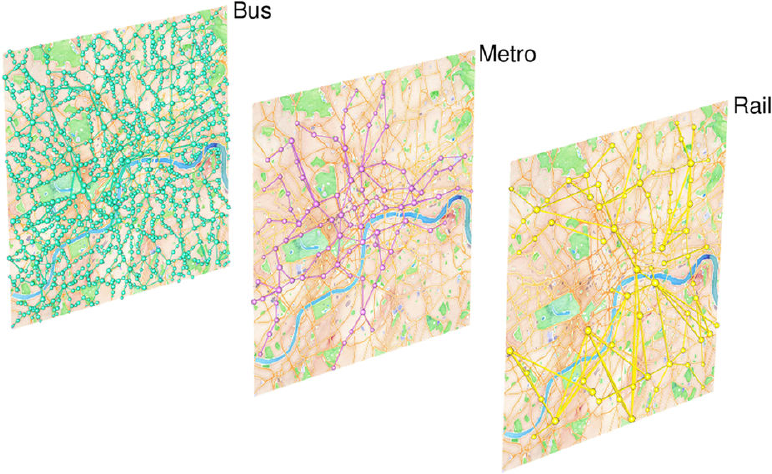}
\caption{An example of a multilayered network approach to mobility. Public transport networks in the London metro area, separated into multiple layers consisting of the bus, subway and rail networks. In this particular instance, one can see that the London bus network is the most ``used" layer. Figure from~\cite{gallotti_2015_multilayer}. }
\label{fig:gallotti_2015_multilayer_2}
\end{figure}

A complete description of mobility must take into account the mutlimodal structure of transport and the transitions between the different modes as alluded to in \sectionname~\ref{sec:multimodal}. A good framework for carrying out such an analysis has been developed in recent years in the context of multilayer or multiplex networks~\cite{boccaletti_2014_structure,kivela_2014_multilayer}. These are networks in which the nodes can be present in one or multiple layers, and each layer by itself contains a set of links or node-node interactions; if a large fraction of nodes in the network are present in all apparent layers, then it is termed multiplex. On the other hand if one finds different flavors of nodes restricted within their own putative layers, then the network is dubbed a multilayered one. Indeed, each layer of the network constitutes a separate network in its own right depending on the context of its interactions. For example, in social networks, layers may correspond to different groupings such as coworkers, family, different cohorts of friends and so on, all of which have within-layer interactions. Information diffusion in the network, however, occurs through each layer and across the full multilayer structure. Translating this to the case of mobility and transportation networks, layers may correspond to different transportation modes (road, subway, airline), while connections between layers constitute the interchanges between these modes. Within this context, the standard way to construct multilayer networks is to associate locations to nodes and flows, and frequency of travel (or simply unweighted simple connections) to links between different transportation modes (see, for example \figurename~ \ref{fig:gallotti_2015_multilayer_2}).

Traditionally, single-mode transport networks have been studied using mono-layers. Examples include metro networks in Beijing \cite{feng_2017_weighted}, Boston \cite{latora_2002_boston}, London \cite{angeloudis_2006_large,guo_2016_london}, Montreal \cite{derrible_2010_complexity}, Paris \cite{angeloudis_2006_large} and Seoul \cite{lee_2008_statistical}; bus networks in China \cite{xu_2007_scaling,chen_2007_study} and Poland \cite{sienkiewicz_2005_statistical}; train networks in Boston, Vienna \cite{seaton_2004_stations} and in India \cite{sen_2003_small}; and combined public transportation networks in $14$ cities around the world \cite{vonferber_2009_public}. In most of these cases, the transport network was generated using different perspectives. One such perspective is the $L$-space representation, where nodes correspond to stations in the transportation map, and are linked together if they are consecutive stops in a transportation line. Similarly, the network can be built in the $P$-space framework, if nodes are stations linked together if they correspond to stops in the \emph{same} line (irrespective of sequence). This latter representation corresponds to a  one-mode projection of a bipartite network~\cite{Newman_book} consisting of stations and lines. (For a graphic explanation of this notation see~\cite{vonferber_2009_public}.) In addition to geographical aspects, the link weights can also incorporate information on the number of passengers in the same route, among other features. 

The fact that a single transport network can be represented in multiple ways, naturally paved the way for the use of multilayer structures \cite{kurant_2006_layered,kurant_2006_extraction}. In this case, each layer contains the expression of the network in a different space, and the multilayer allows for a deeper study of the network accessibility for travelers. As mentioned before, a multilayer network can be also employed to condense the transport information coming from different modes. An example for this can be seen in \figurename~\ref{fig:gallotti_2015_multilayer_2} which shows elements of the transportation network in London, combining the bus, train and metro networks, whereas additional elements including air and ferry can also be incorporated~\cite{gallotti_2015_multilayer}. The more intricate topology of multilayer networks necessarily affects the spreading processes taking place on the system (diffusion and navigation)~\cite{dedomenico_2013_mathematical,gomez_2013_diffusion,kivela_2014_multilayer} that are modeled using the methods discussed in \sectionname~\ref{sec:rw}~and~\ref{sec:poplev}. This leads to new features such as congestion~\cite{sole_2016_congestion}, change in the navigability of the network subject to disruptions~\cite{dedomenico_2014_navigability, battiston_2016_efficient}, produces new dynamic regimes that are absent in monolayers~\cite{radicchi_2013_abrupt} and abrupt reactions to minimal topological changes~\cite{diakonova_2017_dynamical}.

A particular type of multilayer representation of transport networks is worth a separate discussion. In this case, nodes still correspond to locations, and links are assigned according to one of the  space-frameworks ($L$ or $P$), but the weight of the links now account for the temporal duration of trips between the nodes. This network configuration, which some authors refer to as a $PT$-space, appear in each of the layers, where each layer usually corresponds to a single transport mode. With such a representation, the optimal  time for a traveler traversing between a given OD pair, can be calculated using optimal path algorithms across the multilayer structure, containing the full transport network and taking into account intermodality. These types of multilayers have been studied, for example, in British \cite{gallotti_2014_anatomy} and French \cite{alessandretti_2016_user} cities. In the British case, one detects a shift in preference for different transportation modes as well as the overlap between temporally and spatially optimal paths as a function of distance. In the French case, a few particular (temporally optimal) connections were singled out, so-called \emph{efficient connections}, and their occurrence (and hence influence) in the general mobility modes were studied across a number of cities. Using a similar framework, with two layers corresponding to road and metro networks in the cities of London and New York, congestion effects as a function of varying the speed of metro lines was studied~\cite{strano_2015_multiplex}. Finally, it is worth mentioning the point raised in \cite{aleta_2016_multilayer}, where the authors discuss the differences between building the network by considering a layer as a single transport mode or as a single line. The latter description provide a deeper view of the inefficiencies due to line transfer of the passengers, while the former is well-suited to study topological aspects and geographical coverage.



\section{Selected Domains of Application}
\label{sec:applications}

Here we discuss some selected applications of the frameworks, concepts, models and datasets introduced thus far. We organize this section by scale, ranging from single scale (mostly dealing with transportation modes), multi-scale applications with a focus on movement in cities and epidemic spreading, and finally end with some new developments related to virtual scales, i.e mobility patterns seen in online activity.

\subsection{Single-Scale}

\subsubsection{Pedestrian movement}

Large concentrations of people such as those occurring during festivals, sport events or religious ceremonies can present a challenge in terms of service and transport demand as well as public safety. Avalanches in stadiums, processions and even subway stations have occurred in recent decades leading in high casualties~\cite{helbing_2013_pedestrian}. In the face of this, strategies need to be devised for ensuring prompt evacuation in cases of emergency when considering the configuration of new buildings, public spaces and transport systems. Consequently, understanding the empirical patterns governing pedestrian movement and being able to build corresponding models, has the potential to benefit a wide range of disciplines including architecture, urban planning and public service stakeholders among others. Given the broad range of interest, a large amount of research has been devoted to this sphere, over the past few decades. Indeed, several reviews exist on this topic~\cite{helbing_2013_pedestrian,helbing_2001_traffic,zainuddin_2010_characteristics,vicsek_2012_collective,benenson_2014_ten,cao_2015_cyber}, so instead of providing a detailed retrospect, we provide here an overview of the most recent developments. 

Modeling frameworks on pedestrian movement can be broadly classified according to the approach used to consider individual pedestrians. For example, when the aim is crowd control, then usually the number of individuals considered is quite large, and the metrics of interest are crowd velocities or pressures. In such a case, one can define fields at every point in space and time to characterize the state of the system. These fields usually include the density of pedestrians $\rho(\vec{x},t)$, the local velocity $\vec{v}(\vec{x},t)$ or the flow across an exit, $J(\vec{x},t)$. In analogy with fluid dynamics, one sees a continuity equation of the form,\begin{equation}
\label{eq:ped:conv} 
\frac{\partial \rho(\vec{x},t) }{\partial t} = - \nabla (\rho(\vec{x},t) \, \vec{v}(\vec{x},t)),
\end{equation}
stemming from the conservation number of people in the crowd.  
The motion of pedestrians can be assumed to minimize the time spent in the system. This condition is included via the introduction of a field $\phi(\vec{x},t)$ coupled with the density in such a way that the local velocity shows a relation
\begin{equation}
\label{eq:ped:speed} 
\vec{v}(\vec{x},t) = f(\rho) \frac{\nabla \phi}{||\nabla \phi||},
\end{equation}
where the gradient of $\phi$ determines the direction of motion and $f(\rho)$ is a simple function of the density ensuring that the pedestrians speed decreases  as the density increases. Possible choices are a linear relation, $f(\rho) = \rho - \rho_{max}$, keeping $\rho < \rho_{max}$ or a non linear one, $f(\rho) = (\rho - \rho_{max})^2$, among others~\cite{hughes_2002_continuum,helbing_2006_analytical,carrillo_2015_local}. The connection with $\phi$ is established through the function $f(\rho)$ thus,
\begin{equation}
\label{eq:ped:phi} 
||\nabla \phi|| = \frac{1}{f(\rho)},
\end{equation}
leading to the so-called \emph{Hughes model}. In a scenario with individuals exiting a room, it produces two regimes in the pedestrian mobility depending on the global density, one with free flow and another with congestion. Furthermore, it turns out that shock-waves and moving congestion fronts emerge due to the narrowing of the crowd close to the door, leading to intermittent exit flow, avalanches and stop-and-go effects \cite{helbing_2006_analytical}. 

\begin{figure}[t!]
\centering
\includegraphics[width=\textwidth]{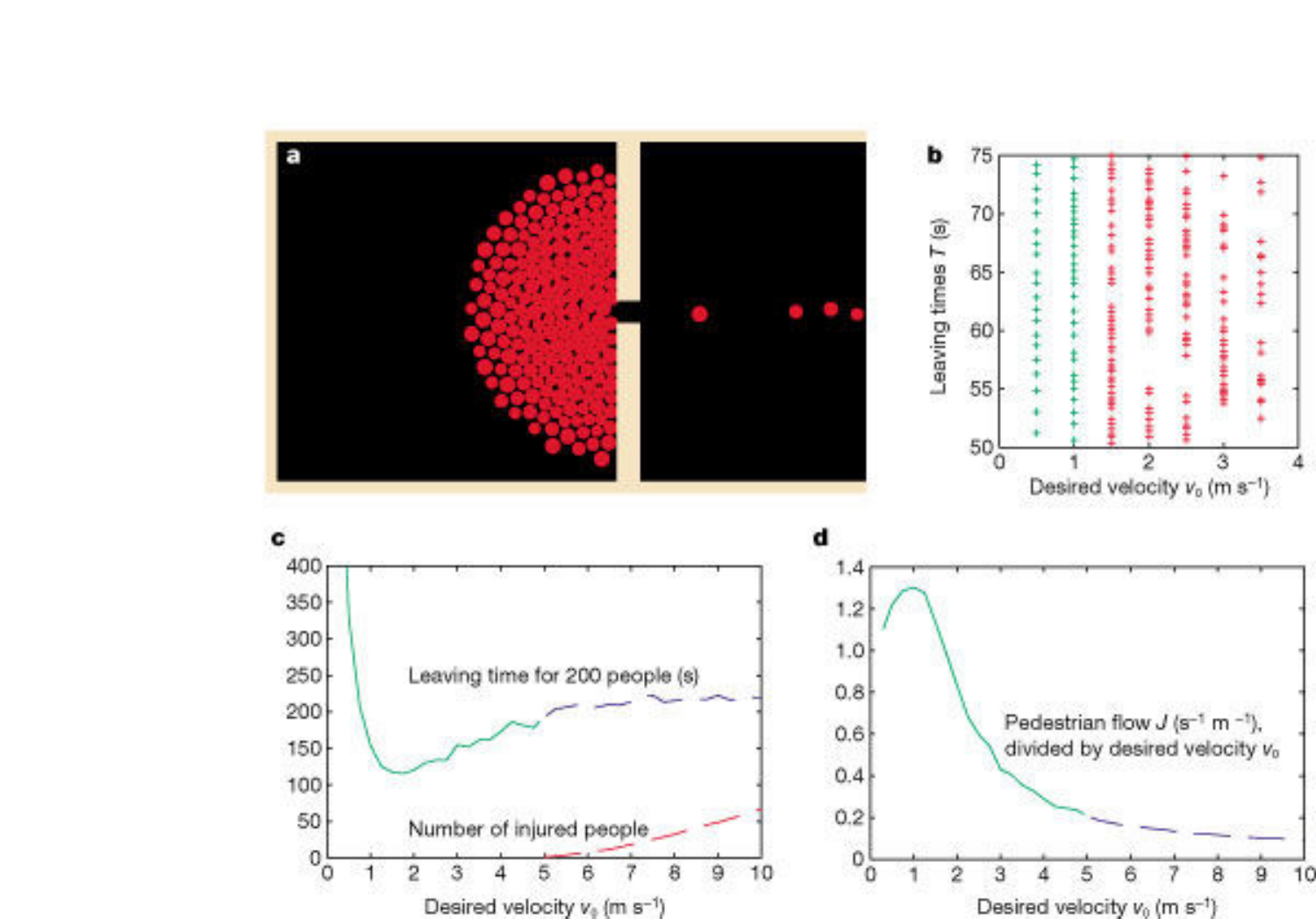}
\caption{Social force model simulation of a crowd trying to leave a room through a narrow door. (a) a representative configuration. (b), the sequence of leaving times of the agents as a function of $v_0$. In (c) and (d), the total evacuation time and the average flow of people as a function of $v_0$. Figure from~\cite{helbing_2000_simulating}. \label{fig:helbing_2000_simulating_1}}
\end{figure}

An alternative approach is to use agent based modeling where the state of every individual (agent) $i$ is simulated in detail. This framework is especially applicable to simulations of heterogeneous populations in which the agents may show different features. The most popular model falling into this category is the so-called \emph{social force model} proposed by Helbing and Moln\'ar \cite{helbing_1995_social, helbing_2001_self}. The idea behind this model is to apply the analogues of Newton's laws of motion to each pedestrian $i$. If the agent's velocity is $v_i$, the location $r_i$ at time $t$, then the acceleration can be expressed as
\begin{equation}   
\label{eq:ped:sfm}  
\frac{d v_i}{dt} = f_i^d + f_i^a + f_i^{rb} + \sum_{j \ne i} f_{ij}^r + \eta(r,t) .
\end{equation}
The first term on the right hand side accounts for the tendency of the individuals to walk at a certain desired speed $v_i^0$. It is usually written as 
\begin{equation}   
\label{eq:ped:des}  
f_i^d = \frac{v_i^0-v_i}{\tau} ,
\end{equation}     
which ensures the recovery of the desired speed in a characteristic reaction time $\tau$. The second term $f_i^a$ is an attraction force depending on the agent's position introduced to guarantee that walkers are compelled to a certain target or direction. The third force, $f_i^{rb}$, is repulsive depending on the agent's location and takes into account the effects of the boundaries and other obstacles on the agent's trajectory. The next term represents a repulsive force felt by the presence of other agents $j$ different from $i$. Finally, $\eta(r,t)$, is an uncorrelated noise term that introduces low levels of random fluctuations to avoid deadlocks.

\begin{figure}[t!]
\centering
\includegraphics[width=0.8\textwidth]{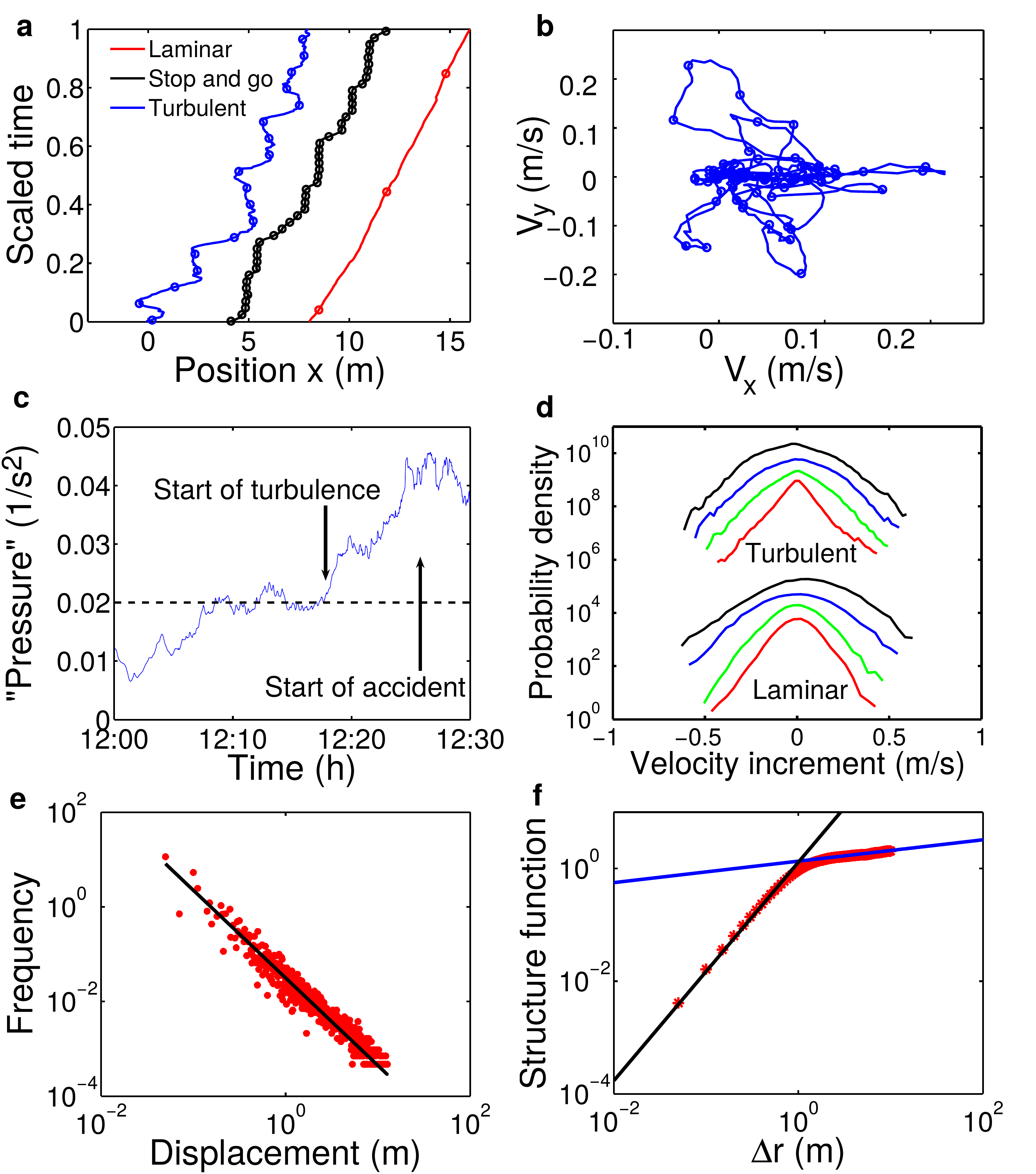}
\caption{Analysis of an avalanche event during the Hajj pilgrimage of 2006. (a) Representative trajectories of the laminar flow, the stop-and-go and the turbulent regimes for individual movements. (b) The velocity in the turbulent regime. (c) The "pressure" as a function of time. (d) Different distributions of speed increments in the two regimes. (e), Distribution of displacements between consecutive stops. (f) The structure function in the turbulent regime. Figure from~\cite{helbing_2007_dynamics}. \label{fig:helbing_2007_dynamics_3}}
\end{figure}

The social force model reproduces several phenomena observed in empirical circumstances. For example, when two groups of pedestrians moving in opposite directions meet in a corridor, unidirectional stripes are formed, or when they meet in a door, the flow displays intermittence due to clogging \cite{helbing_2001_self} (\figurename~\ref{fig:helbing_2000_simulating_1}). Variations of this model have been used to analyze escape patterns from a building in low visibility conditions \cite{helbing_2000_simulating, helbing_2013_pedestrian}, to assess evacuation plans in skyscrapers \cite{sikora_2012_model,parisi_2014_sequential}, ships \cite{chen_2015_modeling} and aircraft \cite{kirchner_2003_friction}. In relation to crowd control, they have also been used to analyze how to mitigate avalanches in the Hajj religious gathering in Saudi Arabia~\cite{helbing_2007_dynamics,johansson_2008_crowd}, in the Notting Hill Carnival in London~\cite{batty_2003_safety} or, along with mobility models for other transport media, to study evacuation of cities and populated areas in case of natural disasters such as flooding ~\cite{lammel_2010_representation,kunwar_2014_large,kunwar_2015_evacuation}. Additionally, these models provide the basic tool to model scenarios such as the formation of mosh and circle pits as a mark of collective motion in heavy music concerts \cite{silverberg_2013_collective}, as well as the risk of disease propagation in massive gatherings \cite{johansson_2012_crowd}.

In the last decade, there has been a large volume of research related to the constituent terms of~\equationname\eqref{eq:ped:sfm}, including empirical calibration of the different coefficients as well as the introduction of a cognitive basis for the different forces~ \cite{zainuddin_2010_characteristics, johansson_2008_crowd}. In high density regimes,  such as those occurring in videos taken of the disaster of the Jamarat Bridge in the Hajj pilgrimage of 2006, the trajectories of single pedestrians become turbulent \cite{helbing_2007_dynamics,johansson_2008_crowd} (\figurename~\ref{fig:helbing_2007_dynamics_3}). The original version of the social force model is unable to  reproduce this turbulent regime and consequently modifications to the force terms in high density regimes, were introduced. One of these modifications is the centrifugal force model, so-called because of its similarity with the classical centrifugal force in rotating bodies, and involves a modification of the mutual repulsion term $ f_{ij}^r $ thus,\begin{equation}
f_{ij}^r  =  -m_i \, K_{ij} \, \frac{V_{ij}^2}{||\vec{r}_{ij}||} \, \vec{e}_{ij}. 
\end{equation}
Here $m_i$ is the mass of pedestrian $i$, $K_{ij}$ is a coefficient enforcing the influence only of pedestrians in front of $i$ on its movements and $\vec{e}_{ij}$ is the unit vector pointing from $i$ to $j$~\cite{yu_2005_centrifugal,yu_2007_modeling} . This modification has the effect of introducing both laminar as well as turbulent regimes, a feature missing in earlier versions of the social force model. Further refinements also included the interactions of individuals based on specific boundaries and shapes of obstacles~\cite{chraibi_2010_generalized}. 

A second modification to the social force model deals with the cognitive capacities and limitations of pedestrians and how they can affect the force terms of the model \cite{moussaid_2009_experimental,moussaid_2011_how}. 
Examples include analysis of the time scales regulating pedestrian interactions~\cite{johansson_2009_constant}, the addition of a field to take into account the (partial) global perception of individuals and the environment under consideration \cite{dietrich_2014_gradient,colombi_2015_moving}, or the role of leaders \cite{degond_2015_time}. The capacity of the pedestrians to foresee the movements of others and the dangers of obstacles have been modeled within this framework in \cite{moussaid_2009_experimental,steffen_2009_modification}. The model has also been adapted to agents with an anisotropic perception of their environment \cite{gulikers_2013_effect} and its instabilities in a context of stop-and-go oscillations have been studied in \cite{chraibi_2014_oscillating,chraibi_2015_jamming}. A systematic stability analysis of its numerical solutions has been provided in~\cite{koster_2013_avoiding} together with a version of the model that notably increases its computational performance. 

In addition to this, there exist a family of models that lie between the macroscopic and microscopic approaches, characterized by discretization of space into cells, and the specification of rules on how agents navigate between these cells. Such \emph{cellular automata} models have the advantage of computational scaleability, as for example, in~\cite{dutta_2014_gpu} where simulations of crowds of size $10^5$, were conducted with running time scaling linearly with the number of individuals. The price to pay for this increased computational performance is the loss of local details within each cell. These models have been used to study clogging in exits, and due to obstacles by adding a ``friction term" between agents \cite{kirchner_2003_friction,yanagisawa_2009_introduction}, pedestrian flow through subsequent bottlenecks \cite{ezaki_2012_pedestrian} and the formation of stripes in junctions with agents moving in different directions \cite{cividini_2013_diagonal}.  Some modifications to these models include, for example, behavioral factors of the agents such as path selection based on the time necessary to reach the destination \cite{kirik_2009_shortest}, heterogeneous agents \cite{sarmady_2010_simulating} and long-range spatial awareness \cite{tissera_2014_simulating}.

\begin{figure}[t!]
\centering
\includegraphics[width=0.9\textwidth]{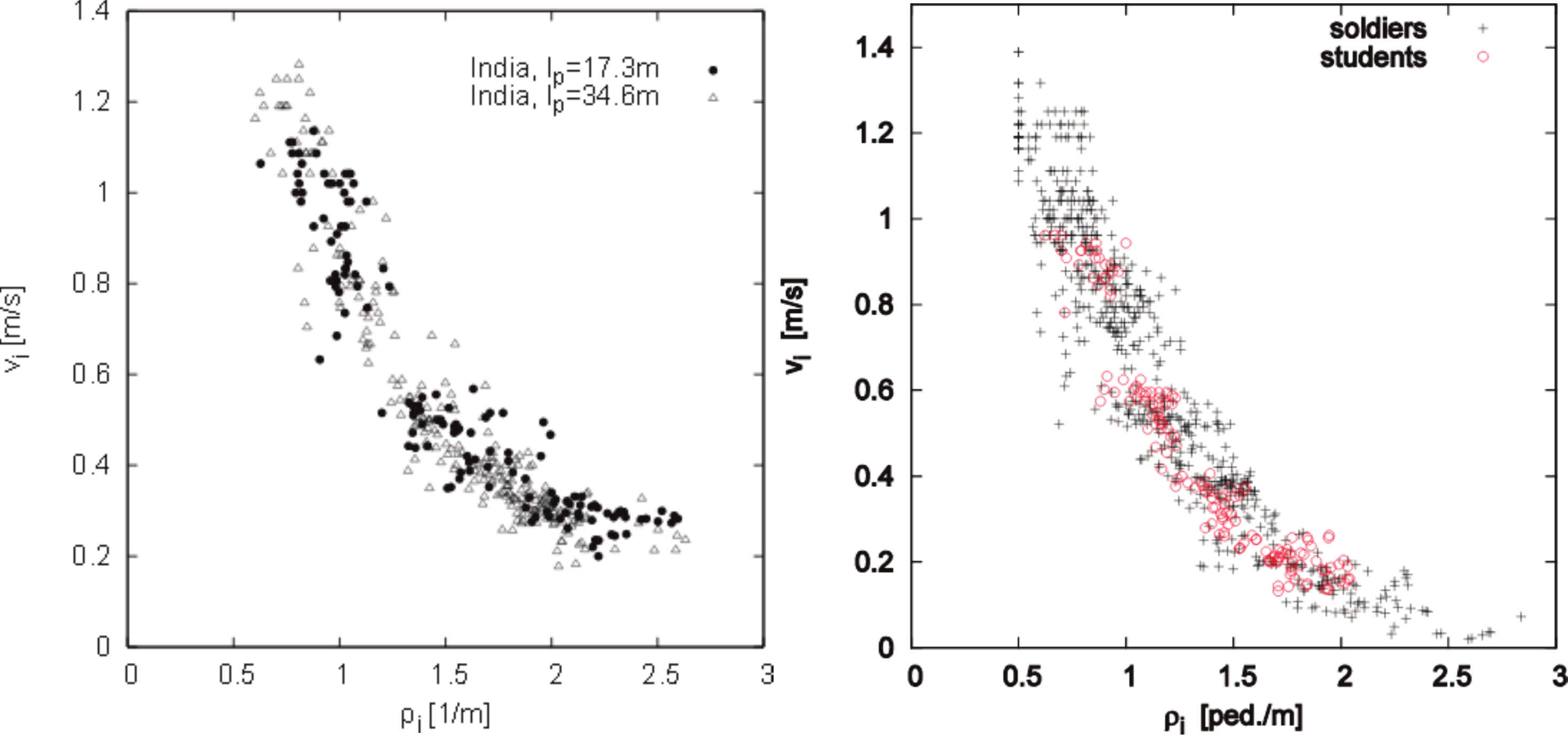}
\caption{Fundamental diagram with the speed as a function of the density of pedestrians for several experiments in a closed loop circuit (ovoid-like). On the left, experiments in India with different loop lenghts, plot originally taken from \cite{chattaraj_2009_comparison}.On the right, experiments in Germany with two set of people: students and soldiers, figure originally taken from \cite{protz_2011_analyzing}. The composite figure comes from~\cite{kretz_2015_social}. \label{fig:kretz_2015_social_2}}
\end{figure}

While some of the models described thus far have been inspired by empirical data obtained from crowds in different contexts, challenges remain in actually monitoring crowds to get accurate and precise measurements on individual movement (see \cite{benenson_2014_ten} for a recent review on methods used for this). 
One way to calibrate models is to take a controlled approach, i.e conduct experiments on a smaller scale, in which groups of people are directed to move under specific constraints. This controlled environment can provide insight into questions such as how pedestrians walk in social groups~\cite{moussaid_2010_walking}, the quantification of pressure and density in different types of bottlenecks and junctions~\cite{zhang_2014_quantification}, the perception of personal space~\cite{ducourant_2005_timing}, the clogging effects and instabilities in narrow areas~\cite{moussaid_2012_traffic,bukacek_2014_experimental} and the asymmetric flow produced by stairs~\cite{corbetta_2015_asymmetric}. One factor that is common to such experiment-based research is the generation and analysis of ``fundamental diagrams" \cite{jelic_2012_properties} (\figurename~\ref{fig:kretz_2015_social_2}), that represent the relationship between the average velocity $v$ and the density $\rho$ of pedestrians, as well the flux through a surface  $J$ (an exit) among other parameters. In the free flow regime, with no obstacles and low density, pedestrians tend to move at an optimal velocity. As the density increases, the velocity decreases until the crowd stops. Similarly, the flux through an exit also decreases with increasing pedestrian concentrations. The relations between these quantities are typically non-linear and are of great relevance for the design of buildings and urban spaces. Due to this, a large effort has been invested in the analysis of these diagrams. The results suggest while qualitatively the functional forms of $v(\rho)$ and $J(\rho)$ have common features across different settings, quantitative differences emerge based on extraneous factors. Some of these are cultural perception of personal space (a comparison between India and Germany was performed in \cite{chattaraj_2009_comparison}), the presence of T-junctions \cite{zhang_2013_experimental}, the individual's perception of open versus closed spaces \cite{zhang_2014_effects} and the bidirectional flow of agents \cite{zhang_2014_comparison}. 

\subsubsection{Air Transportation}
\label{sec:air}

\begin{figure}[t!]
\centering
\includegraphics[width=0.5\textwidth]{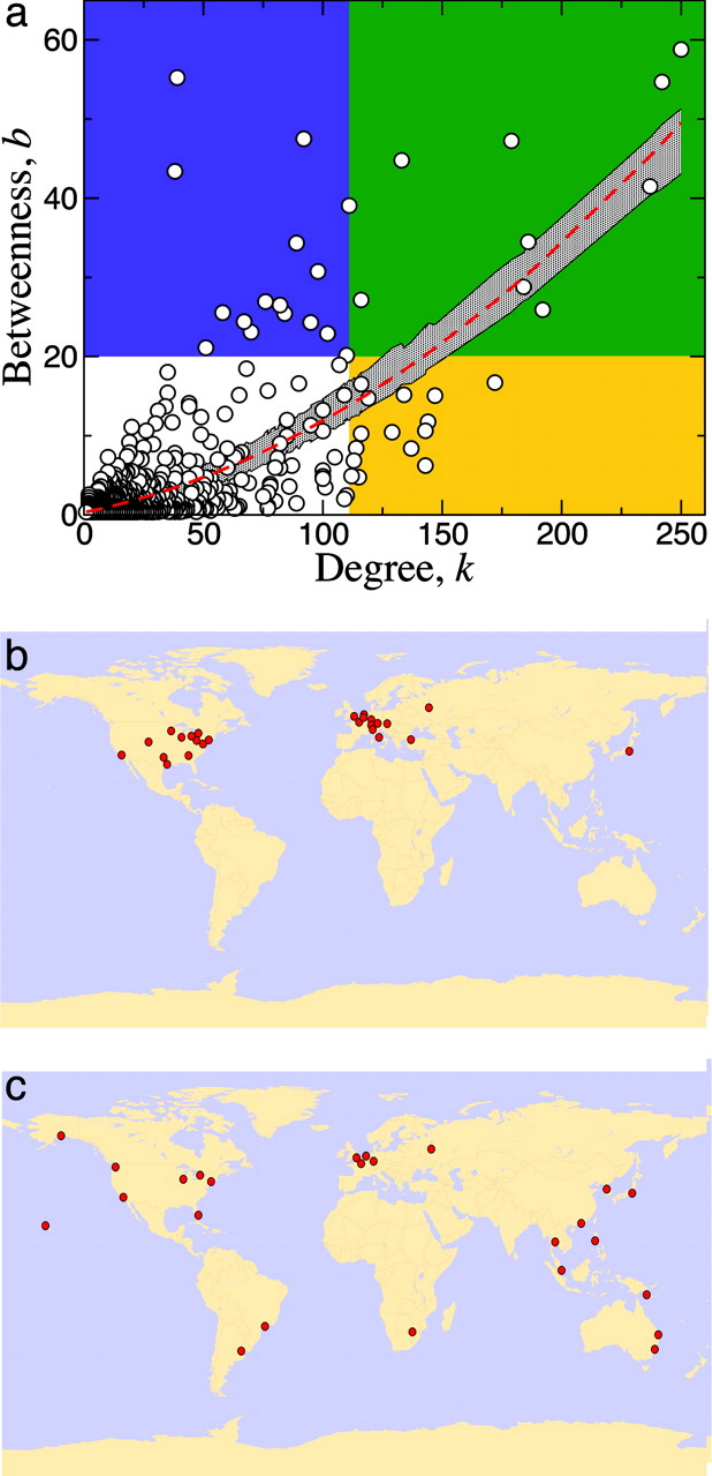}
\caption{(a) The relation between node degree (traffic) and betweenness centrality (measure of load on a node) in the World Airport Network (WAN). (b) The location of the $25$ most connected cities. (c) The top $25$ cities in terms of the load of traffic (betweenness centrality). Figure from~\cite{guimera_2005_worldwide}. \label{fig:guimera_2005_worldwide_2}}
\end{figure}

On the opposite end of the spectrum from pedestrian mobility (which is necessarily short-range) is long-range transportation over thousands of kilometers, which is primarily represented by air transportation. Air transport is of paramount importance for modern global connectivity, playing a central role in the world economy and in the interchange of people, ideas and, unfortunately, also the propagation of diseases. 
While the study of such systems has a long tradition in the engineering community, primarily related to aeronautics and air traffic management~\cite{belobaba_2009_global,cook_2007_european}, recent years has provided further impetus to this line of research with the development of tools based on network science~\cite{Newman_book}. In this setting, initial studies were conducted from a static perspective, where airports are the nodes of the network, with a link between a pair of airports if a direct flight exists between them, discarding any directional information as well as temporal dynamics. The basic statistical features of such networks were analyzed, for instance, in~\cite{guimera_2004_modeling,barrat_2004_architecture,guimera_2005_worldwide} for the World Airport Network (WAN), in ~\cite{li_2004_statistical} for the Chinese network and in~\cite{da_2009_structural} for Brazil. Figure \ref{fig:guimera_2005_worldwide_2} shows the location of the high-degree nodes (maximum traffic) and those with the highest betweenness centrality (bottlenecks) in the WAN. Statistical distributions of the degree (number of destinations of the airports), the traffic (number of passengers per route or per airport) and betweenness were found to be typically heavy-tailed. 

Furthermore, it was found that the networks can be divided into modules with a certain level of self-contained traffic \cite{guimera_2005_worldwide}, although this division is not seasonably stable when the focus is set on smaller geographical scales such as the US network \cite{lancichinetti_2011_finding}. The average number of passengers per route $w_{ij}$ was determined to be a non-linear function of the traffic in airports of the form 
\begin{equation}
w_{ij} \sim \left(k_i\, k_j\right)^\theta,
\label{eq:wij_air}
\end{equation}
where $k_i$ is the traffic at the origin, $k_j$ at the destination and $\theta \approx 1/2$~\cite{barrat_2004_architecture}. An explanation of these measured topological properties has been proposed by simple models based on cumulative advantage~\cite{guimera_2004_modeling,barrat_2004_architecture}.  
The robustness of these networks to disruptions has been considered in terms of a unipartite framework~\cite{verma_2014_revealing} as well as multiplex-like approaches~\cite{cardillo_2013_modeling}, where the network of each airline forms a layer and the airports appearing in the different layers connect the structure. A short review of network analysis on air transportation can be found in \cite{zanin_2013_modelling}.

\begin{figure}
\centering
\includegraphics[width=0.7\textwidth]{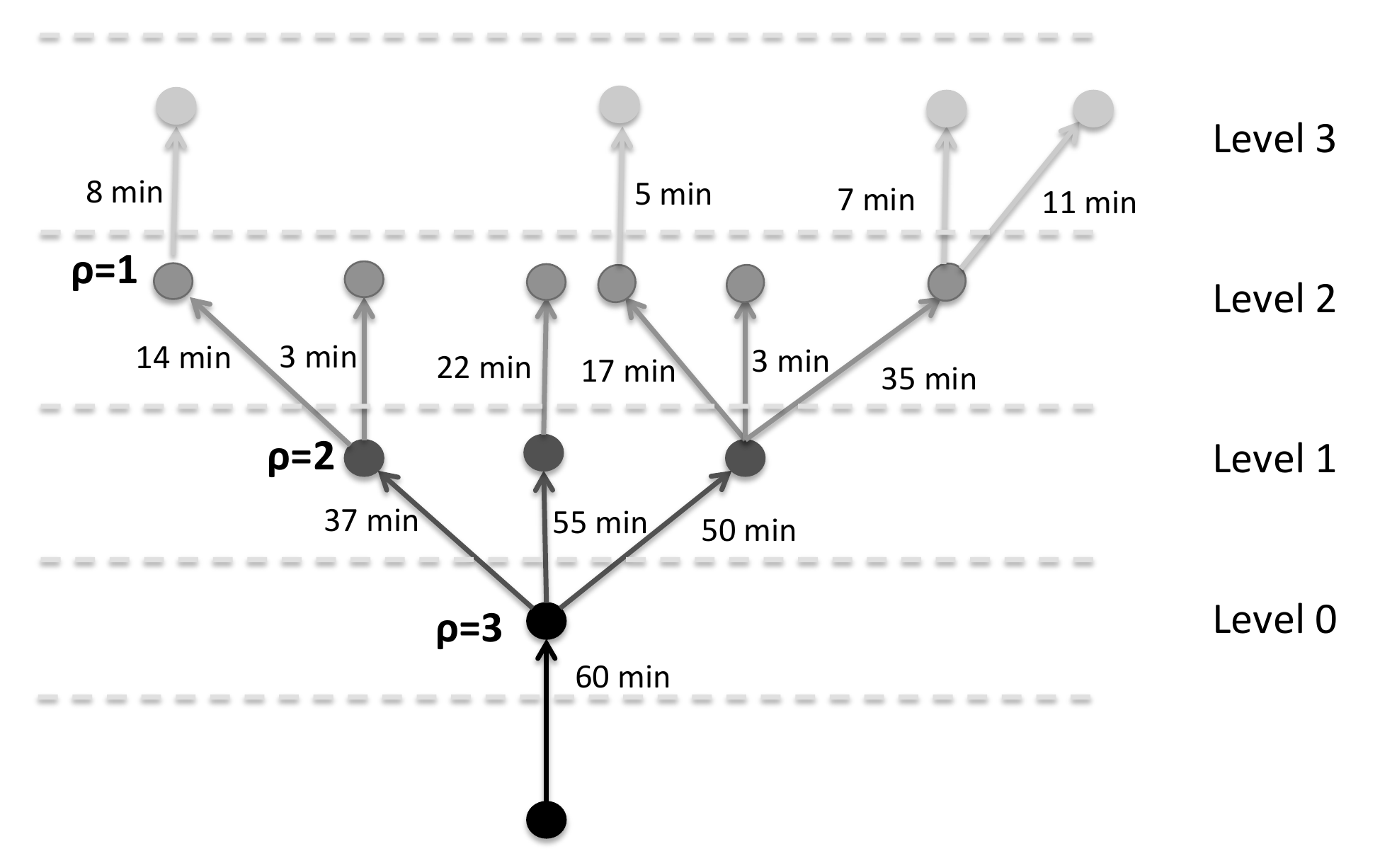}
\caption{A typical configuration of a delay tree. Each node represents an airport and the links represent connections between them. The link weights denote temporal delays (in minutes) and the tree is composed of different levels or generations. Each node has a certain branching number ($\rho$) for the next level.  Figure from~\cite{fleurquin_2014_trees}. \label{fig:fleurquin_2014_trees_2}}
\end{figure}

Of course the static topological framework does not take into account the dynamics occurring on the network. Indeed a key component of air transportation is that goods and people arrive in their destination in time. In particular logistical failures can propagate in the network where travel schedules are not met, with the delay propagating to effect a significant fraction of the system and paralyzing airport operations. This idea was first explored in the context of air transportation in~\cite{beatty_1998_preliminary}, where the concept of \emph{delay multipliers} was introduced. This can be seen as a combination of a delayed flight not only affecting the departure of other flights in the same airport, but also affecting connecting flights down the transportation chain, leading to the formation of avalanches or ``delay-trees"~\cite{fleurquin_2014_trees} as shown in \figurename~\ref{fig:fleurquin_2014_trees_2}. Direct costs of flight delays in Europe amount to more than one billion euros \cite{cook_2011_european}, while in the US direct and indirect costs go beyond $40$ billion dollars \cite{joint_2008_your}. Delays directly affect airlines since they increase operational costs, but as an indirect factor they also bring associated reputational costs \cite{folkes_1987_field,mayer_2003_network}. Passengers, on the other hand, undergo a direct loss of time, which can be further exasperated by lost connections leading to missed business and leisure opportunities. Finally, efforts to recover delays airborne usually imply excess fuel consumption and larger CO$_2$ emissions. 

Given these challenges, it is important to characterize the sources of initial (primary) delays~\cite{rupp_2007_further,ahmadbeygi_2008_analysis} and devise mitigating strategies. Unfortunately this is a highly complex problem with the involvement of numerous (and sometimes unrelated) factors such as weather conditions, technical failures in the aircraft, aircraft rotations, crew and passenger connections, organizational issues in airports, and the mechanisms through which these delays are transferred and amplified in subsequent operations~\cite{bonnefoy_2007_scalability, beatty_1998_preliminary, wang_2003_flight}. Airline schedules typically include a buffer time to absorb these delays \cite{wu_2000_aircraft}, but there are frequent scenarios when this is not nearly enough. The Central Office for Delay Analysis of Eurocontrol (CODA) in Europe and its counterpart in the US, the Bureau of Transportation Statistics of the Department of Transportation, release monthly reports on flight performance including statistics on the major causes of delays and the airport affected. Based on these reports, studies have been conducted in single hubs such as Newark~\cite{allan_2001_analysis} as well as parts of the European \cite{cook_2007_european, jetzki_2009_propagation} and US \cite{mayer_2003_network,churchill_2007_examining,fleurquin_2014_trees} networks. 

\begin{figure}[t!]
\centering
\includegraphics[width=0.9\textwidth]{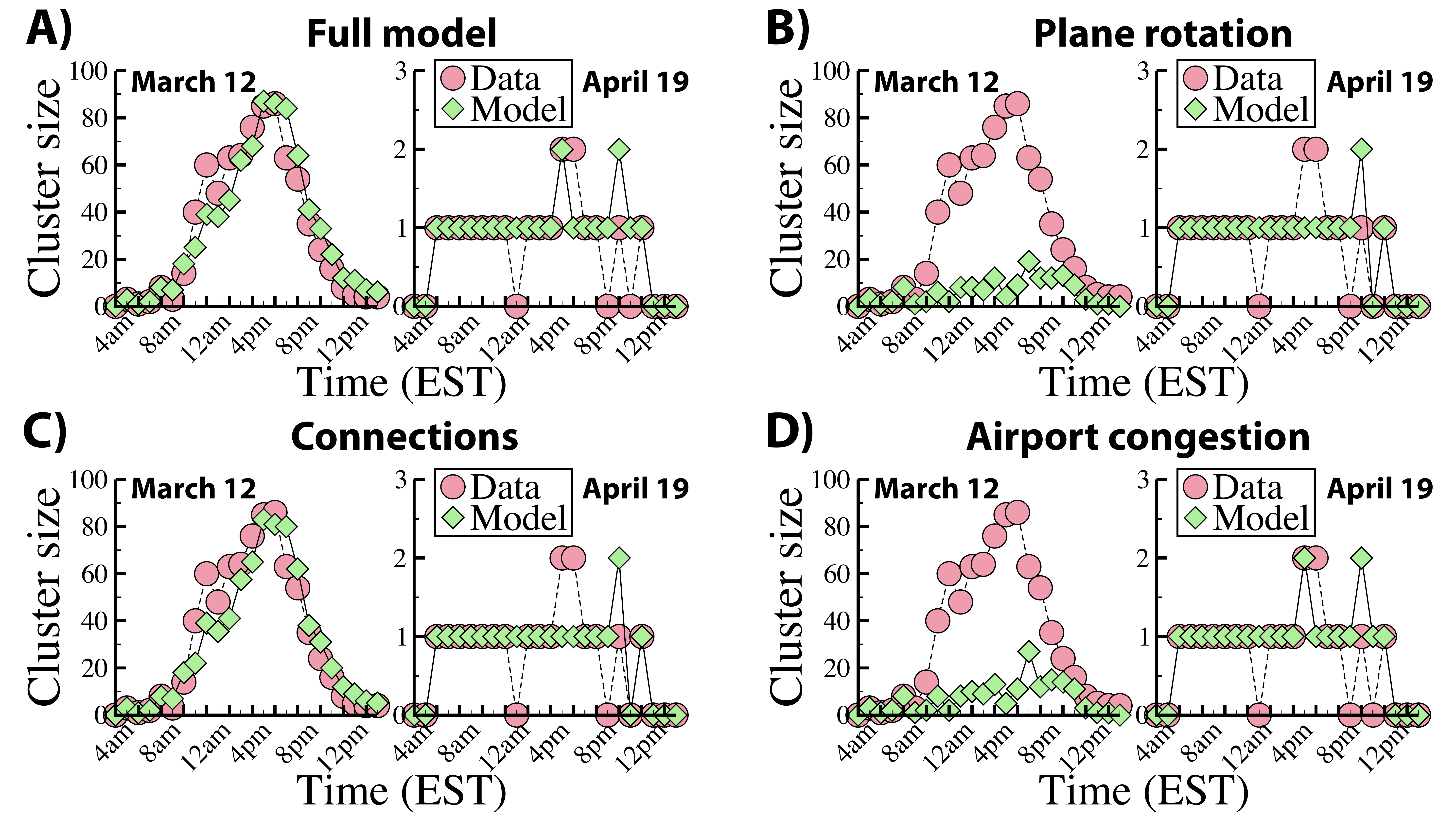}
\caption{Comparison between the congested cluster size as a function of time measured from empirical data. Congested airports are defined as those with a average delay per departing flight of over $29$ minutes in intervals of one hour. The congested cluster is obtained from the largest connected component of the network formed by congested airports connected with direct flights during the day considered. The model has been tested with all the ingredients working or only with some of them to check their importance in delay propagation. (a)-(d) refer to variants of the model taking into account all or some of the factors. Figure from~\cite{fleurquin_2013_systemic}. \label{fig:fleurquin_2013_systemic_4}}
\end{figure}

There have been primarily two approaches considered to model such effects. One class of models probe the network structure to search for weak points, analyzing jamming and congestion phenomena, especially trying to isolate those airports that cause cascades of delays that span a finite-fraction of the network~\cite{lacasa_2009_jamming,wuellner_2010_resilience,ezaki_2014_potential,lordan_2014_study}. This approach has also been extended to multiplex networks of airports and airlines also attempting to identify carriers that are most responsible for delays~\cite{cardillo_2013_modeling}. 
In addition to these more-stylized approaches, there exist other families of models whose objectives are to provide realistic predictions that can be implemented in real-time. Such models are primarily agent-based and incorporate fine-grained details of air operations while sacrificing analytical tractability. Indeed, this has been the main approach adopted in the area of transport engineering related to Air Transport Management (ATM). There are hundreds of contributions on this line mainly presented in sectorial conferences such as the USA/Europe ATM Seminar, the Air Transport Research Society (ATRS) Conference, ICRAT, the SESAR Innovation Days, etc, whose proceedings are accessible online as well as publications in journals such as Journal of Air Transport Management (JATM). The advances in this area cover topics such as the optimization of runway management, the effects of bad weather, sector congestion, how the advent of drone technology affects commercial air transportation, among many other minute details. While a comprehensive listing of all the findings would require a review by itself, we mention a few that are directly related to  delay propagation in networks. Micro-modeling of the connection between air-transport logistics and delays was first explored in~\cite{schaefer_2001_flight,rosenberg_2002_stochastic}. Later a similar framework was used to assess the economic impact of a major disruption in a European hub \cite{janic_2005_modeling}. Also in Europe, within the umbrella of the WPE of the Joint SESAR Undertaking, several projects as NEWO, TREE  and POEM have studied the modeling of delay propagation from different perspectives: the first two from the network and airline managers' and the latter from the passenger's point of view. 

In the same context, data-driven models have been developed to reproduce the delay propagation patterns in the US \cite{pyrgiotis_2013_modelling}. These models were validated against empirical performance data \cite{campanelli_2015_modelling}, showing an acceptable level of accuracy and precision in the prediction of delayed flights, identification of airports displaying major departure-delays, network-wide congestion and generalized bad weather conditions~\cite{fleurquin_2013_systemic}. The models are built with an agent-based approach applied at the level of aircraft, with delays appearing at any point in the operations due to technical, airport organizational or weather-related issues incorporated as initial conditions to the model. The airport management is simulated with a system of queues that depend on the specific network considered (US or Europe). One of the primary finding of such analysis is that the mechanism with the strongest potential to effect delay is passenger and crew connections (\figurename~\ref{fig:fleurquin_2013_systemic_4}). 

\subsubsection{Sea Networks}

\begin{figure}[t!]
\centering
\includegraphics[width=\textwidth]{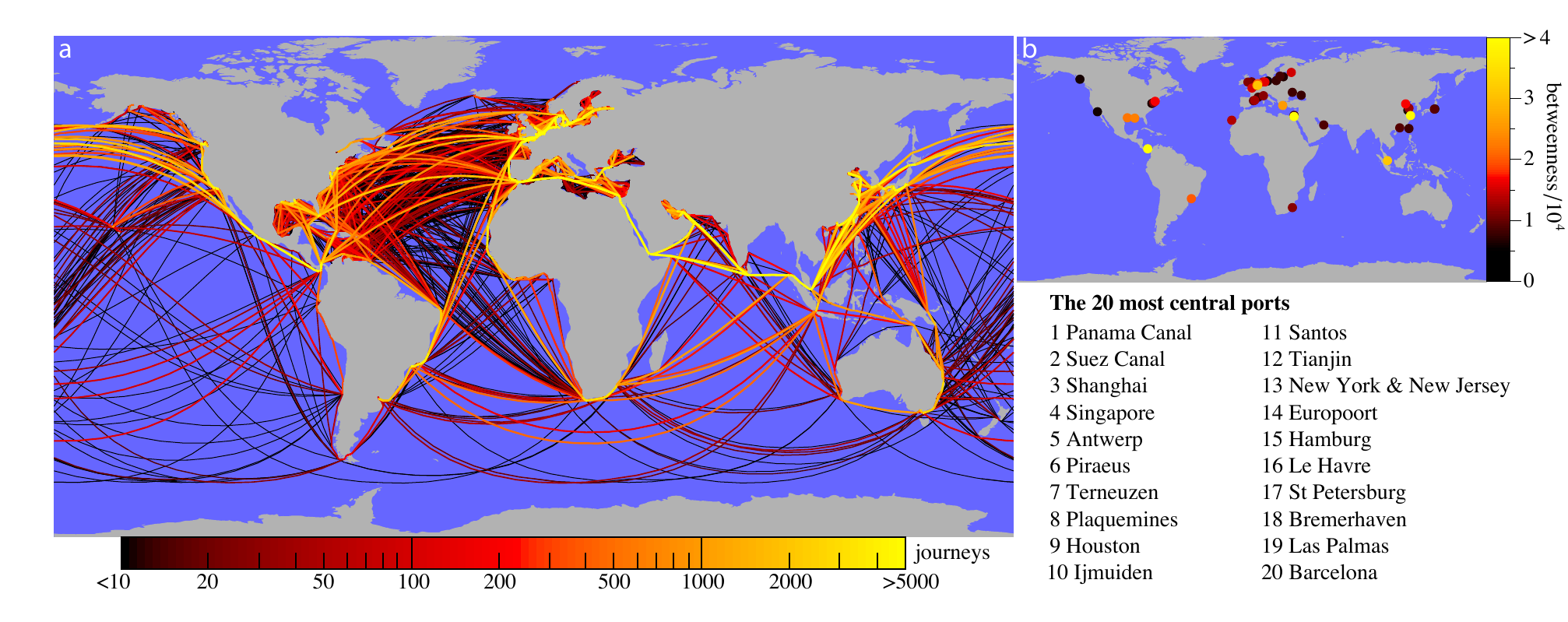}
\caption{(a) The global boat cargo network consisting of ports as nodes, and connections between them when boats navigate from one to the other. The links are colored according to the volume of traffic and their shape is constructed from the geodesic distance between ports. (b) A map of the top $50$ ports in terms of their betweenness centrality. Also listed are the top $20$. Figure from~\cite{kaluza_2010_complex}. \label{fig:kaluza_2010_complex_1}}
\end{figure}

Another example of long-range transportation are sea transportation networks, which have received comparatively less attention than their air traffic counterparts. While historically, sea transportation was the main mode of long-range human mobility, in recent times they mostly involve movement of cargo. Still, for sake of completeness we provide here a brief description on studies conducted on such networks. Typically such networks are studies as representing the ports as nodes, with links connecting them whenever major cargo liners travel from one to other~\cite{fremont_2007_global}. The links typically are given weights that correspond to the number of ships (trips) along the route. The resulting network spans the globe, connecting different continents as shown in~\figurename~\ref{fig:kaluza_2010_complex_1}. Similar to that seen in air transportation networks, statistical distributions of typical topological properties are heavy-tailed, with only a few number of ports accounting for a majority of global traffic, with the majority playing a peripheral role~\cite{kaluza_2010_complex}. The centrality of  ports was studied in \cite{hu_2009_empirical}, where it was found that Singapore, Antwerp, Bushan and Rotterdam top the list in terms of both traffic and load. A similar relationship between connectivity and traffic as in \equationname~\eqref{eq:wij_air} was noted with an exponent of $1.46 \pm 0.1$~\cite{kaluza_2010_complex}. Furthermore, the network can be segmented by the type of ship (containers, oil tankers, etc) in terms of a multiplex network framework \cite{ducruet_2013_network}.  A systematic analysis on the similarities between the world airport network and the boat network was performed in \cite{woolley-meza_2011_complexity}, including a study of the resilience of both networks. From a modeling perspective certain aspects of the flows can be reproduced with a gravity model (\sectionname~\ref{sec:gravity}), although there are some strong limitations. An important additional application of this network, besides the transport of goods, is the study of the invasion of species brought by carriers from different parts of the world \cite{keller_2010_linking,seebens_2013_risk}.

\subsection{Multi-scale}

\subsubsection{Intra urban mobility}

One of the more important and much studied applications of mobility is that seen within cities. Indeed, there are good reasons for this. One is just the availability of more data. 
With rapid urbanization, an increasing part of the global population is living in urban areas.
Large cities are early adopters of new technologies have large populations and consequently more mobile phone users and high densities of 
of phone antennas.
Consequently the spatial resolution of CDR data (Sec.~\ref{sec:cdr}) collected in cities is order of magnitude higher when compared to rural/non-urban areas. 
The same considerations
hold for the data generated from social networking applications, such as
Twitter or Foursquare. Indeed, the usage of these applications has been shown to have socially and spatially 
biases; they are more likely to be used by urban residents, educated
individuals, young people and middle age employed
adults~\cite{adnan_2014_social}. 

\begin{figure}[t!]
\centering
\includegraphics[width=0.7\textwidth]{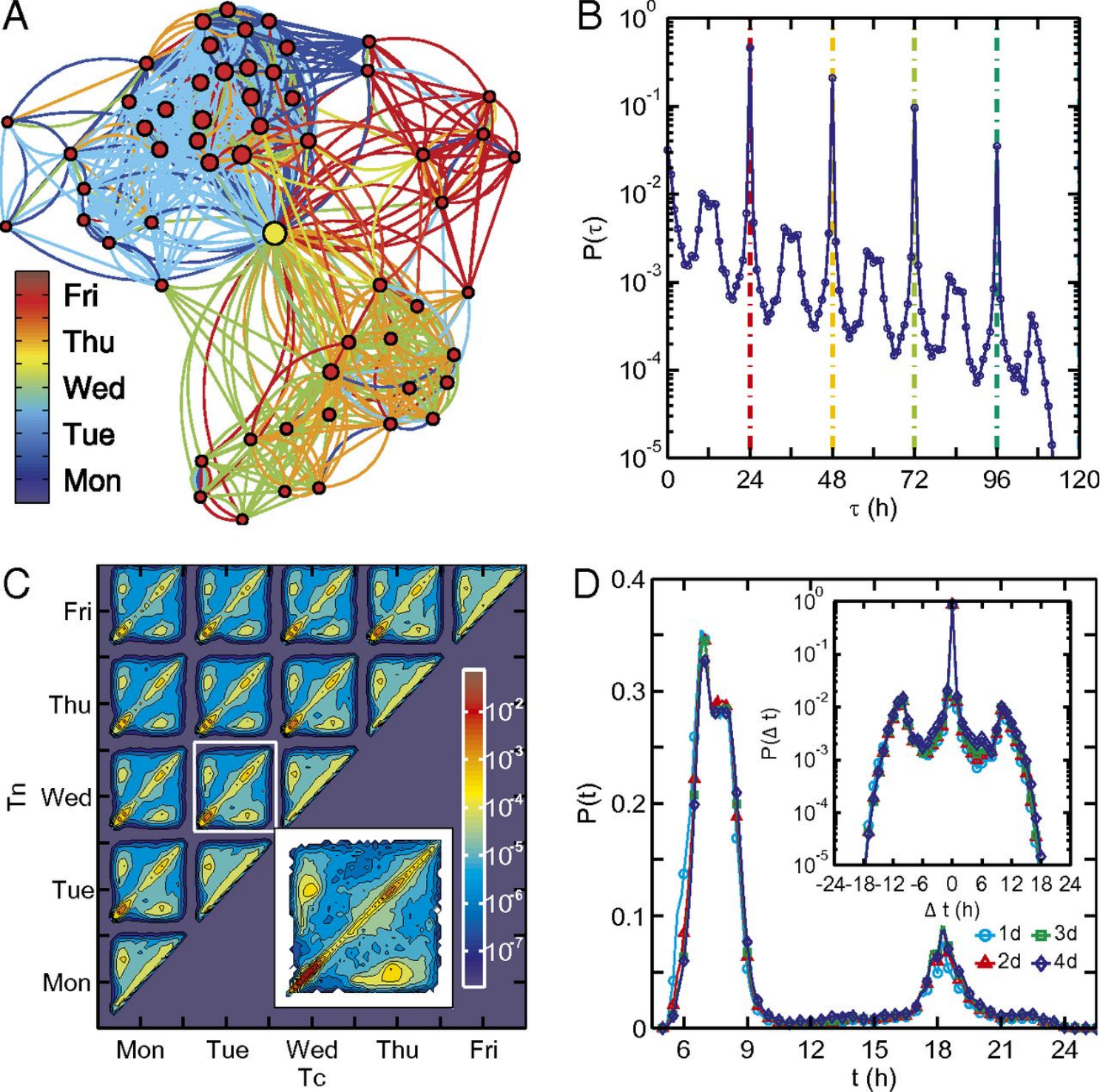}
\caption{Statistics of the co-location times for "familiar strangers". In (a), typical contact network. In (b), inter-encounter distribution times. In (c), joint probability distribution of inter-encounter times. In (d), distribution time between encounters for groups. Figure from~\cite{sun_2013_understanding}.}
\label{fig:sun_2013_understanding_2}
\end{figure}

The other of course is the unique set of challenges that abound from large agglomerations of people resulting in high population density. 
One of these is congestion that has an associated energy
cost as measured for example by C02 emissions. As a large part of human movement  occurs in individual
vehicles, accurate measurements of transportation flows are needed to infer how congestion affects carbon emissions. 
For example, it is known that CO2 emissions
mainly depend on the time spent traveling, and in places of high
density such as cities where congestion appears, traffic jams are
responsible for carbon emissions that
scale super-linearly with the population size \cite{louf_2013_modeling, louf_2014_how}. 
Yet another reason for studying intra urban mobility is related to that fact that cities
typically have  high levels of socioeconomic inequality among their residents, with residential segregation
that is amplified by other forms of segregation, perceived in mobility
practices and facilities that may differ depending on the neighborhood
of residence. Additionally, the multi-modal nature of transport networks
in urban areas (which include privately owned vehicles, buses, metros,
taxis, tramways, bike sharing systems, and pedestrians) gives
rise to specific questions related to the navigability of these
multi-layer networks. For example, the combination of the regularity of daily circadian
rhythms and smart transport card technology provides data
that allows for the study of intriguing phenomena, such as
\emph{familiar strangers}~\cite{sun_2013_understanding}, and can be used to uncover the
properties of ``hidden" temporal networks of the city, formed by
individuals that temporarily co-locate in the same spaces, but do not
know each other (see Fig. \ref{fig:sun_2013_understanding_2}). 

There is an important body of mixed
literature, quantitative and qualitative, on intra-urban mobility
that relies upon longitudinal travel surveys, started more
than 60 years ago. Covering it is beyond the scope of this review, and
possible starting points for those interested may be found in
~\cite{chapin_1974_human, hanson_2004_geography, golledge_1997_spatial}, which
focus on North-American metropolises. Ref. \cite{ratti_2006_mobile} was one of the first papers to explicitly discuss
the possibilities offered by individual digital footprints for urban
analysis. It underlined that such data could be used to gain
understanding of a number of phenomena associated to intra-urban
mobility. Following this, a series of papers highlighted possibilities of real-time visualization and
monitoring of displacements in cities~\cite{calabrese_2006_real,
reades_2007_cellular, calabrese_2011_estimating}.

The first studies that utilized metadata produced by handheld devices
carried by individuals focused on a single city, and essentially consisted of
visualizations and simple aggregated measures, providing information on the
density of individuals using a device in various areas of the city and at
various times of the day.  Metadata provided by mobile phone operators contains information on the nationality of the
phone number, which is also common on social networking
user profiles. Consequently one of the first uses of such databases of individual
geotagged data was to model the movements of foreign visitors in
cities. \cite{girardin_2008_digital} studied the location and mobility
patterns of tourists in Rome, and subsequent studies highlighted the
specific travel patterns of visitors, when compared to the movements
of city residents, notably in
Paris~\cite{olteanu_2012_le,fen-chong_2012_organisation}, Jakarta and
Singapore~\cite{chong_2015_not}.

\begin{figure}[t!]
\centering
\includegraphics[width=0.8\textwidth]{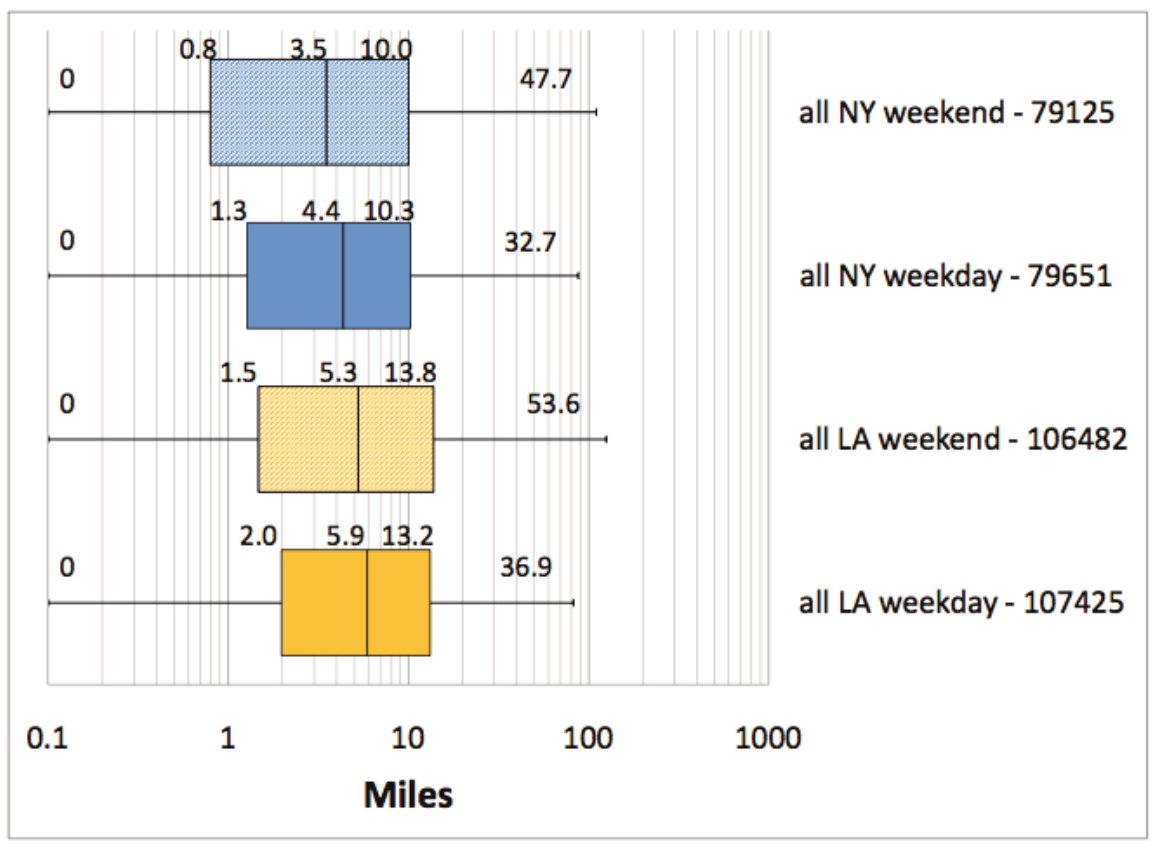}
\caption{Boxplot with the daily mobility ranges in Los Angeles and New York. Figure from~\cite{isaacman_2010_tale}. }
\label{fig:isaacman_2010_tale_4}
\end{figure}

Another popular case study are ones which
focus on the influence an area of residence has on intra-urban
mobility patterns. Most of the first intra-urban studies using ICT
data in these categories have focused on the organization of mobility
in one or two large cities. One of the very first in this category was
~\cite{isaacman_2010_tale}, which used mobile phone CDR data recorded
during a period $D$ of 62 consecutive days in New York and Los
Angeles. They compared aggregated mobility statistics; the
individuals' mean daily range -- defined for each individual as the mean
of the maximal distance traveled each day, $\sum_{d \in
D}{max_{ij}(d(i,j))}/|D|$ -- and the individuals' maximal daily range
-- defined as $max_D(max_{ij}(d(i,j)))$ (see Fig. \ref{fig:isaacman_2010_tale_4}). They showed that residents of
Los Angeles globally commute over further distances, but also
highlighted heterogeneities; some residents in Manhattan commute
farther than their counterparts in Los Angeles and
the neighborhood of residence has an influence on the average journey-to-work commuting
distance. 
Combining mobile phone data and odometer data (a distance measuring device for bikes and automobiles),
\cite{calabrese_2013_understanding} performed similar comparative
measures between the two data sources, highlighting the influence of a
number of built environmental factors on the travel distances
of Boston residents. \cite{desu_2016_effects} went further by
combining mobile phone data and census data which provided information on the
proportion of racial groups in each of the cities' census tracks. They
showed that residential segregation was amplified by a mobility
segregation in the urban areas of Boston and Los Angeles. 

\begin{figure}[t!]
\centering
\includegraphics[width=0.8\textwidth]{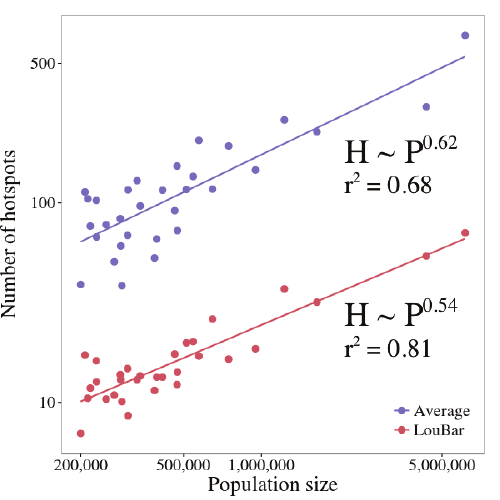}
\caption{Scatter plot with the number of hotspots detected as a function of the population of the cities. The colors of the curves and the points correspond to two different ways of defining the threshold marking an activity cell as hotspot. Figure from~\cite{louail_2014_mobile}. }
\label{fig:louail_2014_mobile_7}
\end{figure}

Focusing on
a single city, some authors have investigated the relation between
morphology and mobility patterns: \cite{roth_2011_structure} used smart cards
data collected over several weeks in the London subway to provide
measures linking the polycentric spatial structure of London and the
organization of mobility patterns in the city. Relying upon mobile phone data collected during 2 months in
31 Spanish urban areas, \cite{louail_2014_mobile} showed how the spatial
organization of hotspots and the shape of cities evolve during the
course of a typical weekday, as residents move inside the cities,
highlighting different typical city forms (see Figure \ref{fig:louail_2014_mobile_7}). Thanks to a large
Foursquare dataset that records GPS positioning data,
\cite{noulas_2012_tale} compared the distribution of 
jump lengths in 34 different cities worldwide. Fitting the
distribution in each city, they conclude that at the city scale, human
displacements cannot be properly fitted by a power-law distribution, a
result in agreement with several other case studies held for
particular cities, which found that the distribution of travel
distances inside an urban area is more properly fitted by an
exponential distribution $P(\Delta r) \propto exp(-\Delta
r/r_0)$. They propose an individual model derived from Stouffer's
intervening opportunities, that states that the important parameter
for explaining destination choice in cities is not the distance by
itself, but rather the density of possible alternative destinations in
a perimeter of distance $d$. Given a set $U$ of places in a city, the
probability of moving from place $u \in U$ to a place $v \in U$ is
formally defined as
\begin{equation}
P_r[u\rightarrow w] \propto \frac{1}{rank_u(v)^a},
\end{equation}
where $rank_u(v) \propto~\mid\{w:d(u,w)<d(u,v)\}\mid $, which can be seen as an urban counterpart to the radiation model
proposed in the same year by \cite{simini_2012_universal} (see section
\ref{sec:models}).

One of the most debated questions in the recent literature on human
mobility has been the shape of the statistical distributions of 
the distances traveled by individuals $P(\Delta r)$, and their
travel times at various spatial and temporal scales and in different
environments. Providing a neat characterization of these
distributions is of extreme importance, and a prerequisite to
modeling. Indeed, the distribution of travel distances $P(\Delta r)$
is a key ingredient that any reasonable model should be able to
reproduce and explain. \cite{gallotti_2015_stochastic} contains a very clear
recap of the efforts of the last decade on this question, and a table
summarizing the parameters values of the fitting distributions that have
been proposed. These distributions have been characterized thanks to
various sources of individual data (dollar bills, mobile phones,
private car GPS, taxis, longitudinal travel surveys, and geolocated
tweets, among others). In most cases, the fit of $P(\Delta r)$ is
consistent with the general form of a truncated power law distribution
\begin{equation}
P(\Delta r) \propto (\Delta r + r_0)^{-\beta} \exp\left(-\frac{\Delta r}{\kappa}\right),
\end{equation}
with $\kappa = 0$ corresponding to the case of a non-truncated
power-law, and $\beta = 0$ corresponding to an exponential
distribution of scale $\kappa$. The majority of papers propose truncated power law fits for $P(\Delta r)$, and
argue that the distribution of displacements follows an underlying L\'evy-flight process.


The studies compared in \cite{gallotti_2015_stochastic} provided very different
results for the parameters values $\beta$, $r_0$ and $\kappa$ of the
general form above, and consequently they proposed different
underlying processes for explaining the shape of this
distribution. To settle the discrepancies, a convincing modeling
framework, able to reproduce a number of stylized facts, and $P(\Delta
r)$ is required, and more research
and cross-checks should be performed. Data accessibility is always an issue in this type of research, so one workaround passes is repeating the same measures on many different datasets, to enable a systematic comparison of the
empirical results as done by~\cite{gallotti_2015_stochastic}.

Apart from this, some of the many aspects studied on intra-urban mobility relate to the typical distance scales of movement ~\cite{bazzani_2010_statistical,gallotti_2015_understanding}; measures of travel duration~\cite{gallotti_2015_understanding}; effects of population size on
the structure and dynamics of mobility~\cite{louf_2014_how,
louail_2015_uncovering}; the effect of city topology on the displacements of its
inhabitants (in terms of travel time, distance, and spatial
organization of flows)~\cite{kang_2012_intra} optimal `strategies' for navigating
multimodal transport networks~\cite{gallotti_2014_anatomy}, the efficiency of transport
itineraries~\cite{lima_2016_understanding}; the level of
complexity of transportation maps, which might exceed cognitive
limits, and therefore explain sub optimal navigation by city residents~\cite{gallotti_2016_lost}.

\subsubsection{Epidemic Spreading}

Perhaps one of the most important applications of studies on human mobility and transportation systems relates to  epidemic spreading. While epidemiology as a discipline has several hundred years of tradition, only recently has it been possible to model the global spreading patterns of diseases in a realistic and reasonably accurate way. Of course, every disease has its own peculiarities, and requires tailored modeling with the inclusion of the specific mechanisms associated with it. Consequently, the literature in this area is extensive, with dedicated monographs and a number of reviews ~\cite{diekmann_2000_mathematical,keeling_2008_modeling,tatem_2014_mapping,pastor-satorras_2015_epidemic}. Here we discuss how recent knowledge of human mobility has influenced the latest generation of epidemiological models.

\begin{figure}
\centering
\includegraphics[width=\textwidth]{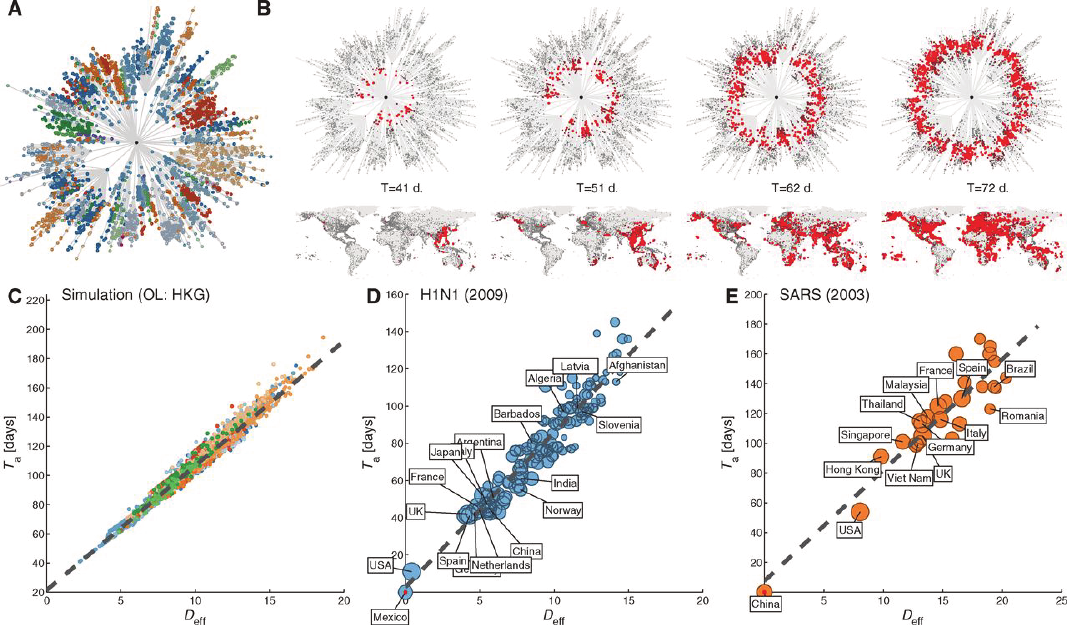}
\caption{In (a), Shortest (most-likely) disease propagation pathways for an epidemic starting in Hong Kong. $D_{eff}$ stands for the "radial distance from the disease origin as defined in \cite{brockmann_2013_hidden}. In (b), time evolution of a simulated pandemic starting in Hong Kong. In (c), epidemic arrival time as a function of $D_{eff}$ in the simulation. In (d) and (e), same analysis but with data for the 2009 H1N1 flu pandemic and for the 2003 SARS outbreak.  Figure from~\cite{brockmann_2013_hidden}. }
\label{fig:brockmann_2013_hidden_2}
\end{figure}

Whilst it has been known for some time that air transportation plays a major role in the global spreading of diseases \cite{rvachev_1985_mathematical,flahault_1992_method}, the implementation of air traffic data in epidemic models has only been possible in recent times with the advances in information technologies and computing resources over the past decade or so~\cite{hufnagel_2004_forecast,grais_2003_assessing,colizza_2006_role},
%
with the majority of studies focusing on the spread of influenza. The flu is a paradigmatic example of contact diseases that spreads annually cross the globe, and for which contagion is airborne with a patient clinical process that allows for relatively simple modeling. Indeed, improved surveillance systems have brought new insights on the temporal and spatial patterns of its propagation \cite{viboud_2006_synchrony,tizzoni_2012_real}. 

Of course, new modeling frameworks are not restricted to the flu but can be also extended to other contact diseases such as SARS \cite{bauch_2005_dynamically}, mosquito-driven infections lile malaria \cite{huang_2013_global} or yellow fever \cite{johansson_2012_crowd}, or even Ebola and other emerging health threats \cite{jones_2008_global,gomes_2014_assessing}. In many such cases, the first instance of the disease in a particular location occurs through the arrival of infected humans from the source location via air transportation. The world airport network (WAN) can thus be considered as the backbone through which new emerging diseases arrive where connections with larger flows of passengers marking the most likely pathways for propagation \cite{brockmann_2013_hidden} (see Fig. \ref{fig:brockmann_2013_hidden_2}). Indeed, within the framework of the WAN (as discussed in Sec.~\ref{sec:air}), it is possible to estimate the potential of every airport as a source for a pandemic \cite{lawyer_2016_measuring}. 

\begin{figure}[t!]
\centering
\includegraphics[width=\textwidth]{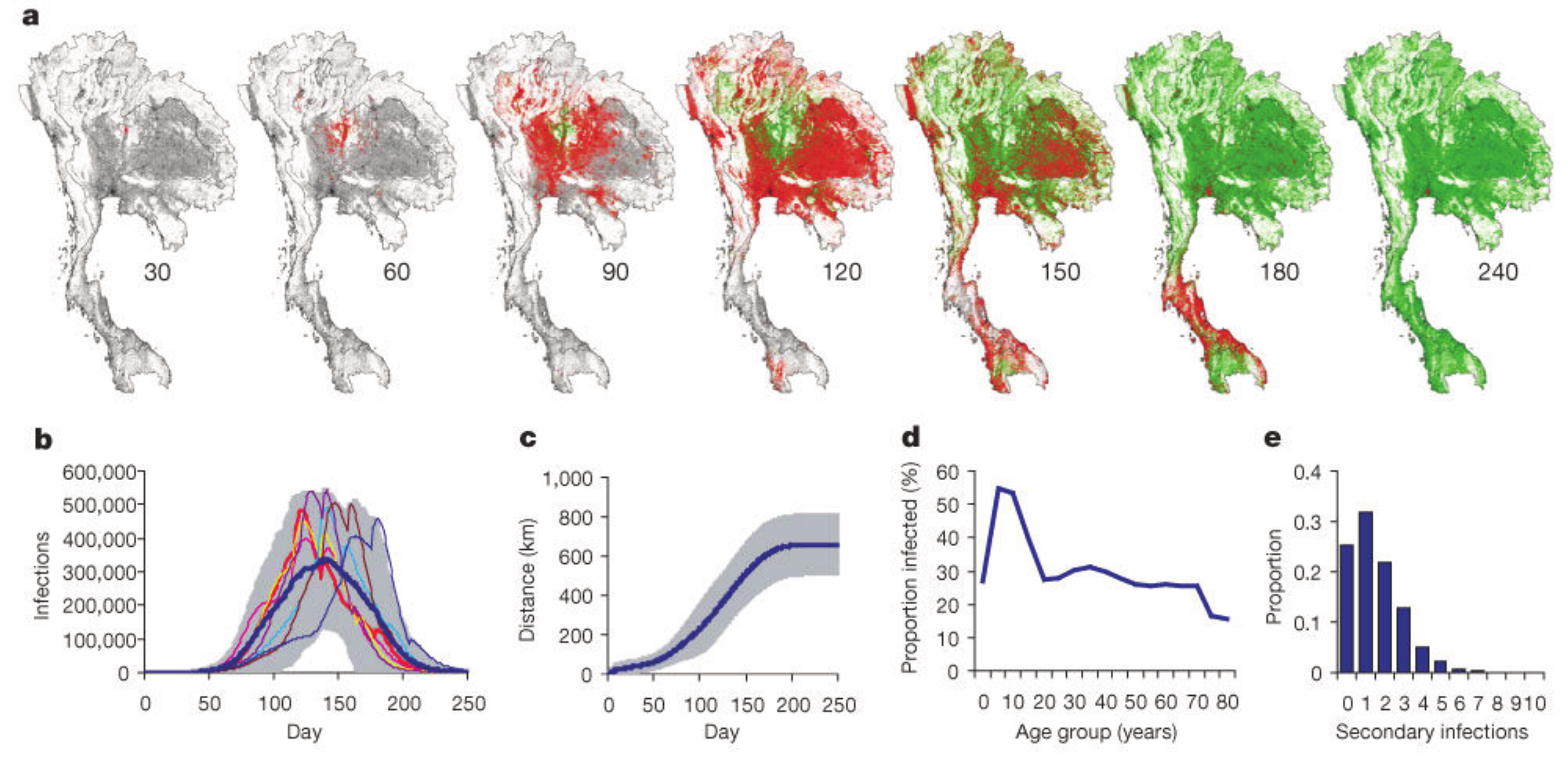}
\caption{In (a), time evolution of a flu epidemic with $R_0 = 1.5$ in the Southeast Asia. Infectious individuals are shown in red, while green ones are those already recovered or removed. In (b), daily incidence of the simulation on average in dark blue, several realizations in different colors and in gray the 95\% confidence interval. In (c), distance from the origin of the disease. In (d), the proportion of infected people by age group. In e, distribution of number of secondary cases produced per infectious individual in the early stages of the outbreak. Figure from~\cite{ferguson_2005_strategies}. }
\label{fig:ferguson_2005_strategies_2}
\end{figure}

Concerning realistic models, two main frameworks have been used in the literature: agent-based modeling and metapopulations. The difference between the two lies in the level of information needed. Agent-based models attempt to generate synthetic population mimicking at an individual (agent) level the realistics aspects related to disease propagation. A detailed knowledge is therefore required of features such as location of residential areas, businesses and schools, leisure activities, population divisions by  age, gender and sizes of households as well as individual mobility. These are typically sourced from census-like surveys and then a synthetic population is created assigning each individual to a household, a work/school place and a characteristic mobility profile. The earliest variants of such models were developed in the 1990's within the Los Alamos project \emph{Transportation Analysis Simulation System} or TRANSIMS. The objective of TRANSIMS was to reproduce via agent-based modeling the daily life of Portland, Oregon, including population, mobility, income levels and other details sourced from the US census \cite{simon_1999_simple}. This was later adapted to study the evolution of epidemics, called \emph{Epidemic Disease Simulation System} or EPISIMS, which was used to study different scenarios to prevent the propagation of smallpox \cite{eubank_2004_modelling}. Similar models were applied to study the propagation of a strain of avian flu, H5N1, in South East Asia, where for instance, epidemic spreading was simulated with the source centered in Thailand (see Fig. \ref{fig:ferguson_2005_strategies_2}), and including neighboring countries at a range of 100 kms containing about $80$ million inhabitants (agents)~\cite{ferguson_2005_strategies,longini_2005_containing}. This was primarily done to assess the effect of several intervention measures (distribution of antivirals and reducing the social contacts) to halt the progress of the disease. Similar models have been applied to even larger populations in the US and UK~\cite{ferguson_2006_strategies,germann_2006_mitigation,halloran_2008_modeling,ajelli_2008_impact}. The level of granularity present in such models, enables minute analysis  such as the effect of closing specific schools, employing different vaccination protocols (say by age) on disease propagation~\cite{germann_2006_mitigation,halloran_2008_modeling}. The arrival of the 2009 flu pandemic provided further impetus to such modeling approaches and agent-based models have been now developed for most of Western European countries \cite{ajelli_2010_comparing,ciofi_2008_mitigation,merler_2009_role,merler_2011_determinants} where a dazzling amount of scenarios have been exhaustively simulated including the risk of spread of a newly created virus strain from a lab following the case of the creation of a new H5N1 influenza strain \cite{merler_2013_containing}. One must note that the mobility aspect in such models relates to the movement of the individuals (agents) and is explicitly implemented both at long and short ranges involving multiple models of (simulated) transportation.

\begin{figure}[t!]
\centering
\includegraphics[width=\textwidth]{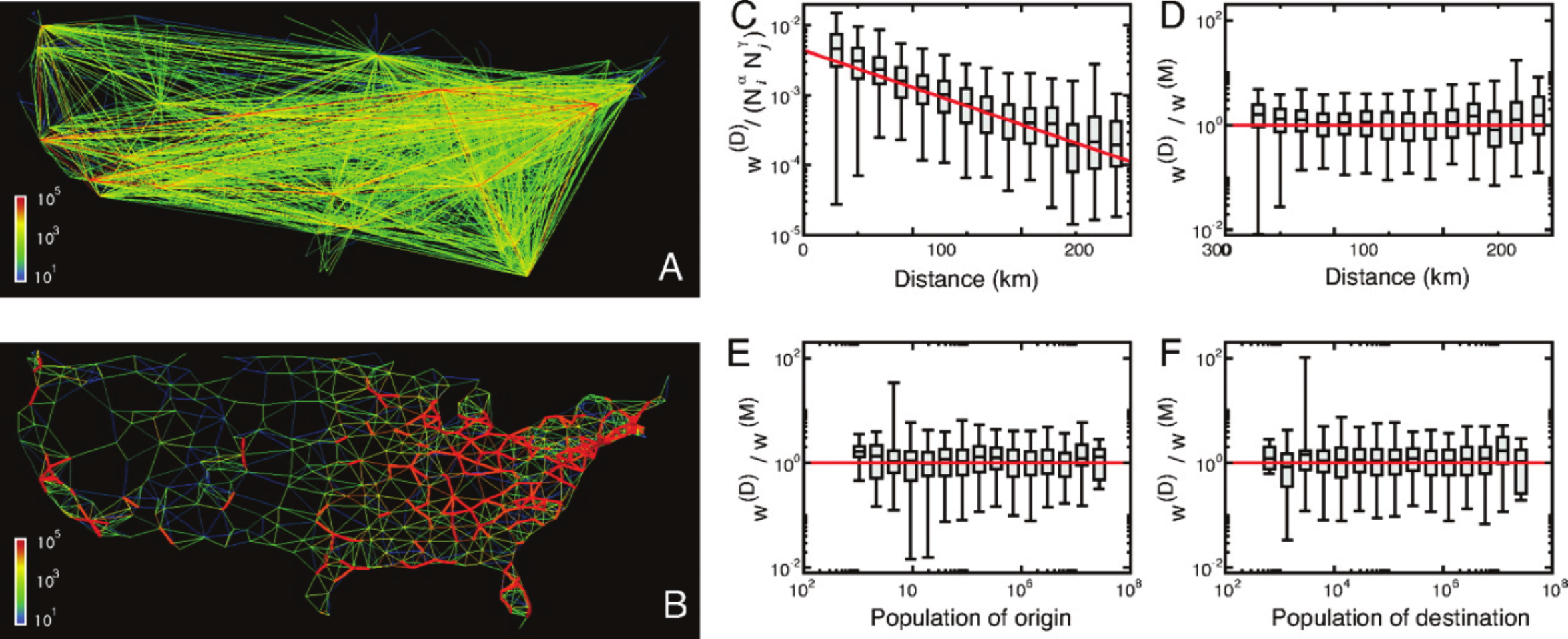}
\caption{In (a), air transport network zoomed in for the US. In (b), commuting network in the same area. In both cases, the link weights are plotted using a heatmap code. In (c)-(f), comparison between flows coming from commuting data $w^D$ and those produced with a gravity model $w^M$ with the aim of extrapolating short-range mobility in zones where the data was not available. Figure from~\cite{balcan_2009_multiscale}. }
\label{fig:balcan_2009_multiscale_1}
\end{figure}

While agent-based models provide the best approximation to reality, they are severely limited by the fact that they require highly detailed input information, the data for which in many cases is noisy, unreliable or just simply unavailable. Furthermore, given the sheer complexity of such models, an analytical treatment is nearly impossible. On the other hand, the metapopulation framework,  balances granularity with analytic tractability, and has easier to obtain data as input parameters~\cite{rvachev_1985_mathematical,sattenspiel_1995_structured,colizza_2007_invasion}. In this model, the population of interest is divided into subpopulations, within which contacts are modeled by a fully-mixed mean-field approach, whereas contacts between subpopulations occur through the (measured) mobility networks. One of the earliest such models is the so-called \emph{Global Epidemic and Mobility model} or GLEaM~\cite{balcan_2009_seasonal}. GLEaM is a global simulation platform that divides the globe in cells of $15\times15$ minutes of arc, which are grouped in basins of attraction around the major transportation hubs, which in most cases are airports. The  mobility is then simulated at two levels: long range air transport, which is calibrated from the world airport network, and short-range mobility obtained from either census commuting data or estimated using the gravity law~\cite{balcan_2010_modeling}. Figure \ref{fig:balcan_2009_multiscale_1} displays the different components of the model.

Most notably, GLEaM was used to study the propagation patterns of the 2009 H1N1 influenza pandemic in \emph{real time}. The pandemic started in a small population of Mexico early in that year \cite{fraser_2009_pandemic}, and its unfolding received attention of mass media media as well as the scientific community, which brought unprecedented details on the characteristics of the first imported cases in several countries. This information allowed for the calibration of the infectivity and the seasonality in GLEaM in early May 2009 \cite{balcan_2009_seasonal}. Once calibrated, the model predicted the disease propagation patterns: essentially, the arrival and prevalence peak times for the rest of the countries of the world, predictions that were later validated, after the full surveillance reports were available in 2012 \cite{tizzoni_2012_real}. Additionally, the model helped estimate the actual initial number of cases present in Mexico in the early stages of the disease~\cite{colizza_2009_estimate}. Ever since, the model has become an important tool for subsequent studies that takes its results as a baseline (see, for example~\cite{brockmann_2013_hidden,lawyer_2016_measuring}) and has been also used to assess the risk of spreading of new emerging health threats~\cite{goncalves_2013_human,gomes_2014_assessing,poletto_2014_assessing}.

The flexibility of the multilevel framework inherent in the model allows for the quantitative assessment of the differences between short-range and long-range traveling on the disease propagation. In particular it was determined that travel restrictions via decrease of the air traffic (or temporary quarantines) can at best delay the disease spread by a few days or weeks. Unless one totally shuts down the network, or severely restricts it (which of course carries a massive economic cost) global propagation is almost guaranteed~\cite{bajardi_2011_human,poletto_2014_assessing}. Newer versions of such models factor in facets of human behavior, such as the tendency of people to restrict contacts and travel when made aware of the existence of a dangerous disease~\cite{meloni_2011_modeling}, as well as heterogeneity in the population (age, gender, socioeconomic indicators) and its effect on travel patterns~\cite{apolloni_2014_metapopulation}. Possible future directions regarding these models include the possibility of feeding them with real-time ICT collected data on mobility \cite{jia_2012_empirical,wesolowski_2014_quantifying,tizzoni_2014_use,lenormand_2015_human} or hybrid frameworks that merge aspects of agent-based and metapopulation modeling~\cite{ajelli_2010_comparing}. 

\subsection{Virtual-Scale}

\subsubsection{Web (Online) Mobility}

Ostensibly, physical human mobility and virtual navigation are indeed activities of very distinct nature, yet there are a number of reasons why the latter can be cast in the context of the former. Indeed, many daily activities that occurred in specific physical locations, such as work, study, leisure, financial transactions, purchasing, are now increasingly migrating towards being conducted online. Moreover, the world wide web (www) has a humanitarian and social role, in the sense that it helps individuals and populations with limited mobility to have access (albeit virtually) to locations, people and information hitherto unreachable. For instance, the disabled \cite{harper_1999_towel,goble_2000_travails,ritchie_2003_promise} and  the elderly \cite{nimrod_2010_seniors,cotten_2013_impact,boxie_2007_using}, at an individual level, for whom the www represents a means of overcoming their physical limitations, or even entire populations, whose mobility is restricted for political or geographical reasons, for which the www is their only access to the outside world \cite{aouragh_2011_confined}. In this section we present an overview of the recent empirical evidence on the statistical similarities between virtual and physical navigation.

\begin{figure}[t!]
\centering
\includegraphics[width=0.9\linewidth]{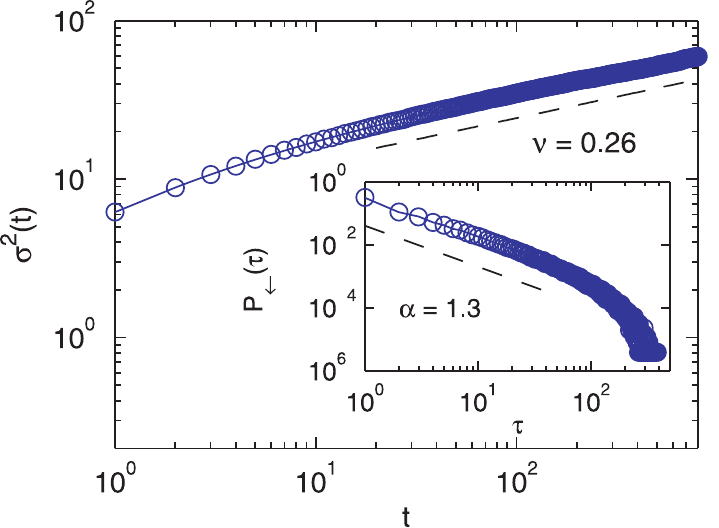}
\caption{Mobility activities within the \emph{Pardus} MMOG. The Mean Squared Displacement (main panel) suggests a subdiffusive process with exponent $\nu=0.26$,  whereas the return probabilities in number of discrete jumps $\tau$ (inset) behave as a power law with exponent $\alpha \approx 1.3$. (adapted from Szell \et \cite{szell_2012_understanding}) }
\label{fig:szell2012understandingfig5}
\end{figure}

In the context of this parallel between virtual and physical mobility, we differentiate two classes of activities that although related, have practical and functional distinctions, that is, mobility in virtual spaces (for example, World of Warcraft, Second Life and The Sims)  and navigation  in virtual \emph{spaceless} environments (for example, browsing the WWW or navigating through the interface of a device). The difference between these two classes of activities is given by the fact that in the first there is an explicit notion of space such that  displacements between different locations come with a temporal cost, often proportional to some  distance measure between them \cite{shen_2013_human}.

\begin{figure}[t!]
\centering
\includegraphics[width=0.9\linewidth]{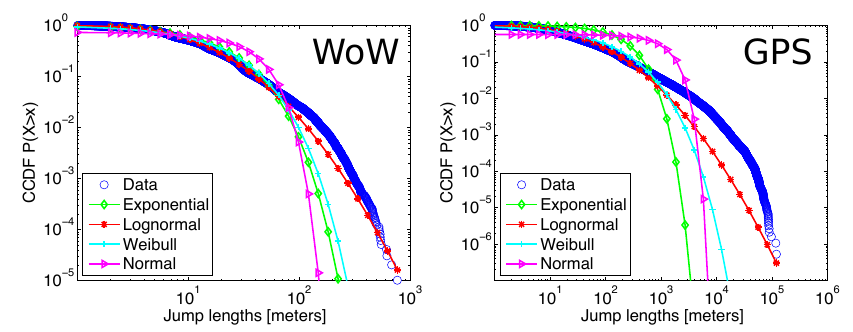}\\
\includegraphics[width=0.9\linewidth]{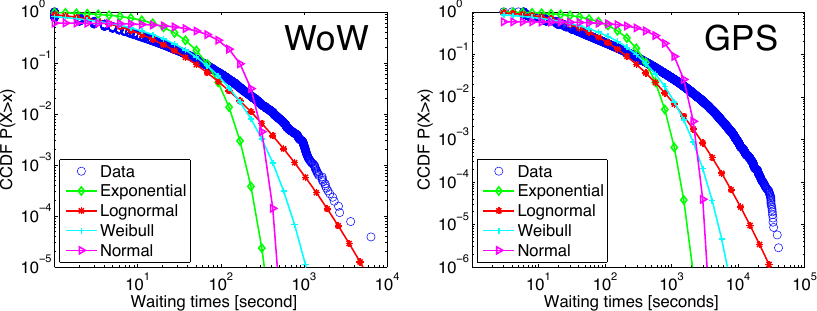}

\caption{Jump lengths (top) and waiting-times (bottom) distributions within the virtual environment of World of Warcraft (left) and physical mobility from GPS traces (right) (adapted from~\cite{shen_2013_human})}
\label{fig:shen2014characterizationfig2}
\end{figure}

An empirical analysis of the behaviors and trajectories in the virtual space of the online game \emph{Pardus} suggests a subdiffuse behavior in online mobility \cite{szell_2012_understanding}, similar to that observed in physical mobility \cite{song_2010_modelling,gonzalez_2008_understanding} ( \figurename \ref{fig:szell2012understandingfig5}). Additionally, it appears that the return time (in number of discrete steps) of users of the Pardus game follows a power-law distribution with an exponential cut-off (inset of \figurename \ref{fig:szell2012understandingfig5}).
These similarities between virtual and physical mobility were later empirically corroborated by other similar datasets. A comparative analyses between traces of World of Warcraft (WoW) players and GPS traces~\cite{shen_2014_characterization} revealed that the distributions of  jump lengths and waiting times are both approximated by lognormal distributions (\figurename \ref{fig:shen2014characterizationfig2}). 

\begin{figure}[t!]
\centering
\includegraphics[width=0.9\linewidth]{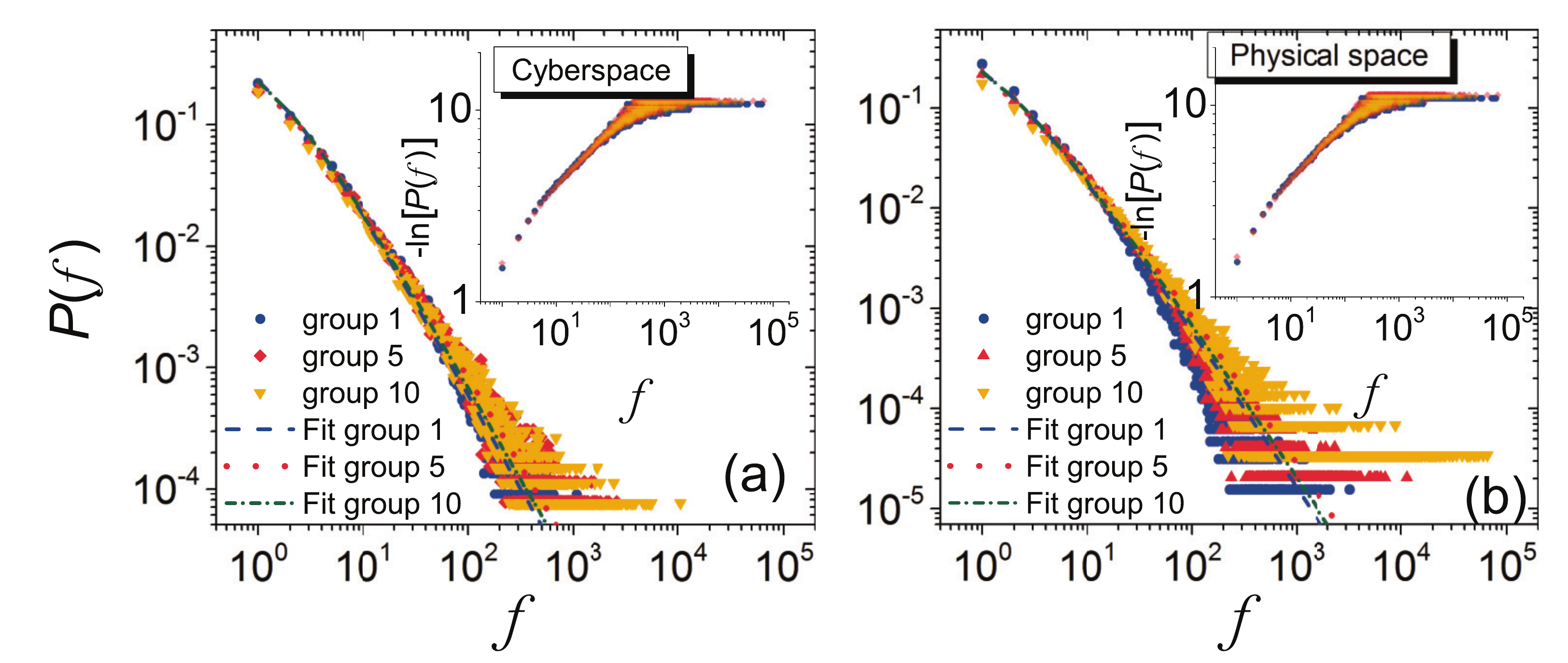}
\caption{Visitation frequencies distribution in online (a) and physical (b) spaces (adapted from Zhao \et \cite{zhao_2014_scaling}). }
\label{fig:zhao2014scalingfig2}
\end{figure}

Furthermore, the analyses of CDRs along with the Web activities of 20,000 mobile phone users~\cite{zhao_2014_scaling} suggest that visitation frequencies in both activities have a stretched exponential distribution (see \figurename \ref{fig:zhao2014scalingfig2}). 
However, Web traces from mobile phone users tend to overrepresent a particular set of activities such as fact finding, information gathering and communication \cite{cui_2008_exploring}. More recently, the analyses of complete Web browsing activities of 582 users, has revealed that it is possible to profile Web users based on their \emph{exploratory vs. exploitative} behaviors in their browsing patterns~\cite{barbosa_2016_returners}. More surprisingly, however, is the fact that the distribution of \emph{returners} and \emph{explorers} profiles are remarkably similar to the one observed in human mobility~\cite{pappalardo_2015_returners}.

The observed similarities between physical and virtual navigation is beginning to draw increased attention from the scientific community. Of course the analogy can oly be taken so far, as compared to physical movement, virtual activity is relatively recent (on an evolutionary time scale) and much remains to be understood. It is reasonable, however, to believe that online activity is driven by the same psychological mechanism as \emph{off-line} behaviors, now adapted to new environments or contexts. In fact, such intuition has increasingly found scientific support, both from a theoretical and empirical point of view, such as evidence of foraging behaviors in online shopping~\cite{hantula_2008_online} and information consumption on the Web~\cite{pirolli_1999_information,stenstrom_2008_online}, similar to that observed in human resource acquisition behaviors~\cite{stephens_2007_foraging}.

\section{Conclusions}
\label{sec:conclusion}
In this work we discussed the state-of-the art in the field of human mobility. After a short historical description of the field, we focused on current work starting with the type of data utilized in most research projects, followed by the description of metrics and models of individual mobility and population mobility. We illustrated a list of selected applications of human mobility modeling, demonstrating the insights the field can bring to the solution of real-world problems. 

It should be clear from the survey that the study, understanding, and modeling of human mobility is of primacy to many areas of society. At a deeper level the current research raises a fundamental question regarding human cognitive limitation. Among the many features described here, one in particular stands out: Why are we likely to return to frequently-visited locations or recently-visited locations? Does it relate to our ability to remember where we have been? Is it a innate characteristic of individuals or something that can be learned (\ie nature vs. nurture)? The same can be said about regularities whereby human behavior is divided between those that are explorers or returners in their mobility pattern \cite{pappalardo_2015_returners}. Indeed, this begs the interesting question, as to whether this observed division is an inherent property of society, say acting as a stabilizing influence? Indeed, this dichotomy  is also found in other context, such as social learning, where it has been suggested that a population cooperating to solve a complex task can attain a better performance if it is composed by a mixture of explorers, individuals who try new solutions, and exploiters, individuals that have less propensity to abandon the optimally extant solutions~ \cite{rendell_2011_cognitive}. 
More generally, the exploration vs exploitation dilemma is a fundamental issue in optimal decision-making and is analyzed in economics (multi-armed bandits) and computer science (reinforcement learning) and finds natural applications in marketing and advertisement, where recommendation algorithms determine whether people are likely to adopt familiar or unfamiliar products. All these problems require an understanding of human dynamics and many of these areas can possibly be studied using approaches inspired from research in human mobility. With the data we continue to collect as a society, and given the methodology developed to explain regularities in human mobility, we now have an additional set of mathematical tools to characterize the regularities in human activities more broadly.

A different structural dimension linked to human mobility relates to the invariant that humans are linked to each other forming social networks. Studies of social networks aim at grasping the multidimensional properties of the structures formed by entities (e.g., persons and organizations) and the ties connecting them together \cite{wasserman_1994_social}. In fact, the intersection between human mobility and social networks goes beyond the obvious analogy that humans are social and their movement may be related to them being social~\cite{wellman_1979_networks,tilly_1990_transplanted,liljeros_2001_web,tassier_2008_social,barabasi_2013_network}. The field of Social Network research is a huge enterprise in its own right and elucidating the connection between social networks and human movement is a nascent effort that should be explored in more depth. 

An interesting research frontier in human mobility in the near future will likely concern the shift from traditional vehicles to autonomous, self-driving, vehicles. The diffusion of autonomous vehicles will transform private and public transportation, with sweeping consequences for society, the economy and the environment.
How exactly will these transformations happen and how will mobility habits change in response to these new technologies?
How can such vehicles be controlled and how does one plan routes in order to optimize fuel and time consumption, reducing congestion and pollution?
Answers to these important questions, may well arise from a fuller understanding of human mobility and behavior at the individual and population (collective) levels. 

Finally, it should be clear from this survey that interdisciplinarity and cross-fertilization is a {\it condicio sine qua non} to the success of the field. Computer scientists, physicists, social scientists, environmentalists, engineers, government officials and many other professions need to work together to develop and implement solutions to many of the problems society faces today that could benefit from understanding mobility (and dynamics) such as crime, urban planning, energy consumption, social integration, to name but a few.

\section*{Acknowledgements}
HB, GG, RM and MT were partially supported by the US Army Research Office under Agreement Number W911NF-17-1-0127. FS is supported by EPSRC First Grant EP/P012906/1. JJR received funding from the Spanish Ministry of Economy, Industry and Competitiveness (MINEICO) and FEDER (EU) under the project ESOTECOS (FIS2015-63628-C2-2-R), and also from the European Commission H2020 project BigData4ATM (699260) under the SESAR Joint Undertaking.

\section{References}

\bibliographystyle{elsarticle-num} 
\bibliography{bib_harmonized}

\appendix
\renewcommand*{\thesection}{\Alph{section}}

\section*{Appendix}
\section{Modeling Frameworks and Algorithms}
\label{sec:comp}

As we advance our understanding of the fundamental mechanisms governing human traveling behaviors, mobility models become increasingly more refined. Therefore, implementing those models requires more specialized tools such as efficient graph libraries, advanced spatial capabilities (GIS) and statistical routines, to name a few. Luckily, libraries for such purposes are widely available for the most popular languages, allowing researchers to focus their efforts (and grant resources) onto more relevant parts of their contributions.
In this section we present several practical aspects related to the modeling and simulation of human traveling behaviors. Such details do not correspond necessarily to implementation details but rather to many of the basic building blocks from which we can construct all of the mobility models we presented in this survey.

\subsection{Modeling Frameworks}
\label{sec:frameworks}

Computational models are becoming increasingly important in contemporary research, with applications ranging widely in scope, technique, and goals. Strictly speaking, any simulation model implemented and executed on a computer can be considered a computational model, from numerical models of complex dynamical systems to recurrent neural networks in modern applications of artificial intelligence. 
What all computer models have in common is the fact that they are described in such a way that can be unambiguously translated to low-level code to run on a computer. Methodological and implementation details vary in several dimensions and must be carefully defined considering the characteristics of the system or process being modeled, the objectives of the experiments, and the expected results.

It is important to emphasize that a mobility model can be composed of smaller modeling blocks such as random number generators, optimizers, individual-level dynamics to name but a few. The actual macrolevel dynamics results therefore from the combination of these smaller processes. For instance, a CTRW is composed of waiting-time and jump-length distribution models. Often times, more complex mobility models incorporate ingredients such as social interactions \cite{grabowicz_2014_entangling,toole_2015_coupling} and environmental features \cite{eash_1984_development,erlander_1990_gravity,goncalves_1993_development,simini_2012_universal}. The soundness of a mobility model will thus rely heavily on the validity of those sub-components in representing the processes being modeled. Not surprisingly, many of the most groundbreaking models were based, at one level or another, on physical mechanisms and first principle formulations.
However, what all those mobility models have in common is the fact that they are based on Monte Carlo methods, in the sense that they rely on generating repeated random samples as an approximation for the behavior of the system (in this case, human displacements). 

Computational models can come in different flavors, varying in many aspects such as modeling approach and complexity, and are used to simulate human mobility to validate the analytical models by comparing data generated from the simulation with real world traces. Simulations can be grouped into three major modeling approaches: numerical, particle-based and agent-based models (ABM). Also, there are simulation of discrete events and  simulation based on system dynamics \cite{zeigler_2000_theory}. The former consists in conducting entities through processes or queues in order to analyze the general behavior exhibited by a system over time. In the latter, simulations are performed by analyzing the behavior of variables in systems of differential equations \cite{epstein_1997_nonlinear}. Historically, the first simulations were numerical in nature and although effective, in many cases the results of the simulations of the models did not have good adherence to reality, especially in cases where many abstractions are needed (i.e., intentional disregard of aspects of systems to simplify the model) \cite{lane_2000_you}.
The rise in popularity of ABM models, especially to model human mobility and social phenomenas, is due to the fact that they are more suitable when the simplicity of a Monte Carlo method is not sufficient or appropriate to represent more complex entities such as adaptive processes or local interactions. Also ABMs are especially useful when second-order dynamics or emergent phenomena are involved. In the context of human mobility, elements such as past-events memory, social networks and spatial capabilities make ABM the natural modeling approach.

Agent-based models are a class of computational models that emphasizes interactions of entities to assess their effects on a system. The key point is that simple rules and micro-level interactions lead to the emergence of complex macroscopic regularities and organizations that represent the observable outcomes to allow the study of the dynamics of individual behavior. 
Computational agents, according to Wooldridge and Jennings \cite{wooldridge_1995_intelligent}, normally have the following characteristics:
   \begin{description}
   \item [Autonomy] The agents' operation must take place without the direct control of its actions or internal states;
   \item [Social behavior] Agents must interact with other agents through some kind of standard communication;
   \item [Reactivity] Agents must be able to understand the environment and respond to it;
   \item [Initiative] In addition to reacting to the environment, agents are also able to take the initiative of an action, changing their behavior so as to fulfill a certain purpose.
 \end{description}
 
 Based on these aspects, simulations carried out with agent-based models allow for a deep investigation on the system from microlevel behaviors to macrolevel phenomena. Thus, agent-based models, depending on the kind of model to be designed, may show several benefits over other modeling techniques. According to Bonabeau \cite{bonabeau_2002_agent}, they may be summarized as follows: 
 \begin{description}
 	\item[Production of emergent phenomena] resulting from the interactions of individual entities. 
 	\item[Possibility to offer a natural way to describe systems] that are composed by interacting entities (e.g., stock markets).
 	\item[Be flexible] because agents can be modeled from different levels and approaches, endowing them with different kinds of behaviors, degrees of rationality  or learning skills.
 \end{description}

\subsection{Algorithms}
\label{sec:algo}

In this sections we provide the pseudo-code of the algorithms required to implements some of the mobility models. We also provide pseudo code to generate random numbers sampled from specific probability distributions. The general framework we use to model individual mobility is agent based as introduced in \sectionname\ref{sec:frameworks}.

\subsubsection{Individual Mobility}
 The ubiquity of cellular phones leads to the availability of large data sets capturing spatio-temporal traces of human trajectories. In the context of mobility modeling, the mathematical framework of \emph{Random Walks (RW)}, a class of stochastic random path models, has been extremely successful at reproducing some of the statistical properties of movements at a long-term limit. The mechanism at the core of random walks consists of a series of successive random events alternating displacements (i.e. movements) and changes of direction (\sectionname\ref{sec:rw}).
 
\begin{figure}[ht]
\centering
	\subfigure[5000 steps]{
	\includegraphics[width=0.48\columnwidth]{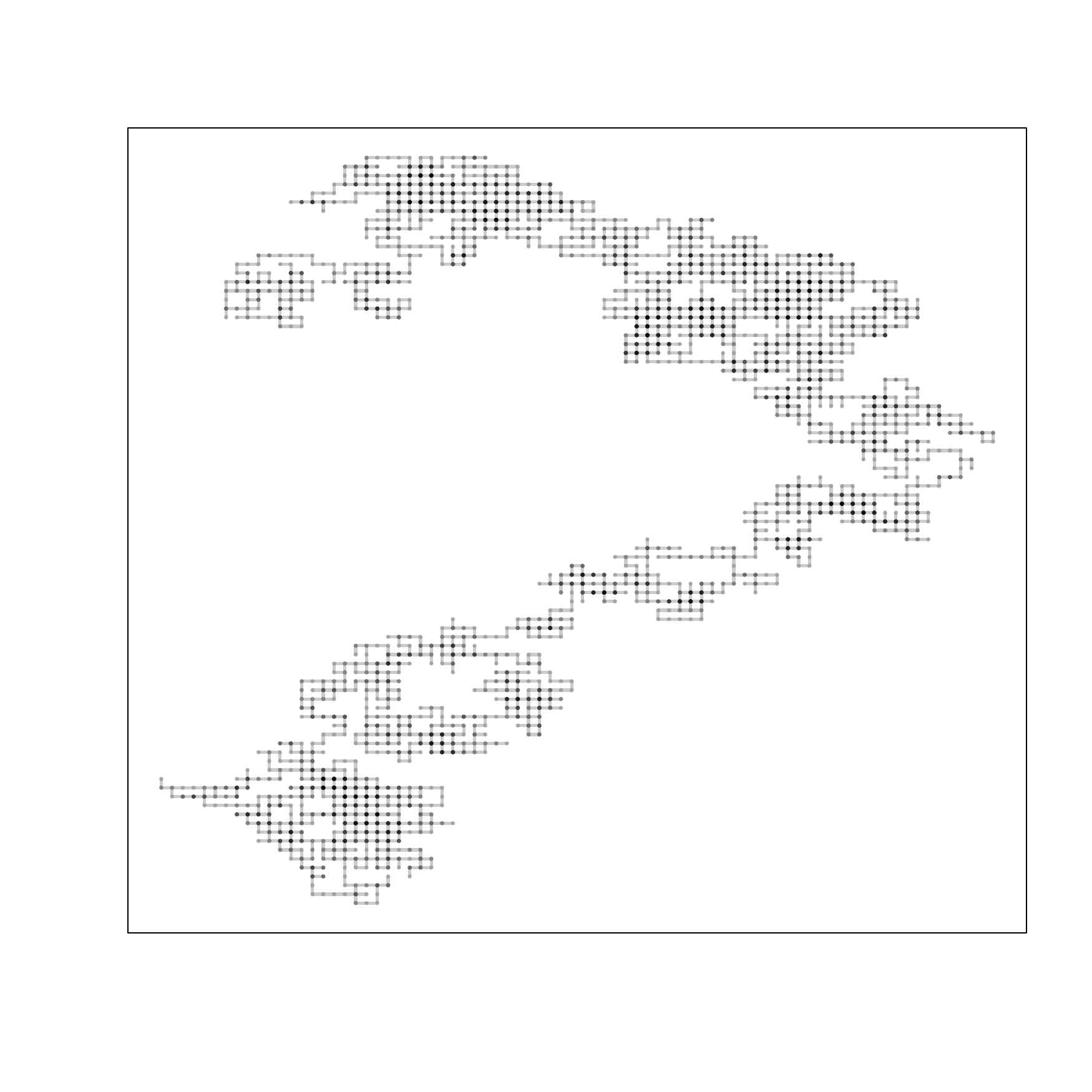}
	\label{subfig:rw5}
	}
	\subfigure[25000 steps]{
	\includegraphics[width=0.48\columnwidth]{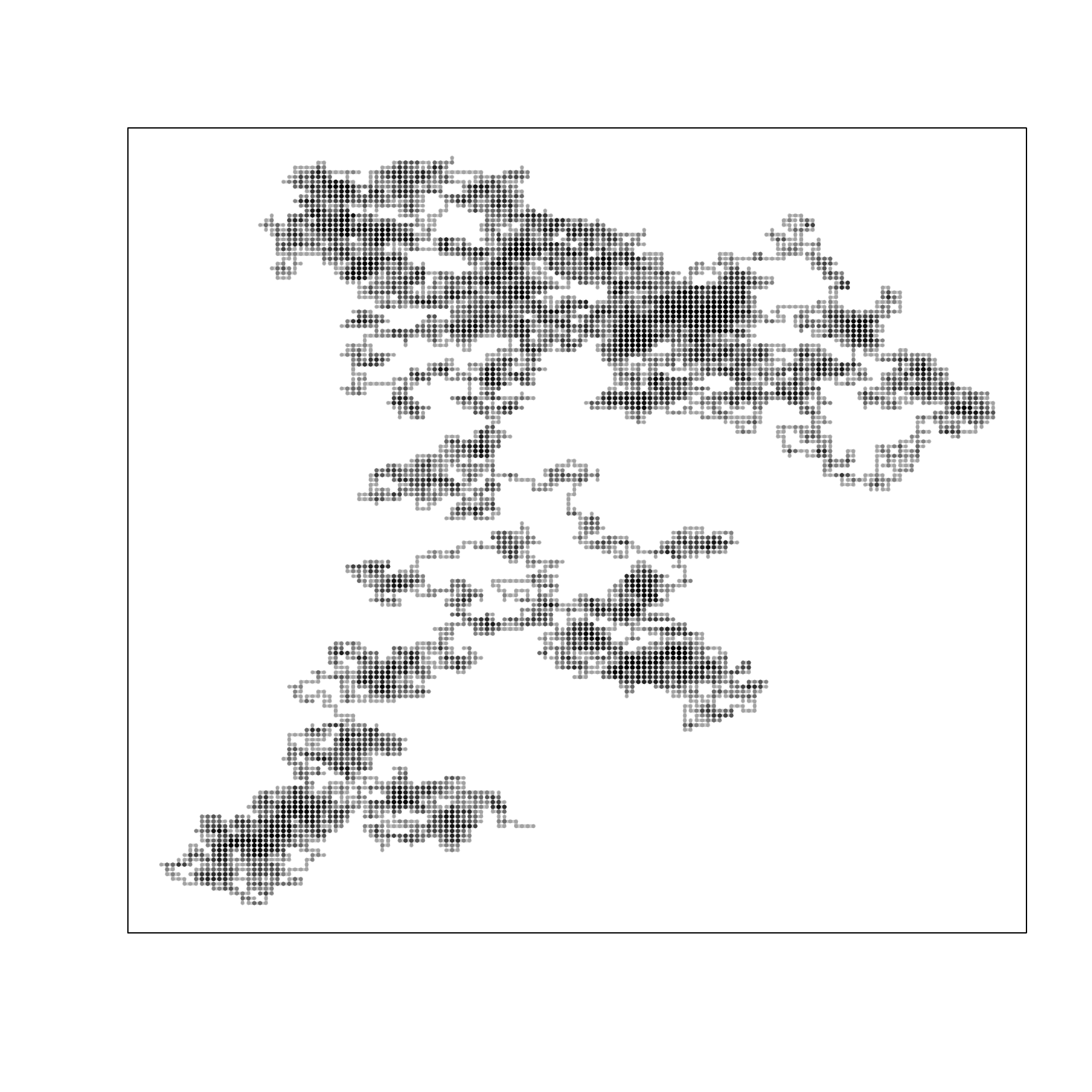}
	\label{subfig:rw25}
	}
\caption{Illustration of random walks in two dimensions. In \ref{subfig:rw5}, we have 5000 discrete steps while in \ref{subfig:rw25}, 25000 discrete steps. In the figure, more traversed lines are represented by darker lines.}
\label{fig:rw}
\end{figure}

\paragraph{Random Walk}
A simple random walk model has been described in \sectionname\ref{sec:brownian}.
A way to simulate a random walk on a computer is to let the agent move to a $D-$dimensional lattice. We can simulate a simple random walk of $N$ steps on a 2-dimensional lattice (see \figurename\ref{fig:rw}) by starting from the origin $(0,0)$ of the space, and then iteratively choose one of the four possible directions available (up, down, left, right), and move of one unit of length in that direction. The process stops when we iterated $N$ times. The procedure is illustrated in Algorithm \ref{alg:RW}. Note that It is crucial to choose each direction with the same chance or the random walk will be biased to some direction. For such purpose, modern computing devices and operating systems provides specific functions. In our case, we assume we have a function \texttt{RANDOM(min,max)} that is able to generate a uniformly distributed random number in the range $[\texttt{min},\texttt{max})$. Such function is trivial enough to be implemented on any platform or any programming language.

\begin{algorithm}
\caption{Simple 2D Random Walk}
\label{alg:RW}
\begin{algorithmic}[1]
\Procedure{Random\textendash Walk}{N}
\State $x \gets 0$
\State $y \gets 0$
\State $n \gets 0$
	\While{$n < N$ }
		\State $r \gets \text{RANDOM(0,4)}$
		\If {$r < 1$} \Comment Up
			\State $y \gets y + 1$
		\ElsIf {$r < 2$} \Comment Down
			\State $y \gets y - 1$
		\ElsIf {$r < 3 $} \Comment Right
			\State $x \gets x + 1$
		\ElsIf {$r < 4$} \Comment Left
			\State $x \gets x - 1$
		\EndIf
	\State $n \gets n+1$
	\EndWhile
\EndProcedure
\end{algorithmic}
\end{algorithm}

\begin{figure}[htbp]
\centering
	\subfigure[5000 steps]{
	\includegraphics[width=0.48\textwidth]{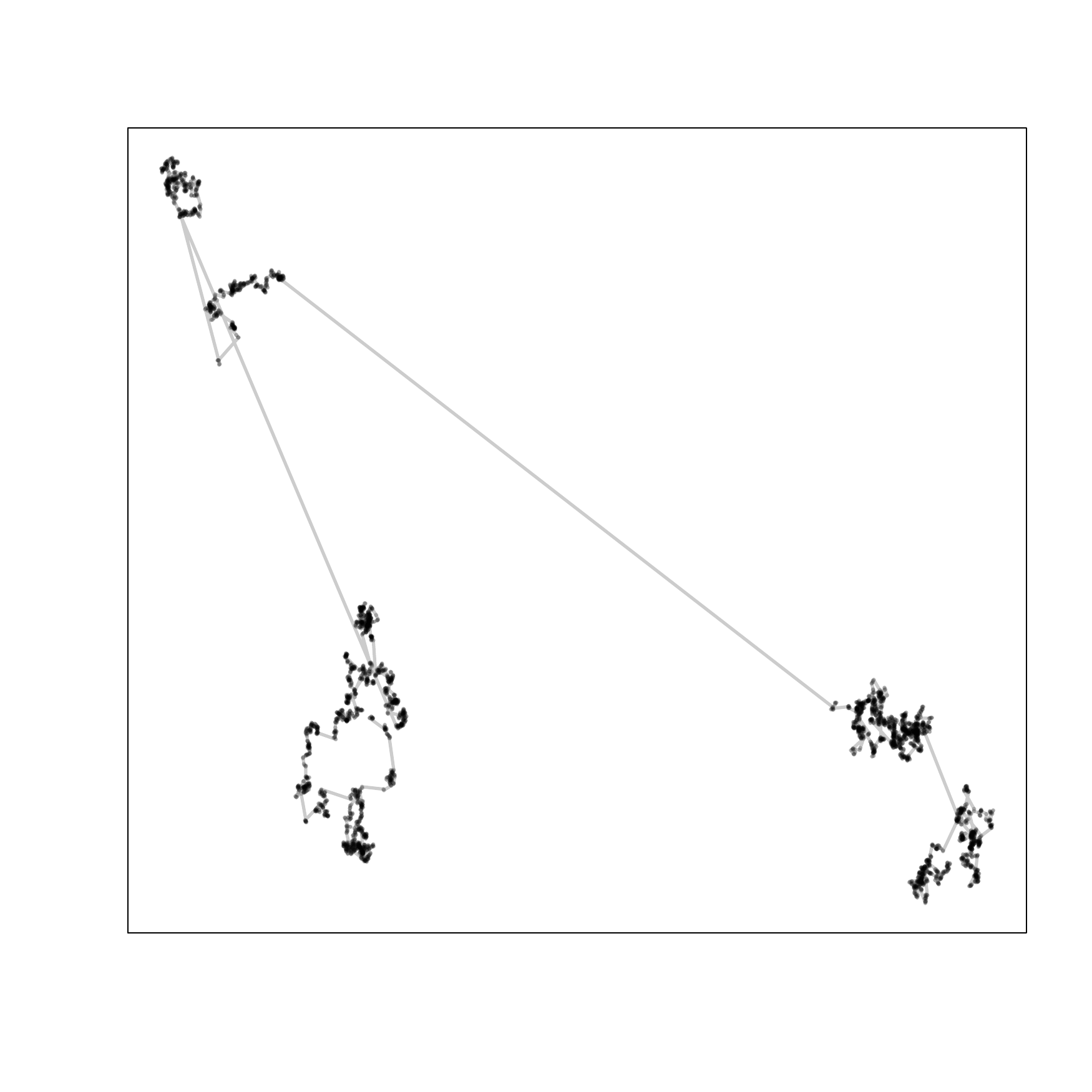}
	\label{subfig:lf5}
	}
	\subfigure[25000 steps]{
	\includegraphics[width=0.48\textwidth]{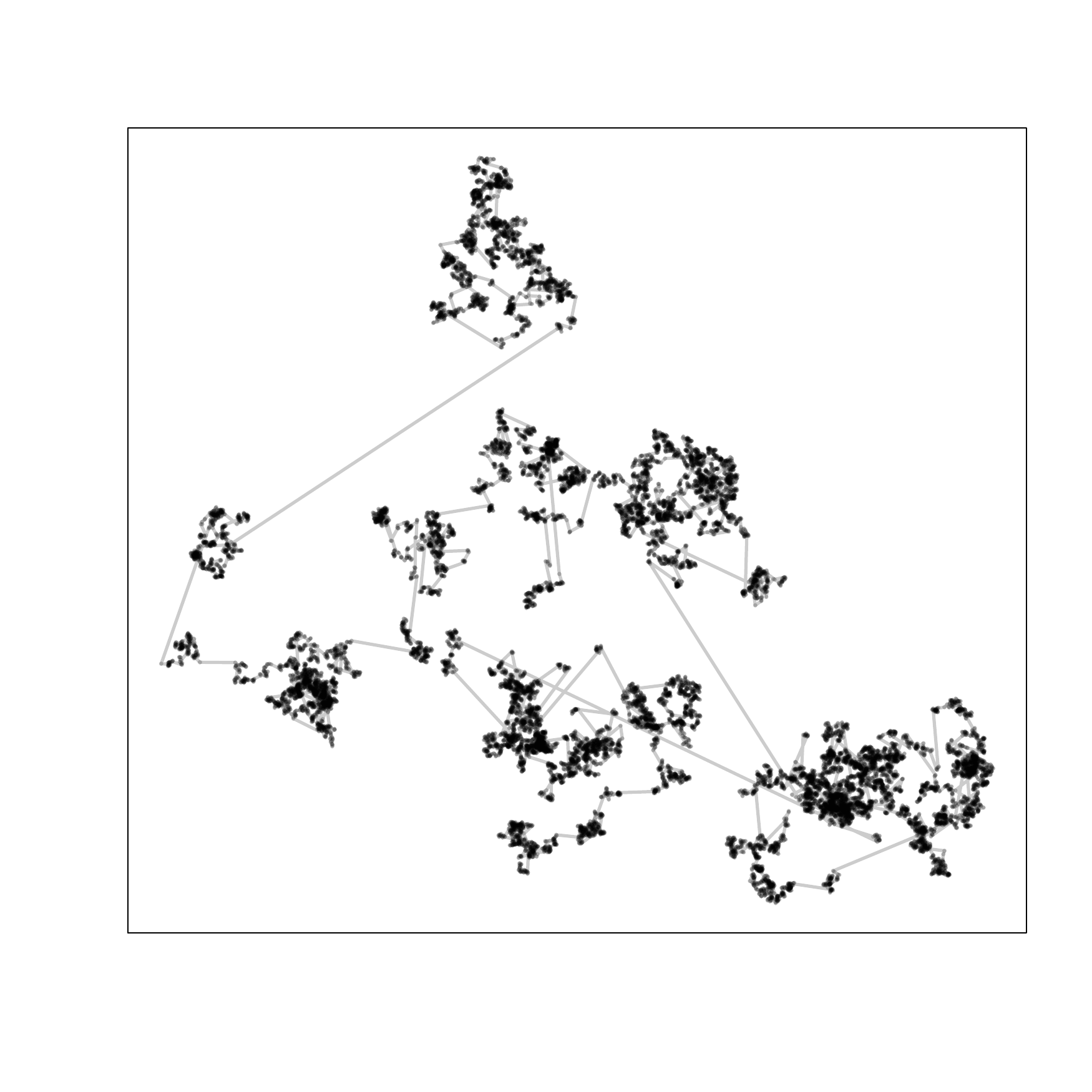}
	\label{subfig:lf25}
	}
\caption{Sample trajectories of L\'evy flights in two dimensions for $\beta = 0.6$ and step length $\Delta r \geq 1$. In \ref{subfig:lf5}, 5000 steps while in \ref{subfig:lf25}, 25000 discrete steps.}
\label{fig:lf}
\end{figure}

\paragraph{L\'evy Flight}
L\'evy Flight have been described in details in \sectionname\ref{sec:levy}. The main difference with the simple random walk is the type of distribution used to generate the jump length at each step, which follows a heavy-tailed distribution (e.g., power law). Therefore the L\'evy flight is a succession of short jumps and some occasional long jumps (\figurename\ref{fig:lf}).
When trying to implement a L\'evy flight on a computer there are several issue to take in account:
\begin{itemize}
	\item If there are RNG functions available to draw a number from a distribution, usually the power law distribution is not supported.
	\item For a power law distribution we need to specify $\alpha$ and $x_{\text{min}}$ values.
	\item Continuous power law distribution is easy to deal with, but the discretization process may introduce an error.
\end{itemize}
To generate a random number drawn from a power law distribution we can use the transformation method \cite{william_1997_numerical}. Usually, on a computer we have a source of random reals $r$ uniformly distributed in the interval $[0,1)$ that are provided by a pseudo-random number generator. In this case we can obtain a random number $x$ drawn from a continuous probability density $p(x)$ with a cumulative distribution function $P$ by obtaining the functional inverse of the CDF \cite{clauset_2009_power}:
\begin{equation}
x = P^{-1}(1-r)
\end{equation}
where 
\begin{equation}
P(x) = \left(\frac{x}{x_{\text{min}}}\right)^{-\alpha+1}
\end{equation}
for a continuous power law, and
\begin{equation}
P(x) = \frac{\zeta(\alpha,x)}{\zeta(\alpha,x_{\text{min}})}
\end{equation}
for a discrete power law. Therefore we obtain
\begin{equation}
x = x_{\text{min}}(1 - r)^{-1/(\alpha-1)}
\end{equation}
for the continuous case, which is straightforward to implement in any programming languages. However, the discrete form of $P(x)$ cannot be inverted in closed form, therefore we solve $P(x)=\sum_{x'=x}^{+\infty} p(x') = 1-r $ instead by a combination of doubling up starting from $x=x_{\text{min}}$, which ensure  that $1-r \in [P(x),P(2x))$, and binary search (see Algorithm \ref{alg:power}). Note that we need to evaluate the generalized Riemann zeta function $\zeta$ to compute $P(x)$ which might be computationally expensive, and it is usually done by using third party scientific libraries. When speed is important, techniques like dynamic programming can greatly improve algorithm speed by storing the result of previous evaluations that therefore need not to be computed again. Furthermore, if accuracy is not of great importance, which is the case for most applications where $x_{\text{min}} > 5$, we can approximate the discrete power law to a continuous one. We first generate power law distributed numbers $y \ge x_{\text{min}} - \frac{1}{2}$ and then round it off to the nearest integer $x = \big\lfloor y + \frac{1}{2} \big\rfloor$ \cite{clauset_2009_power}:
\begin{equation}
	x = \left\lfloor \left( x_{\text{min}} - \frac{1}{2} \right) \Bigg( 1 - r \Bigg)^{-1/(\alpha - 1)} + \frac{1}{2} \right\rfloor
\end{equation}
It is straightforward to notice that the approximation is bigger for smaller value of $x$ and therefore it is maximum when $x = x_{\text{min}}$. The error between the approximated distribution an the discrete one when $x = x_{\text{min}}$ can be computed as:
\begin{equation}
\Delta p = 1 - \left ( \frac{ x_{\text{min}} + \frac{1}{2}}{ x_{\text{min}} - \frac{1}{2}} \right)^{-\alpha+1} - \frac{x_{\text{min}}}{\zeta(\alpha,x_{\text{min}})}
\end{equation}
\begin{algorithm}
\caption{Discrete Power Law Random Number Generator}
\label{alg:power}
\begin{algorithmic}[1]
\Procedure{Power\textendash Law}{$\alpha, x_{\text{min}}$}
\State $r \gets \text{RNG}(0,1)$
\State $x_1 \gets x_{\text{min}}$
\State $x_2 \gets x_{\text{min}}$

\Repeat \Comment Determine Boundaries of Solution
	\State $x_1 \gets x_2$
	\State $x_2 \gets 2 \times x_1$
\Until {$P(x_2) < 1 - r$}

\Repeat \Comment Binary Search of Solution
	\State $mid \gets  x_1 +  (x_2 - x_1) / 2 $
	\If {$ P(mid) < 1 - r$}
		\State $x_1 \gets mid$
	\Else
		\State $x_2 \gets mid$
	\EndIf
\Until {$x_2 - \lfloor x_1 \rfloor \ge 1$}
\State \Return {$\lfloor x_1 \rfloor$}
\EndProcedure
\end{algorithmic}
\end{algorithm}
Once we are able to generate power law distributed random numbers, we can modify the algorithm of the simple random walk to implement a L\'evy walk by choosing a direction randomly and uniformly, an angle $\theta$, and the length $r$ of the jump (in this case from a continuous power law distribution). In this case we move away from a simple lattice to let the coordinates to be in the Euclidean plane $\mathbb{R}^2$ (Algorithm \ref{alg:Levy}). 

\begin{algorithm}
\caption{L\'evy Walk}
\label{alg:Levy}
\begin{algorithmic}[1]
\Procedure{Levy\textendash Walk}{N}
\State $x \gets 0$
\State $y \gets 0$
\State $n \gets 0$
	\While{$n < N$ }
		\State $\theta \gets \text{RANDOM}(0,360)$
		\State $r \gets \text{POWER-LAW}(\alpha, x_{\text{min}})$
		\State $x \gets x + r \cos \theta$
		\State $y \gets y + r \sin \theta$
		\State $n \gets n+1$
	\EndWhile
\EndProcedure
\end{algorithmic}
\end{algorithm}

\paragraph{Continuos-Time Random Walk (CTRW)}
A limitation of the RW models introduced so far is that time intervals between two successive steps are constant (\sectionname\ref{sec:ctrw}).
In order to simulate a CTRW we can modify Algorithm \ref{alg:Levy} to include the wait time $\tau$ between jumps (Algorithm \ref{alg:CTRW}). Notice that the wait time and jump length need not to be drawn from power law distributions, but CTRW with power laws has been shown to be useful to model human mobility \cite{brockmann_2006_scaling}. Moreover, the power laws can have different parameters.

\begin{algorithm}
\caption{Continuos-Time Random Walk}
\label{alg:CTRW}
\begin{algorithmic}[1]
\Procedure{CTRW}{N}
\State $x \gets 0$
\State $y \gets 0$
\State $n \gets 0$
	\While{$n < N$ }
		\State $\theta \gets \text{RANDOM}(0,360)$
		\State $r \gets \text{POWER-LAW}(\alpha, x^{r}_{\text{min}})$
		\State $x \gets x + r \cos \theta$
		\State $y \gets y + r \sin \theta$
        \State $\tau \gets \text{POWER-LAW}(\beta, x^{\tau}_{\text{min}})$ \Comment Wait Phase
		\While{$\tau > 0$} 
			\If {$\tau >= 1$}
				\State $\text{WAIT}(1)$
				\State $\tau \gets \tau - 1$
			\Else
				\State $\text{WAIT}(\tau)$
			\EndIf
		\EndWhile
		\State $n \gets n+1$
	\EndWhile
\EndProcedure
\end{algorithmic}
\end{algorithm}

\paragraph{Exploration-Preferential Return (EPR)}

\begin{figure}[htbp]
\centering
	\subfigure[Individual Mobility Model]{
	\includegraphics[width=0.48\textwidth]{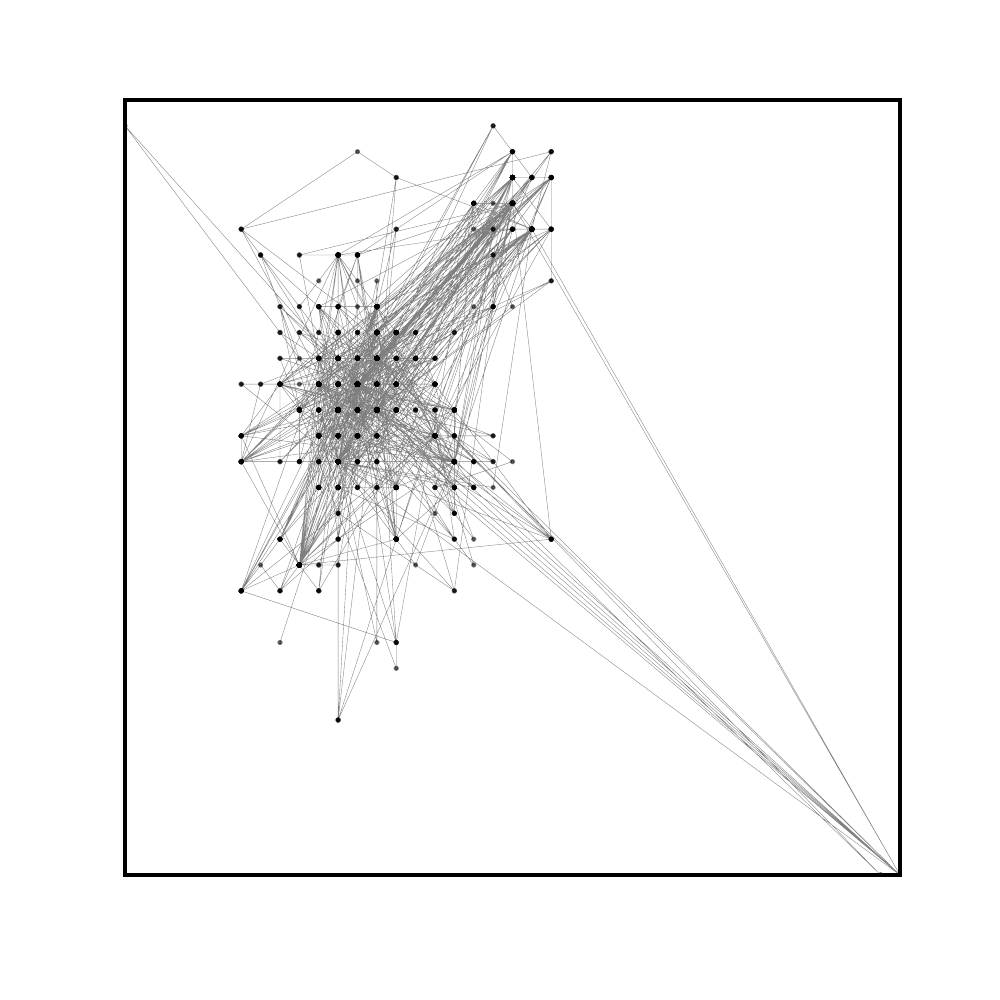}
	\label{subfig:epr}
	}
	\subfigure[Recency Model]{
	\includegraphics[width=0.48\textwidth]{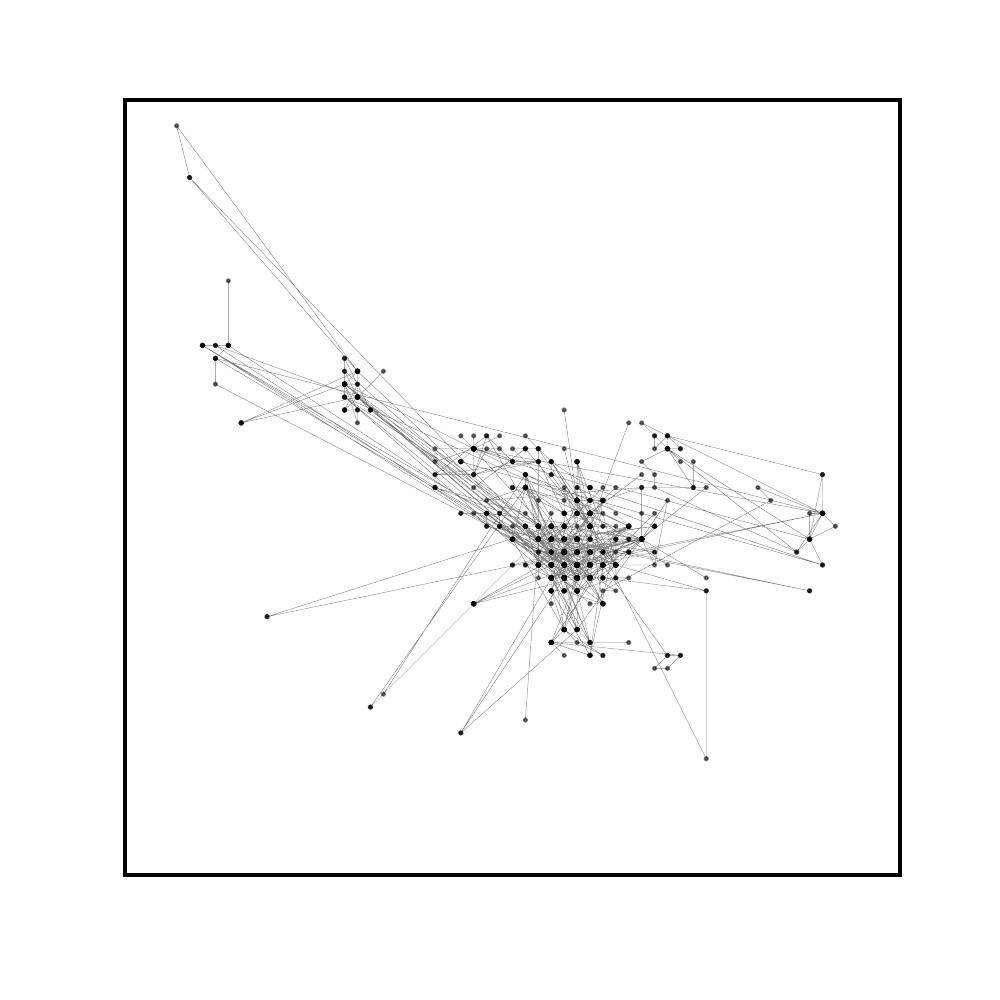}
	\label{subfig:recency}
	}
\caption{The darker is the color of the dot, the more the location has been visited. \ref{subfig:epr} In the Individual Mobility (IM) model proposed by Song \et, a user goes back to frequently visited locations with increasing probability, therefore the color of the visited locations is mostly black. Furthermore, most of the visited locations are very close to the initial location from where the user movement started. \ref{subfig:recency} In the recency model, a user can decide to go back to a recently visited location even though it has been visited only once or few times, therefore the visitation frequency is more evenly distributed. Furthermore, the spatial pattern shows several clusters further from the starting location, similarly to what happens with L\'evy flights.}
\label{fig:eprre}
\end{figure}

Although there is an extensive literature supporting the idea that human trajectories indeed follow reproducible scaling laws, the predictions of L\'evy flight and CTRW models systematically deviate from empirical results. Song \et \cite{song_2010_modelling} proposed two key mechanisms, unique to human trajectories, that were missing from existing models: \emph{Exploration} and \emph{Preferential Return} (see \sectionname\ref{sec:epr}). Another aspect that differs from previous study is the use of power law with exponential cutoff for the jump length and wait time (\figurename\ref{subfig:epr}).  Power law with cutoff does not have a closed-form expression. However, we can generate a random number drawn from a power law with parameters $x_{\text{min}}, \alpha$ and an exponential cutoff $K$ with the following process \cite{clauset_2009_power}: generate an exponentially distributed random number $x = x_{\text{min}} - \frac{1}{\lambda} \ln \left(1-r\right)$, where $r$ is a uniformly distributed number in the interval $[0,1)$. Then, accept or reject it with probability $p$ or $1 - p$ respectively, where $p = \left(\frac{x}{x_{\text{min}}}\right)^{-\alpha}$. Repeating the process until a number $x$ is accepted will guarantee that it will follow the desired distribution. This procedure is illustrated by Algorithm \ref{alg:pc}. Note that in the pseudo-code $x$ and $y$ are lists, therefore the operator $+$ has the meaning of concatenation of the two operands.
\begin{algorithm}
\caption{Power Law with Exponential Cutoff Random Number Generator}
\label{alg:pc}
\begin{algorithmic}[1]
\Require $\alpha > 1$
\Require $x_{\text{min}} \ge 1$
\Procedure{Power\textendash Law\textendash Cutoff}{$\alpha, x_{\text{min}, ~K}$}
        \State $x \gets \{\}$
        \State $y \gets \{\}$
        \State $n \gets 1$
        \State $\lambda \gets 1 / K$
       	\While {True}
        	\For {$i\gets 0, 10*n$} \Comment Generate exponentially distributed numbers
        		\State $y \gets y + \left\{ x_{\text{min}} - \left( \frac{1}{\lambda} \right) * \ln \left(1-\text{RANDOM(0,1)}\right) \right\}$
        	\EndFor
           	\State $y_{\text{temp}} \gets \{\}$
           	\For {$i\gets 0, 10*n$} \Comment Accept  with probability $p = \left(\frac{x}{x_{\text{min}}}\right)^{-\alpha}$
                \If {$\text{RANDOM(0,1)} < \left(\frac{y_i}{x_{\text{min}}}\right)^{-\alpha}:$}
                    \State $y_{\text{temp}} \gets y_{\text{temp}} + \left\{y_i\right\}$
                \EndIf
            \EndFor
           	\State $y \gets y_{\text{temp}}$
           	\State $x \gets x + y$
           	\State $q \gets \text{LENGTH}(x) - n$
           	\If {$q == 0$} 
               	\State break
			\EndIf
           	\If {$q > 0$} \Comment Make sure numbers are random
                \State $r \gets \{0, 1, 2, \cdots, \text{LENGTH}(x)\}$
               	\State $\text{SHUFFLE}(r)$

               	\State $x_{\text{temp}} \gets \{\}$
               	\For {$j \gets 0, \text{LENGTH}(x)$}
                    \If {$j \notin \{r_0, r_1, \cdots, r_q\}$}
                        \State $x_{\text{temp}} \gets x_{\text{temp}} + \{x_j\}$
                    \EndIf
                \EndFor
               	\State$x \gets x_{\text{temp}}$
               	\State break
			\EndIf            
            \If {$q<0$:} 
               	\State $y \gets \{\}$
            \EndIf
		\EndWhile            
		\State \Return $x_0$
\EndProcedure
\end{algorithmic}
\end{algorithm}
Finally, if we define $S$ as the set of previously visited locations, $F_i$ the number of times (frequency) that the location $i$ has been visited, and $\rho$ and $\gamma$ the parameters that control the probability $P_{\text{new}}$ of an exploration jump then the EPR random walk is implemented by Algorithm \ref{alg:epr}. Note that while the $x$ and $y$ coordinates vary continuously, the space is actually divided in discrete ``locations'', like Voronoi patches or squares of a grid. Therefore, different points in the plane may be part of the same location; to this purpose we use the function \texttt{CURRENT-LOCATION(x,y)} to obtain the location associated with the current coordinates of the agent.

\begin{algorithm}
\caption{Exploration-Preferential Return (EPR) Random Walk}
\label{alg:epr}
\begin{algorithmic}[1]
\Procedure{EPR}{N}
\State $x, y, n, S \gets 0$
	\While{$n < N$ }
    	\State $i \gets \text{CURRENT-LOCATION(x,y)}$
    	\If {$i \in S$}
        	\State $F_i \gets F_i + 1$
        \Else
        	\State $S \gets S + i$
            \State $F_i \gets 1$
        \EndIf
        
        \State $\theta \gets 0$
        \If {$RANDOM(0,1) < P_{\text{new}}$} \Comment Exploration Step
        	\State $\theta \gets \text{RANDOM}(0,360)$
			\State $r \gets \text{POWER-LAW-CUTOFF}(\alpha, x^{r}_{\text{min}}, K_{r})$

        \Else \Comment Frequency Based Return Step
        	\State $r, \theta \gets \text{FREQUENCY-RETURN()}$
        \EndIf
        \State $x \gets x + r \cos \theta$
		\State $y \gets y + r \sin \theta$
        
    	\State $\tau \gets \text{POWER-LAW-CUTOFF}(\beta, x^{\tau}_{\text{min}}, K_{\tau})$  \Comment Wait Phase
		\While{$\tau > 0$}
			\If {$\tau >= 1$}
				\State $\text{WAIT}(1)$
				\State $\tau \gets \tau - 1$
			\Else
				\State $\text{WAIT}(\tau)$
			\EndIf
		\EndWhile
		\State $n \gets n+1$
	\EndWhile
\EndProcedure
\Function {Frequency-Return}{}
	\State $tot \gets \sum_{j \in S} F_j$
    \State $throw \gets \text{RANDOM(0,tot)}$
    \State $partial \gets 0$
    \For {$j \in S$}
       	\State $partial \gets partial + F_j$
        \If{$partial < throw$}
           	\State $r \gets \text{EUCLID-DIST(i,j)}$
            \State $\theta \gets \arctan \left( \frac{j.y - i.y}{j.x - i.x} \right)$
            \State \Return $r,\theta$
        \EndIf
    \EndFor
\EndFunction
\end{algorithmic}
\end{algorithm}

\paragraph{Recency}
A further improvement over the EPR model has been introduced by Barbosa \et \cite{barbosa_2015_effect}. The main observation is that frequently visited locations are also recently visited locations (\figurename\ref{subfig:recency}). However, recently visited locations might not always be frequently visited, because it might be that a location has been visited for the first time (\sectionname\ref{sec:recency}). The algorithm to implement the recency mobility model is exactly the same as for EPR, but when deciding the for a return jump we can choose either a frequency based return with probability $\alpha$ or a recency based return with probability $1-\alpha$, where $\alpha$ is a parameter extracted from empirical data (Equation (\ref{eq:recency}), Algorithm \ref{alg:recency}).
\begin{algorithm}
\caption{Function to Choose a Recently Visited Location}
\label{alg:recency}
\begin{algorithmic}[1]
\Function {Recency-Return}{}    
    \State $k \gets \text{ROUND(POWER-LAW($\nu$))}$ \Comment select the rank from a zipfian law
	\While{$k \ge \text{SIZE($S$)}$}
    	\State $k \gets \text{ROUND(POWER-LAW($\nu$))}$
	\EndWhile
  	\State 	$S \gets \text{SORT($S$)}$ \Comment order locations according to the visiting time
    \State $j \gets S_k$
    \State $r \gets \text{EUCLID-DIST(i,j)}$
    \State $\theta \gets \arctan \left( \frac{j.y - i.y}{j.x - i.x} \right)$
    \State \Return $r,\theta$
\EndFunction
\end{algorithmic}
\end{algorithm}

\paragraph{Social Models}

It is intuitive to understand that the mobility of socially related individuals is correlated; for example friends often hang out and visit places together. Therefore, mobility models that include social network links and proximity based interactions were developed (See \sectionname\ref{sec:social}). Two prominent examples are the models described in \cite{grabowicz_2014_entangling} and \cite{toole_2015_coupling}, which can be seen as variations of the preferential return model (\sectionname\ref{sec:epr}), where the preferential return is governed by the social strength between the users instead of the visitation frequency of the locations. Given the previously presented building blocks and algorithms, it should be trivial at this point to implement these models, therefore the explicit algorithms are omitted.

In \cite{grabowicz_2014_entangling}, at every step each agent chooses to visit a social contact (i.e., another agent with whom the agent has a link) with probability $p_v$. Otherwise, it chooses a new location with the complementary probability $1 - p_v$ following a L\`evy-like flight. After the movement, the agent creates a new link with probability $p$ in the ``neighborhood'', that is the nearby agents, and probability $p_c$ with a random agent. An agent can easily keep track of the social links with the other agents by updating a list of the agent's ids it meets. 

In \cite{toole_2015_coupling}, the first part of the mobility model is identical to the preferential return model, where the agent returns to a previously visited location with probability $1 - \rho S^{-\gamma}$ or visit a new location with probability $\rho S^{-\gamma}$. However, it adds an additional ``social'' step. The actual choice of the location to visit, either new or returning, is made based on the social influence with probability $\alpha$ or, with probability $1 - \alpha$, just the individual preference (i.e., proportionally to the location visitation frequency). In case the location is chosen according to a social contact, the social contact $j$ is chosen with a probability directly proportional to the current mobility similarity between the two agents. The mobility similarity between two agents $i$ and $j$ is defined as the cosine similarity in the location visiting profiles:
\begin{displaymath}
cos \theta_{i,j} = \frac{\mathbf{v_i}\cdot\mathbf{v_j}}{|\mathbf{v_i}||\mathbf{v_j}|}
\end{displaymath}
where $\mathbf{v_i}$ and $\mathbf{v_j}$ represent the total visits made by the agents to each location in the simulation space.

\subsubsection{Population Mobility}
A simple way to formalize the flow of individuals from one location to another is to divide the area of interest into different zones labeled as $i = 1, \dots, n$ and to count the number of individuals going from location $i$ to location $j$, $\forall i$. These numbers $T_{ij}$ are the elements of the so called \emph{Origin-Destination} ($OD$) matrix (\sectionname\ref{sec:odmatrix})
\begin{equation}
OD(t) = 
\begin{bmatrix}
    T_{11} & T_{12} & \dots  & T_{1m} \\
    T_{21} & T_{22} & \dots  & T_{2m} \\
    \vdots & \vdots & \ddots & \vdots \\
    T_{n1} & T_{n2} & \dots  & T_{nm}
\end{bmatrix}
\end{equation}
Such matrix defines a directed and weighted network, and in the general case is time-dependent. Note that usually $n = m$ and $T_{ij} = 0, \forall i = j$.
The $OD$ matrix is very different from a ``segment'' measurement, where we can easily count the number of individuals going through a link of the transportation system under consideration (e.g., one of the flights of the airline network, a segment of road, etc.). In contrast, the $OD$ matrix is usually extremely difficult and costly to obtain and measure. There are obviously many factors which control the origin-destination matrix: land use, location of industries and residential areas, accessibility, etc.

In order to simulate the $OD$ matrix using the multi-agent modeling framework we need to define a process that drives the movement of the agents from one location to another. Let's consider a discrete-time agent simulation for which at every time tick $t$ we know the elements $T_{ij}$ of the matrix $OD(t)$. $T_{ij}$ can be considered as a frequency count, therefore we can transform the elements of the $OD$ matrix by computing the relative frequency $P_{ij}$:
\begin{displaymath}
P_{ij} = \frac{T_{ij}}{\sum_i \sum_j T_{ij}}
\end{displaymath}
$P_{ij}$ represents the probability of an agent to move from location $i$ to location $j$ at a specific instant in time $t$. If we want to run a simulation for an amount of time $\tau$, for each time instant $t$ the simulation algorithm normalizes the matrix $OD(t)$, then loop through each element $P_{ij}$ of the normalized matrix $OD$ and move an agent, \text{MOVE-AGENT(i,j)}, from $i$ to $j$ with the probability $P_{ij}$ (Algorithm \ref{alg:od}). This framework is general and applies to any mobility model that uses origin destination matrices such as the gravity models familiy (\sectionname\ref{sec:gravity}), intervening opportunities, and radiation model. However each model may be different in the way that the elements $T_{ij}$ are computed.
It is also interesting to note that if we represent the movement of the agents over e.g., a map layer, then the flux of agents over the connection between two locations should represent and estimate of the intensity of the traffic.

\begin{algorithm}
\caption{Origin Destination Matrix Agent Simulation}
\label{alg:od}
\begin{algorithmic}[1]
	\State $t \gets 0$
	\While{$t < \tau $}
		\State $OD \gets \text{NORMALIZE($OD(t)$)}$	\Comment normalize the $OD$ matrix
		\For {$i \gets 1, n$}	
    		\For {$j \gets 1, n$}
    			\If {$\text{RANDOM(0,1)} < P_{ij}$}
            		\State \text{MOVE-AGENT(i,j)}
            	\EndIf
        	\EndFor
    	\EndFor
        \State $t \gets t + 1$
    \EndWhile
\Function {Normalize}{OD}
	\State $\text{totalSum} \gets 0$
    \For {$i \gets 1, n$}
    	\For {$j \gets 1, n$}
    		\State $\text{totalSum} \gets \text{totalSum} + T_{ij}$
        \EndFor
    \EndFor
    \For {$i \gets 1, n$}
    	\For {$j \gets 1, n$}
    		\State $T_{ij} \gets T_{ij} / \text{totalSum}$
        \EndFor
    \EndFor
    \State \Return $OD$
\EndFunction
\end{algorithmic}
\end{algorithm}

\newpage

\end{document}